\documentclass[final,1p,times]{elsarticle}

\usepackage{bm}
\usepackage{color}
\usepackage{array}
\usepackage{graphicx}
\usepackage{multirow}
\usepackage{booktabs}
\usepackage{fancyvrb}
\usepackage{upquote}

\usepackage{geometry}

\newcolumntype{P}[1]{>{\centering\arraybackslash}p{#1}}
\newcolumntype{M}[1]{>{\centering\arraybackslash}m{#1}}
\def\nn{\nonumber}
\newcommand{\be}{\begin{equation}}
\newcommand{\ee}{\end{equation}}
\newcommand{\een}{\end{subequations}}
\newcommand{\ben}{\begin{subequations}}
\newcommand{\beq}{\begin{eqalignno}}
\newcommand{\eeq}{\end{eqalignno}}

\newcommand{\lsim}{\mathrel{\mathop{\kern 0pt \rlap
      {\raise.2ex\hbox{$<$}}}\lower.9ex\hbox{\kern-.190em $ \sim$}}}
\newcommand{\gsim}{\mathrel{\mathop{\kern 0pt
      \rlap{\raise.2ex\hbox{$>$}}}\lower.9ex\hbox{\kern-.190em $\sim$}}}

\newcommand{\CO}{\mathcal{O}}

\newcommand{\erf}{\mbox{erf}}

\newcommand{\todo}[1]{{\color{red} \ifmmode\else[todo]\fi #1}}

\usepackage{amssymb}

\journal{Astroparticle Physics}

\begin{document}

\begin{frontmatter}



\title{Present and projected sensitivities of Dark Matter direct
  detection experiments to effective WIMP-nucleus couplings}


\author{Sunghyun Kang}
\ead{francis735@naver.com}
\author{Stefano Scopel}
\ead{scopel@sogang.ac.kr}
\author{Gaurav Tomar}
\ead{tomar@sogang.ac.kr}
\author{Jong--Hyun Yoon}
\ead{jyoon@sogang.ac.kr}
\address{Department of Physics, Sogang University, 
Seoul, Korea, 121-742}

\begin{abstract}
  Assuming for Weakly Interacting Massive
  Particles (WIMPs) a Maxwellian velocity distribution in the Galaxy
  we explore in a systematic way the relative sensitivity of an
  extensive set of existing and projected Dark Matter (DM) direct
  detection experiments to each of the 14 couplings that parameterize
  the most general non-relativistic (NR) effective Hamiltonian allowed
  by Galilean invariance for a contact interaction driving the elastic
  scattering off nuclei of WIMPs of spin 1/2. We perform our analysis
  in terms of two free parameters: the WIMP mass $m_{\chi}$ and the
  ratio between the WIMP-neutron and the WIMP-proton couplings
  $c^n/c^p$. We include the modified signal spectral shape due to
  non--standard interactions when it is needed in the determination of
  the bound, such as in the case of background subtraction or of the
  application of the optimal--interval method. For each coupling, in
  the $m_{\chi}$--$c^n/c^p$ plane we provide contour plots of the most
  stringent 90\% C.L. bound on the WIMP--nucleon cross section and
  show the experiment providing it.  We also introduce {\verb NRDD_constraints },
  a simple interpolating code written in Python
  that allows to obtain the numerical value of the bound as a function
  of the WIMP mass $m_{\chi}$ and of the coupling ratio $c^n/c^p$ for
  each NR coupling.  We find that 9 experiments out of the 14 present
  Dark Matter searches considered in our analysis provide the most
  stringent bound on some of the effective couplings for a given
  choice of $(m_{\chi},c^n/c^p)$: this is evidence of the
  complementarity of different target nuclei and/or different
  combinations of count--rates and energy thresholds when the search
  of DM is extended to a wide range of possible interactions.
\end{abstract}


\end{frontmatter}



\section{Introduction}
\label{sec:introduction}

Up to 27\% of the total mass density of the Universe~\cite{planck} and
more than 90\% of the halo of our Galaxy are believed to be composed
of Dark Matter (DM). The properties of such invisible component are
yet unknown since DM has been only observed through gravity so
far. However, to comply with the Cold Dark Matter (CDM) paradigm of
Galaxy formation and to be a viable thermal relic, in one of its most
popular scenarios DM is believed to be composed of Weakly Interacting
Massive Particles (WIMPs) with a mass in the GeV-TeV range and
weak--type interactions with ordinary matter. Such small but non
vanishing interactions can drive WIMP scatterings off nuclear targets,
and the measurement of the ensuing nuclear recoils in low--background
detectors (direct detection) represents the most straightforward way
to detect them.  Indeed, a large worldwide effort is currently under
way to observe WIMP-nuclear scatterings, but, with the exception of
the DAMA collaboration \cite{dama_1998, dama_2008,dama_2010,dama_2018}
that has been observing for a long time an excess compatible to the
annual modulation of a DM signal, many other experiments using
different nuclear targets and various background--subtraction
techniques have failed to observe any WIMP signal so
far~\cite{xenon_2018, panda_2017, kims_2014, cdmslite_2017,
  super_cdms_2017, coupp, picasso, pico60_2015, pico60, cresst_II,
  cdex, damic, ds50}.

The expected WIMP interaction scale happens to fit nicely to what is
also believed to be the cut--off scale of the Standard Model (SM),
beyond which new physics is expected to come on shell in order to
stabilize the Higgs vacuum, and indeed most of the explicit
ultraviolet completions of the SM contain WIMP exotic states that are
viable DM candidates and for which detailed predictions for
WIMP--nuclear scattering can be worked out. Crucially, this allows to
determine how the WIMP interacts with different targets, and to
compare in this way the sensitivity of different detectors to a given
WIMP candidate, with the goal of choosing the most effective detection
strategy.

A typical example of this approach is the direct search for the
Supersymmetric neutralino, whose cross section off nuclei is usually
driven either by a Higgs-- or squark--exchange propagator, leading to
a Spin Independent (SI) interaction that is the same for protons and
neutrons (isoscalar) and scales with the square of the atomic mass
number:

\begin{equation}
  \sigma_{\chi N}\propto \left [  c^p Z + (A-Z) c^n\right ]^2,
\label{eq:si}
\end{equation}

\noindent with $A$ the nuclear mass number, $Z$ the nuclear charge and
$c^{p,n}$ the WIMP couplings to protons and neutrons, with $c^n$=$c^p$.

If indeed the neutralino--nucleus interaction amplitude is given by
Eq.(\ref{eq:si}) the corresponding cross section is non-vanishing off
any target and highly enhanced for heavy nuclei. Since expected
signals are very low due to the very tight present constraints, very
large exposures are required, as well as extremely low background
levels that nowadays can only be achieved using discrimination
techniques to distinguish nuclear recoils from natural
radioactivity. This has naturally led to a very strong drive in the
physics community to develop large--mass dual--phase (liquid and
gaseous) xenon detectors \cite{xenon_2018,lux,LZ} that indeed already
provide today and are expected to provide in the future the most
stringent bounds on this type of interaction.

Notice, however, that the current leading position of xenon detectors
in the direct search of DM rests on the specific assumption of
Eq. (\ref{eq:si}) for the scaling law of the WIMP--nucleon cross
section with different targets. Two trivial counter--examples that
show how there are viable DM--nucleus interactions for which xenon
might not be the optimal target to detect DM are provided by the case
of isospin--violating models
\cite{isospin_violation1,isospin_violation2} where the ratio between
the WIMP--proton and the WIMP--neutron couplings is tuned to
$c^n/c^p\simeq Z/(Z-A)\simeq -0.7$ to suppress the WIMP--xenon
interaction amplitude, and by particles such as the Higgsino or a
Majorana neutrino that couple to ordinary matter through a $Z$--boson
propagator leading to a Spin--Dependent (SD) WIMP--nucleon
interaction:

\begin{equation}
{\cal L}_{int}\ni
c^p\vec{S}_{\chi}\cdot \vec{S}_p+c^n\vec{S}_{\chi}\cdot \vec{S}_n,
\label{eq:sd}
\end{equation}

\noindent where $\vec{S}_{\chi}$, $\vec{S}_n$ and $\vec{S}_p$ are the
spins of the WIMP, the neutron and the proton, respectively. 

In the case of an isospin--violating SI interaction, since the $A/Z$
ratio of all stable nuclear targets, including those used in direct
detection, are not too far from unity, tuning $c^n/c^p$ to suppress
the response of xenon inevitably leads also to a suppression of the
WIMP scattering rates off all other targets. As a consequence, this
would not only imply a different hierarchy among the sensitivity of
different detector materials, but also an overall loss of sensitivity
of present and future direct DM searches to the physics beyond the SM
underlying DM.  On the other hand, in the case of a spin--dependent
interaction the relative sensitivity of different targets to the
interaction (Eq.\ref{eq:sd}) is completely different from
(Eq.\ref{eq:si}), since nucleon spins inside nuclei are not coherently
enhanced. This implies that, at variance with the SI interaction, the
SD one has no preference for heavy targets, so that the leading edge
of xenon detectors compared to other targets is in general
reduced. Moreover, isotopes with spin correspond to only about 47\% of
the overall target number in natural xenon, and since they have an
even number of protons one has $\sum_p \vec{S}_p\rightarrow$0 implying
a strongly suppressed sensitivity to the $c^p$ coupling. As a
consequence, the sensitivity of proton--odd targets such as those in
fluorine detectors can be better than xenon when $c^n\ll c^p$.

In the present paper we wish to extend the discussion above, providing
an assessment of the overall present and future sensitivity of direct
detection experiments to WIMPs with the {\it most general} scaling law
for WIMP--nucleus scattering, besides the SI and the SD one. In
particular this task is achievable without fixing a specific
high--energy extension of the SM since the WIMP--nucleus cross section
can be parameterized in terms of the most general non--relativistic
effective theory complying with Galilean symmetry, including a
possible explicit dependence of the scattering cross section on the
transferred momentum and of the WIMP incoming
velocity~\cite{dobrescu_eft,reece_eft,haxton1,haxton2}.  This
approach, in which scales related to the spontaneous breaking of the
chiral symmetry of QCD are integrated out, is alternative to
incorporating the QCD constraints from chiral
symmetry~\cite{chiral_eft}. In our analysis, we will adopt for the
velocity distribution $f(\vec{v})$ of the incoming WIMPs a standard
thermalized non--relativistic gas described by a Maxwellian
distribution. In particular, compared to other phenomenological
analyses existing in the literature on WIMP--nucleus effective
interactions
\cite{Catena_Gondolo_global_limits,Catena_Gondolo_global_fits} in the
present paper we wish to calculate updated constraints on the
WIMP-nucleon cross section in the non--relativistic effective theory
and provide a comparative discussion of the reach of different
experiments to the various effective operators in order to show their
complementarity in a transparent way.

Our approach will be relatively straightforward: we will consider an
extensive list of present (XENON1T~\cite{xenon_2018},
PANDAX-II~\cite{panda_2017}, KIMS~\cite{kims_2014},
CDMSlite~\cite{cdmslite_2017}, SuperCDMS~\cite{super_cdms_2017},
COUPP~\cite{coupp}, PICASSO~\cite{picasso}, PICO-60 (using a CF$_3$I
target ~\cite{pico60_2015} and a C$_3$F$_8$ one \cite{pico60})
CRESST-II \cite{cresst_II,cresst_II_ancillary}, DAMA (modulation data)
\cite{dama_1998, dama_2008,dama_2010,dama_2018}, DAMA0 (average count
rate) \cite{damaz}), CDEX \cite{cdex} and DarkSide--50 (DS50)
\cite{ds50}) and future DM direct detection experiments (LUX--ZEPLIN
(LZ)~\cite{LZ}, PICO-500 \cite{pico500} and COSINUS \cite{cosinus})
and the most general WIMP--nucleus effective Lagrangian for a WIMP
particle of spin 1/2 scattering elastically off nuclei. Then,
systematically assuming dominance of one of the possible interaction
terms, we will provide for each of them a two--dimensional plot where
the contours of the most stringent 90\% C.L. upper bounds\footnote{For
  each experiment we apply a statistical treatment similar to that
  used for the published result, see \ref{app:exp}.}  to an
appropriately defined WIMP--nucleon effective cross section
$\sigma_{\cal N}$ (that is related to the usual one in the case of
interactions with a non--vanishing long--range asymptotic component
such as the usual SI and SD cases) are shown as a function of the two
parameters $m_{\chi}$ (WIMP mass), and $c^n/c^p$. Moreover, in the
same $m_{\chi}$--$c^n/c^p$ plots, regions depicted with a different
color will allow to determine which experiment provides the most
stringent constraint for that particular choice of parameters.  To
summarize our results for each coupling we will then provide as a
function of the WIMP mass the maximal range spanned by the most
constraining 90\% C.L. exclusion plot on the WIMP--nucleon cross
section as a function of the WIMP mass when the ratio $c^n/c^p$ is
varied.  In~\ref{app:program} we will also introduce {\verb NRDD_constraints },
a simple interpolating code written in Python
that allows to obtain the numerical value of the most constraining
limit on the effective cross section defined in
Eq.(\ref{eq:conventional_sigma_nucleon}) as a function of the WIMP
mass $m_{\chi}$ and of the coupling ratio $c^n/c^p$ for each NR
coupling.

In the present analysis we discuss one of the NR couplings at a time
because they are the most general building blocks of the low--energy
limit of any ultraviolet theory, so that an understanding of the
behaviour of such couplings is crucial for the interpretation of more
general scenarios containing the sum of several NR
operators\footnote{Nevertheless, it is always possible to conceive a
  linear combination of relativistic operators leading to a single NR
  operator, since the number of the former is larger than that of the
  latter, although this might require a tuning of the
  couplings. }. However, our results can be used also to estimate an
approximate upper bound on the cross section in the case of the
presence of more than one NR operator. The procedure to do so is
discussed in Section \ref{sec:mixing}. Our analysis is somewhat
complementary to that of Ref.~\cite{Schneck_eft}, where present WIMP
direct detection experimental sensitivities are discussed for a limited
number of non--relativistic operators and nuclear targets, but
interferences among different operators are included in the discussion.

The paper is organized as follows. In Section \ref{sec:eft} we
summarize the non--relativistic Effective Field Theory (EFT) approach
of Refs.\cite{haxton1,haxton2} and the formulas we use to calculate
expected rates for WIMP--nucleus scattering; Section
\ref{sec:analysis} is devoted to our quantitative analysis; in Section
\ref{sec:mixing} we show how our results can be applied to the case of
more than one NR operator. We will provide our conclusions in Section
\ref{sec:conclusions}. In \ref{app:wimp_eft} we provide for
completeness the WIMP response functions for the non--relativistic
effective theory while in \ref{app:exp} we provide the details of each
experiment included in the analysis.
\ref{app:nuclear_response_functions} describes our treatment of the
nuclear response functions for those isotopes for which a full
calculation is not available in the literature. Finally,
in~\ref{app:program} we introduce {\verb NRDD_constraints }, a simple
interpolating code written in Python that allows to retrieve the
numerical value of the limits on the effective WIMP--nucleon cross
section discussed in Section \ref{sec:analysis} and whose contour
plots are shown in Figs.~\ref{fig:c1_plane}--\ref{fig:c15_plane}.

\section{Summary of WIMP rates in non--relativistic effective models}
\label{sec:eft}

Making use of the non--relativistic EFT approach of
Refs. \cite{haxton1,haxton2} the most general Hamiltonian density
describing the WIMP--nucleus interaction can be written as:

\begin{eqnarray}
{\bf\mathcal{H}}({\bf{r}})&=& \sum_{\tau=0,1} \sum_{j=1}^{15} c_j^{\tau} \mathcal{O}_{j}({\bf{r}}) \, t^{\tau} ,
\label{eq:H}
\end{eqnarray}

\noindent where:

\begin{eqnarray}
  \CO_1 &=& 1_\chi 1_N; \;\;\;\; \CO_2 = (v^\perp)^2; \;\;\;\;  \CO_3 = i \vec{S}_N \cdot ({\vec{q} \over m_N} \times \vec{v}^\perp); \nn\\
  \CO_4 &=& \vec{S}_\chi \cdot \vec{S}_N;\;\;\;\; \CO_5 = i \vec{S}_\chi \cdot ({\vec{q} \over m_N} \times \vec{v}^\perp);\;\;\;\; \CO_6=
  (\vec{S}_\chi \cdot {\vec{q} \over m_N}) (\vec{S}_N \cdot {\vec{q} \over m_N}) ;\nn \\
  \CO_7 &=& \vec{S}_N \cdot \vec{v}^\perp;\;\;\;\;\CO_8 = \vec{S}_\chi \cdot \vec{v}^\perp;\;\;\;\;\CO_9 = i \vec{S}_\chi \cdot (\vec{S}_N \times {\vec{q} \over m_N}); \nn\\
  \CO_{10} &=& i \vec{S}_N \cdot {\vec{q} \over m_N};\;\;\;\;\CO_{11} = i \vec{S}_\chi \cdot {\vec{q} \over m_N};\;\;\;\;\CO_{12} = \vec{S}_\chi \cdot (\vec{S}_N \times \vec{v}^\perp); \nn\\
  \CO_{13} &=&i (\vec{S}_\chi \cdot \vec{v}^\perp  ) (  \vec{S}_N \cdot {\vec{q} \over m_N});\;\;\;\;\CO_{14} = i ( \vec{S}_\chi \cdot {\vec{q} \over m_N})(  \vec{S}_N \cdot \vec{v}^\perp );  \nn\\
  \CO_{15} &=& - ( \vec{S}_\chi \cdot {\vec{q} \over m_N}) ((\vec{S}_N \times \vec{v}^\perp) \cdot {\vec{q} \over m_N}).
\label{eq:ops}
\end{eqnarray}

\noindent In the above equation $1_{\chi N}$ is the identity operator,
$\vec{q}$ is the transferred momentum, $\vec{S}_{\chi}$ and
$\vec{S}_{N}$ are the WIMP and nucleon spins, respectively, while
$\vec{v}^\perp = \vec{v} + \frac{\vec{q}}{2\mu_{\chi {\cal N}}}$ (with
$\mu_{\chi {\cal N}}$ the WIMP--nucleon reduced mass) is the relative
transverse velocity operator satisfying $\vec{v}^{\perp}\cdot
\vec{q}=0$. Following Refs.\cite{haxton1,haxton2} in the following we
will not include the operator ${\cal O}_2$ in our analysis. For a
nuclear target $T$ the quantity $(v^{\perp}_T)^2 \equiv
|\vec{v}^{\perp}_T|^2$ can also be written as:

\begin{equation}
(v^{\perp}_T)^2=v^2_T-v_{min}^2.
\label{eq:v_perp}
\end{equation}

\noindent where:

\begin{equation}
v_{min}^2=\frac{q^2}{4 \mu_{T}^2}=\frac{m_T E_R}{2 \mu_{T}^2},
\label{eq:vmin}
\end{equation}

\noindent represents the minimal incoming WIMP speed required to
impart the nuclear recoil energy $E_R$, while $v_T\equiv|\vec{v}_T|$
is the WIMP speed in the reference frame of the nuclear center of
mass, $m_T$ the nuclear mass and $\mu_{T}$ the WIMP--nucleus reduced
mass. Moreover $t^0=1$, $t^1=\tau_3$
denote the $2\times2$ identity and third Pauli matrix in isospin
space, respectively, and the isoscalar and isovector (dimension -2)
coupling constants $c^0_j$ and $c^{1}_j$, are related to those to
protons and neutrons $c^{p}_j$ and $c^{n}_j$ by
$c^{p}_j=(c^{0}_j+c^{1}_j)$ and $c^{n}_j=(c^{0}_j-c^{1}_j)$.

In the following we will only consider a contact effective interaction
between the WIMP and the nucleus, i.e., we will assume the
coefficients $c_j^{\tau}$ as independent on the transferred momentum
$q$. However when the latter is comparable to the pion mass a pole is
known to arise in the case of pseudoscalar and axial interactions
\cite{bishara}.  This may affect our estimation of the sensitivity by
less than an order of magnitude for operators ${\it O}_{6}$ and ${\it
  O}_{10}$ when the WIMP and the target mass are heavy (specifically,
for xenon in XENON1T and PANDAX--II and for iodine in PICO-60). Since
such effect depends on the particular relativistic model the NR theory
descends from~\cite{sogang_eft_rev} and its impact is anyway limited,
we have neglected it in our analysis.

The expected rate in a given visible energy bin $E_1^{\prime}\le
E^{\prime}\le E_2^{\prime}$ of a direct detection experiment is given
by:

\begin{eqnarray}
R_{[E_1^{\prime},E_2^{\prime}]}&=&M\mbox{T}\int_{E_1^{\prime}}^{E_2^{\prime}}\frac{dR}{d
  E^{\prime}}\, dE^{\prime}, \label{eq:start}\\
 \frac{dR}{d E^{\prime}}&=&\sum_T \int_0^{\infty} \frac{dR_{\chi T}}{dE_{ee}}{\cal
   G}_T(E^{\prime},E_{ee})\epsilon(E^{\prime})\label{eq:start2}\,d E_{ee}, \\
E_{ee}&=&q(E_R) E_R \label{eq:start3},
\end{eqnarray}

\noindent with $\epsilon(E^{\prime})\le 1$ the experimental
efficiency/acceptance. In the equations above $E_R$ is the recoil
energy deposited in the scattering process (indicated in keVnr), while
$E_{ee}$ (indicated in keVee) is the fraction of $E_R$ that goes into
the experimentally detected process (ionization, scintillation, heat)
and $q(E_R)$ is the quenching factor, ${\cal
  G_T}(E^{\prime},E_{ee}=q(E_R)E_R)$ is the probability that the
visible energy $E^{\prime}$ is detected when a WIMP has scattered off
an isotope $T$ in the detector target with recoil energy $E_R$, $M$ is
the fiducial mass of the detector and T the live--time of the data
taking. For a given recoil energy imparted to the target the
differential rate for the WIMP--nucleus scattering process is given
by:

\be
\frac{d R_{\chi T}}{d E_R}(t)=\sum_T N_T\frac{\rho_{\mbox{\tiny WIMP}}}{m_{\chi}}\int_{v_{min}}d^3 v_T f(\vec{v}_T,t) v_T \frac{d\sigma_T}{d E_R},
\label{eq:dr_de}
\ee

\noindent where $\rho_{\mbox{\tiny WIMP}}$ is the local WIMP mass density in the
neighborhood of the Sun, $N_T$ the number of the nuclear targets of
species $T$ in the detector (the sum over $T$ applies in the case of
more than one nuclear isotope), while

\be
\frac{d\sigma_T}{d E_R}=\frac{2 m_T}{4\pi v_T^2}\left [ \frac{1}{2 j_{\chi}+1} \frac{1}{2 j_{T}+1}|\mathcal{M}_T|^2 \right ],
\label{eq:dsigma_de}
\ee

\noindent and, assuming that the nuclear interaction is the sum of the
interactions of the WIMPs with the individual nucleons in the nucleus:

\begin{equation}
  \frac{1}{2 j_{\chi}+1} \frac{1}{2 j_{T}+1}|\mathcal{M}_T|^2=
  \frac{4\pi}{2 j_{T}+1} \sum_{\tau=0,1}\sum_{\tau^{\prime}=0,1}\sum_{k} R_k^{\tau\tau^{\prime}}\left [c^{\tau}_j,(v^{\perp}_T)^2,\frac{q^2}{m_N^2}\right ] W_{T k}^{\tau\tau^{\prime}}(y).
\label{eq:squared_amplitude}
\end{equation}

\noindent In the above expression $j_{\chi}$ and $j_{T}$ are the WIMP
and the target nucleus spins, respectively, $q=|\vec{q}|$ while the
$R_k^{\tau\tau^{\prime}}$'s are WIMP response functions (that we
report for completeness in Eq.(\ref{eq:wimp_response_functions}))
which depend on the couplings $c^{\tau}_j$ as well as the transferred
momentum $\vec{q}$ and $(v^{\perp}_T)^2$. In equation
(\ref{eq:squared_amplitude}) the $W^{\tau\tau^{\prime}}_{T k}(y)$'s
are nuclear response functions and the index $k$ represents different
effective nuclear operators, which, crucially, under the assumption
that the nuclear ground state is an approximate eigenstate of $P$ and
$CP$, can be at most eight: following the notation in
\cite{haxton1,haxton2}, $k$=$M$, $\Phi^{\prime\prime}$,
$\Phi^{\prime\prime}M$, $\tilde{\Phi}^{\prime}$,
$\Sigma^{\prime\prime}$, $\Sigma^{\prime}$,
$\Delta$,$\Delta\Sigma^{\prime}$. The $W^{\tau\tau^{\prime}}_{T
  k}(y)$'s are function of $y\equiv (qb/2)^2$, where $b$ is the size
of the nucleus. For the target nuclei $T$ used in most direct
detection experiments the functions $W^{\tau\tau^{\prime}}_{T k}(y)$,
calculated using nuclear shell models, have been provided in
Refs.~\cite{haxton2,catena} under the assumption that the dark matter
particle couples to the nucleus through local one--body interactions
with the nucleons. In our analysis we do not include two--body
effects~\cite{two_body_1,two_body_2} which are only available for a
few isotopes and can be important when the one--body contribution is
suppressed. 

In the present paper, we will systematically consider the possibility
that one of the couplings $c_{j}$ dominates in the effective
Hamiltonian of Eq. (\ref{eq:H}). In this case it is possible to
factorize a term $|c_j^p|^2$ from the squared amplitude of
Eq.(\ref{eq:squared_amplitude}) and express it in terms of the {\it
  effective} WIMP--proton cross section\footnote{With the
  definition of Eq.(\ref{eq:conventional_sigma}) the WIMP--proton SI
  cross section is equal to $\sigma_p$, and the SD WIMP--proton cross
  section to 3/16 $\sigma_p$.}:

\begin{equation}
\sigma_p=(c_j^p)^2\frac{\mu_{\chi{\cal N}}^2}{\pi},
  \label{eq:conventional_sigma}
\end{equation}

\noindent (with $\mu_{\chi{\cal N}}$ the WIMP--nucleon reduced mass)
and the ratio $r\equiv c_j^n/c_j^p$. It is worth pointing out here
that among the generalized nuclear response functions arising from the
effective Hamiltonian of Eq. (\ref{eq:H}) only the ones corresponding to $M$
(SI interaction), $\Sigma^{\prime\prime}$ and $\Sigma^{\prime}$ (both
related to the standard spin--dependent interaction) do not vanish for
$q\rightarrow$0, and so allow to interpret $\sigma_p$ in terms of a
long--distance, point--like cross section. In the case of the other
interactions $\Phi^{\prime\prime}$, $\Phi^{\prime\prime}M$,
$\tilde{\Phi}^{\prime}$, $\Delta$ and $\Delta\Sigma^{\prime}$ the
quantity $\sigma_p$ is just a convenient alternative to directly
parameterizing the interaction in terms of the $c_j^p$ coupling.

Finally, $f(\vec{v}_T)$ is the WIMP velocity distribution, for which
we assume a standard isotropic Maxwellian at rest in the Galactic rest
frame truncated at the escape velocity $u_{esc}$, and boosted to the
Lab frame by the velocity of the Earth. So for the former we assume:

\begin{eqnarray}
  f(\vec{v}_T,t)&=&N\left(\frac{3}{ 2\pi v_{rms}^2}\right )^{3/2}
  e^{-\frac{3|\vec{v}_T+\vec{v}_E|^2}{2 v_{rms}^2}}\Theta(u_{esc}-|\vec{v}_T+\vec{v}_E(t)|),\\
  N&=& \left [ \erf(z)-\frac{2}{\sqrt{\pi}}z e^{-z^2}\right ]^{-1},  
  \label{eq:maxwellian}
  \end{eqnarray}

\noindent with $z=3 u_{esc}^2/(2 v_{rms}^2)$. In the isothermal sphere
model hydrothermal equilibrium between the WIMP gas pressure and
gravity is assumed, leading to $v_{rms}$=$\sqrt{3/2}v_0$ with $v_0$
the galactic rotational velocity.

With the exception of DAMA, all the experiments included in our
analysis are sensitive to the time average of the expected rate for
which $<v_E>$=$v_{\odot}$ and $v_{\odot}$=$v_0$+12 (accounting for a
peculiar component of the solar system with respect to the galactic
rotation).  In the case of DAMA, the yearly modulation effect is due to
the time dependence of the Earth's speed with respect to the Galactic
frame, given by:

\begin{equation}
|\vec{v}_E(t)|=v_{\odot}+v_{orb}\cos\gamma \cos\left [\frac{2\pi}{T_0}(t-t_0)
  \right ],
  \end{equation}

\noindent where $\cos\gamma\simeq$0.49 accounts for the inclination of
the ecliptic plane with respect to the Galactic plane, $T_0$=1 year
and $v_{orb}$=2$\pi r_{\oplus}/(T_0)\simeq$ 29 km/sec ($r_{\oplus}$=1
AU neglecting the small eccentricity of the Earth's orbit around the
Sun).

In our analysis for the two parameters $v_0$ and $u_{esc}$ we take
$v_0$=220 km/sec \cite{v0_koposov} and $u_{esc}$=550 km/sec
\cite{vesc_2014}. Our choice of parameters corresponds to a WIMP
escape velocity in the lab rest frame $v_{esc}^{lab}\simeq$ 782 km/s.
To make contact with other analyses, for the dark matter density in
the neighborhood of the Sun we use $\rho_{\mbox{\tiny WIMP}}$=0.3,
which is a standard value commonly adopted by experimental
collaborations, although observations point to the slightly higher
value $\rho_{\mbox{\tiny WIMP}}$=0.43 \cite{rho_DM_salucci_1,
  rho_DM_salucci_2}. Notice that direct detection experiments are only
sensitive to the product $\rho_{\mbox{\tiny WIMP}}\sigma_p$, so the
results of the next Section can be easily rescaled with
$\rho_{\mbox{\tiny WIMP}}$.

\section{Analysis}
\label{sec:analysis}

\begin{figure}
\begin{center}
\includegraphics[width=0.49\columnwidth]{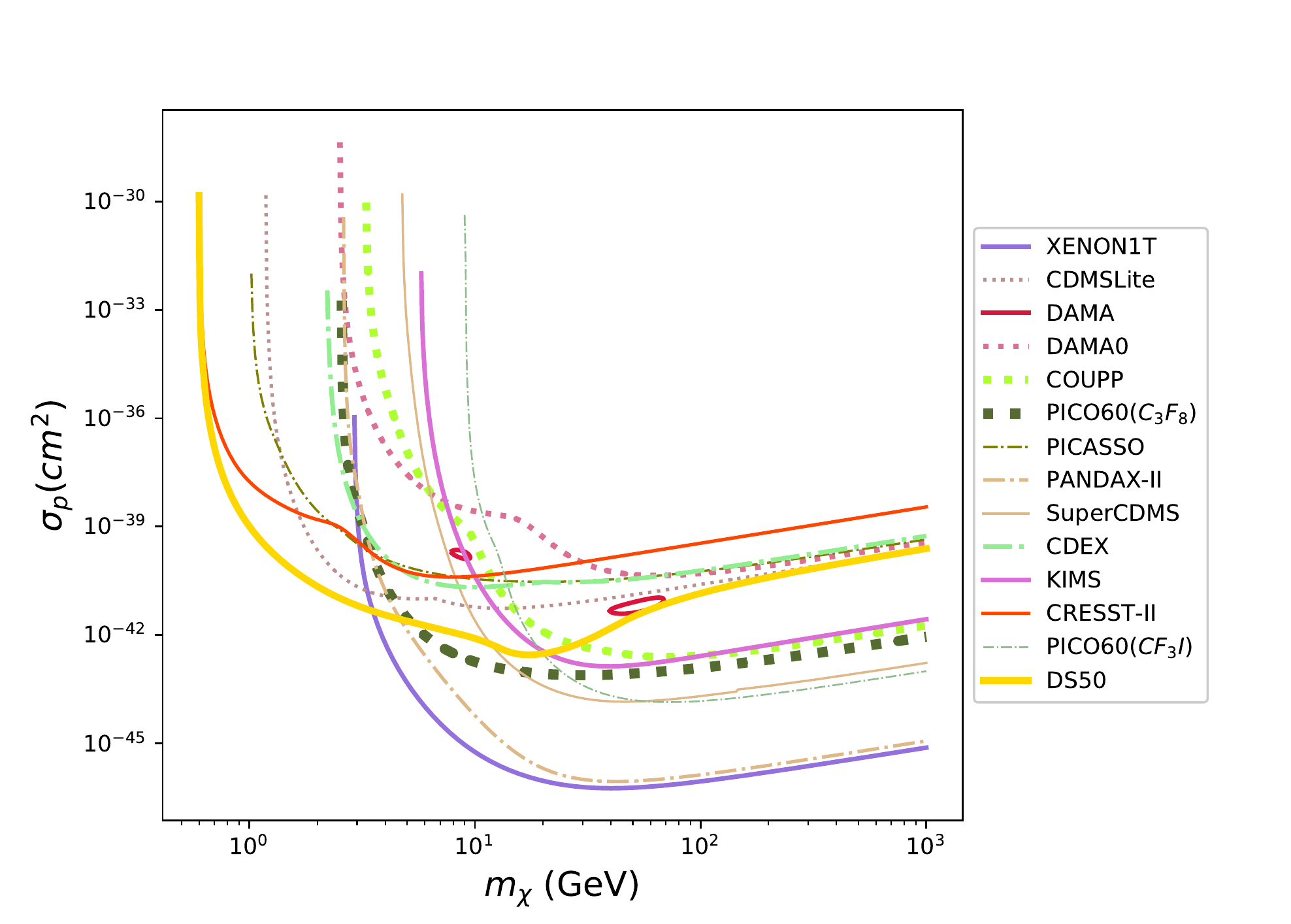}
\includegraphics[width=0.49\columnwidth]{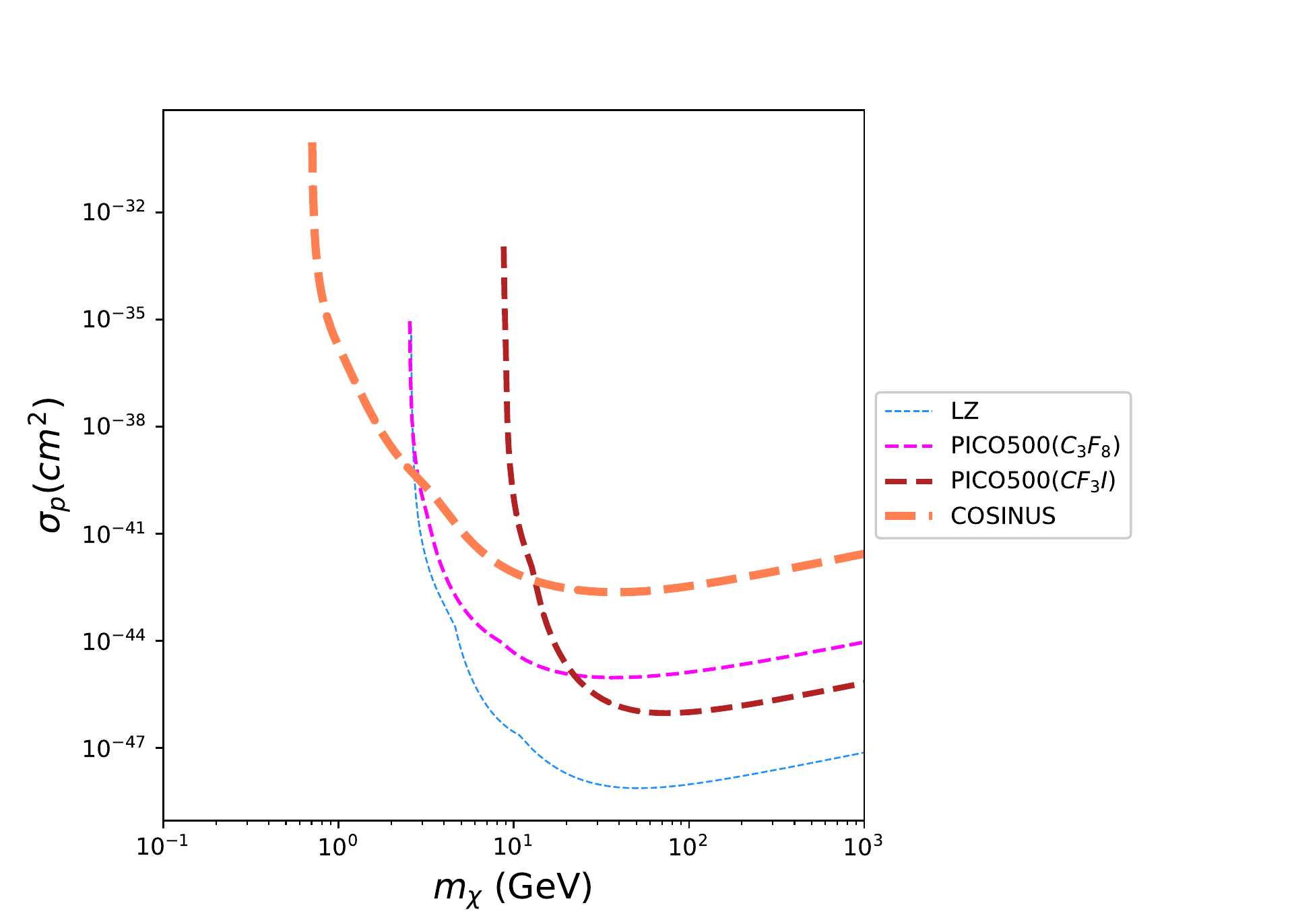}
\end{center}
\caption{Current (left) and future (right) 90\% C.L. exclusion plots
  to the effective WIMP--proton cross section $\sigma_p$ of
  Eq. (\ref{eq:conventional_sigma}) for the SI interaction of
  Eq. (\ref{eq:si}) corresponding to the ${\cal O}_1$ operator in
  Eq.(\ref{eq:H}) for the isoscalar case $c_1^p$=$c_1^n$. The figure
  shows the constraints from the full set of experiments that we
  include in our analysis, which consists in the latest available data
  from 14 existing DM searches, and the estimated future sensitivity
  of 4 projected ones (LZ, PICO-500 (C$_3$F$_8$), PICO-500 (CF$_3$I)
  and COSINUS). The closed solid (red) contour represents the 5--sigma
  DAMA modulation amplitude region, while we indicate with DAMA0 the
  upper bound from the DAMA average count--rate. Notice that after the
  release of the DAMA/LIBRA-phase2 result~\cite{dama_2018} a
  spin--independent isoscalar ($c^n/c^p$=1) interaction does not
  provide anymore a good fit to the modulation effect, while it still
  does for different values of $c^n/c^p$ and for other effective
  couplings \cite{dama_2018_sogang}.}
\label{fig:c1_summary}
\end{figure}

The current 90\% C.L. exclusion plots to the effective WIMP--proton
cross section $\sigma_p$ of Eq. (\ref{eq:conventional_sigma}) for the
SI interaction of Eq.(\ref{eq:si}) (corresponding to the ${\cal O}_1$
operator in Eq.(\ref{eq:H})) are shown for the isoscalar case
$c_1^p$=$c_1^n$ and for the full set of the DM search experiments that
we include in our analysis in Fig.\ref{fig:c1_summary}. The plot
includes the latest available data from a total of 14 existing
experiments, and the estimated future sensitivity of 4 projected
ones. The details of our procedure to obtain the exclusion plots are
provided in \ref{app:exp}. 

The relative sensitivity of different detectors is determined by two
elements: the thresholds $v_{min}^{th}$ of different experiments
expressed in terms of the WIMP incoming velocity, and the scaling law
of the WIMP--nucleus cross section off different targets.

The former element explains the steep rise of all the exclusion plot
curves at low WIMP masses, which corresponds to the case when
$v_{min}^{th}$ approaches the value of the escape velocity in the lab
rest frame, and is sensitive to experimental features close to the
energy threshold that are typically affected by uncertainties, such as
efficiencies, acceptances and charge/light yields. With the
assumptions listed in \ref{app:exp}, among the experiments included in
our analysis the ones with the lowest velocity thresholds turn out to
be DS50, CRESST--II, CDMSlite and CDEX. In particular, for
$m_{\chi}$=1 GeV we have $v_{min,DS50}^{th}\simeq$ 450 km/s,
$v_{min,CRESST-II}^{th}\simeq$ 480 km/s (for scatterings off oxygen),
$v_{min,CDMSlite}^{th}\simeq$ 910 km/s, $v_{min,CDEX}^{th}\simeq$ 1600
km/s. Assuming $v_{esc}^{lab}\simeq$ 782 km/s (see the previous
Section) this implies that in our analysis only DS50 and CRESST-II
(for effective interactions for which argon and oxygen have a
non--vanishing nuclear response function) are sensitive to
$m_{\chi}\lsim$ 1 GeV.  On the other hand CDMSlite and CDEX are
sensitive to slightly higher masses (for instance, for $m_{\chi}$=2
GeV $v_{min,CDMSlite}^{th}\simeq$
460 km/s, $v_{min,CDEX}^{th}\simeq$ 850 km/s, while for $m_{\chi}$=3
GeV $v_{min,CDEX}^{th}\simeq$ 580 km/s). The velocity threshold is a
purely kinematical feature that does not depend on the type of
interaction and that favors experiments with the lowest $v_{min}^{th}$
at fixed $m_{\chi}$.

\begin{figure}
\begin{center}
  \includegraphics[width=0.44\columnwidth]{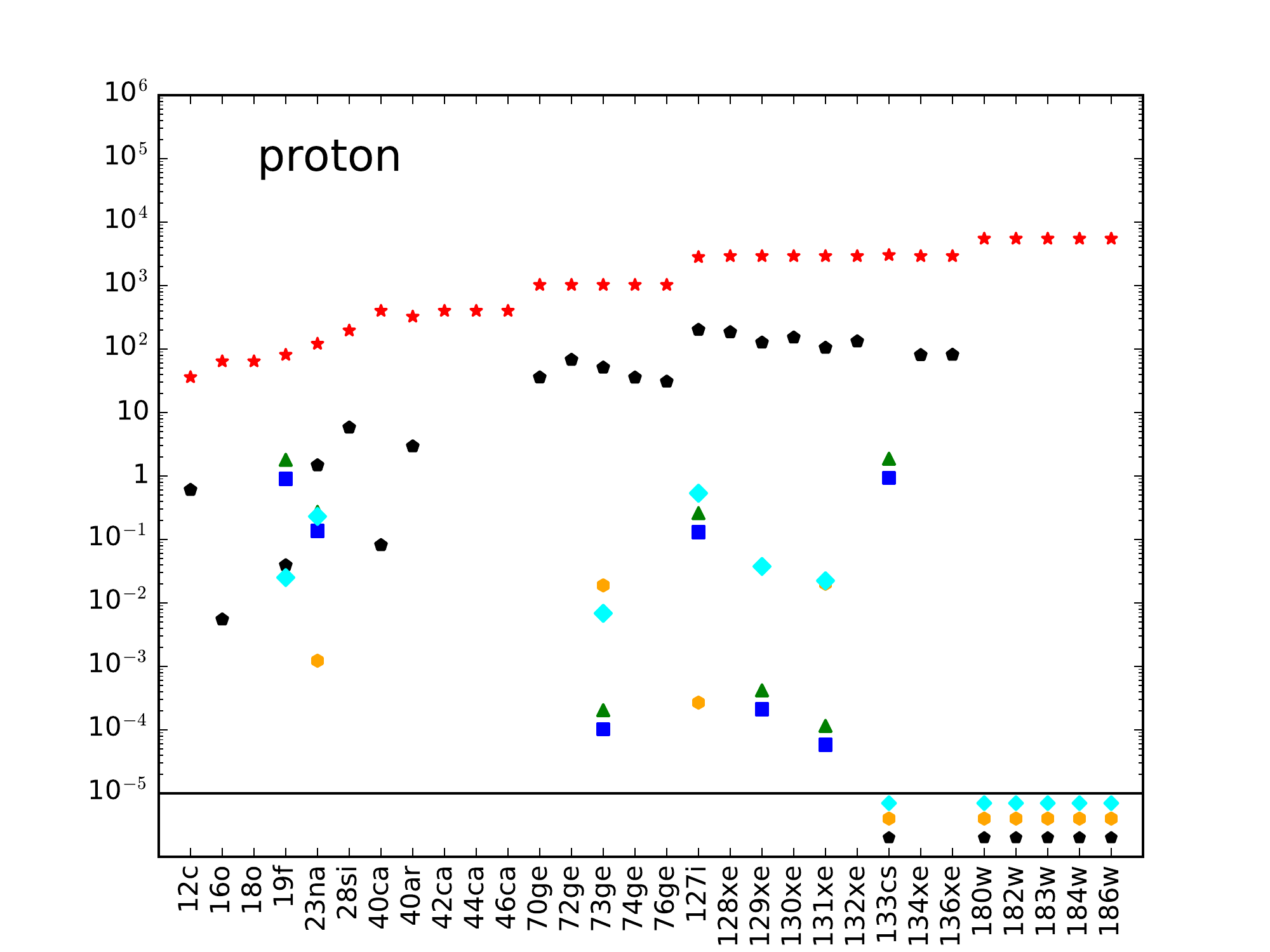}
  \includegraphics[width=0.44\columnwidth]{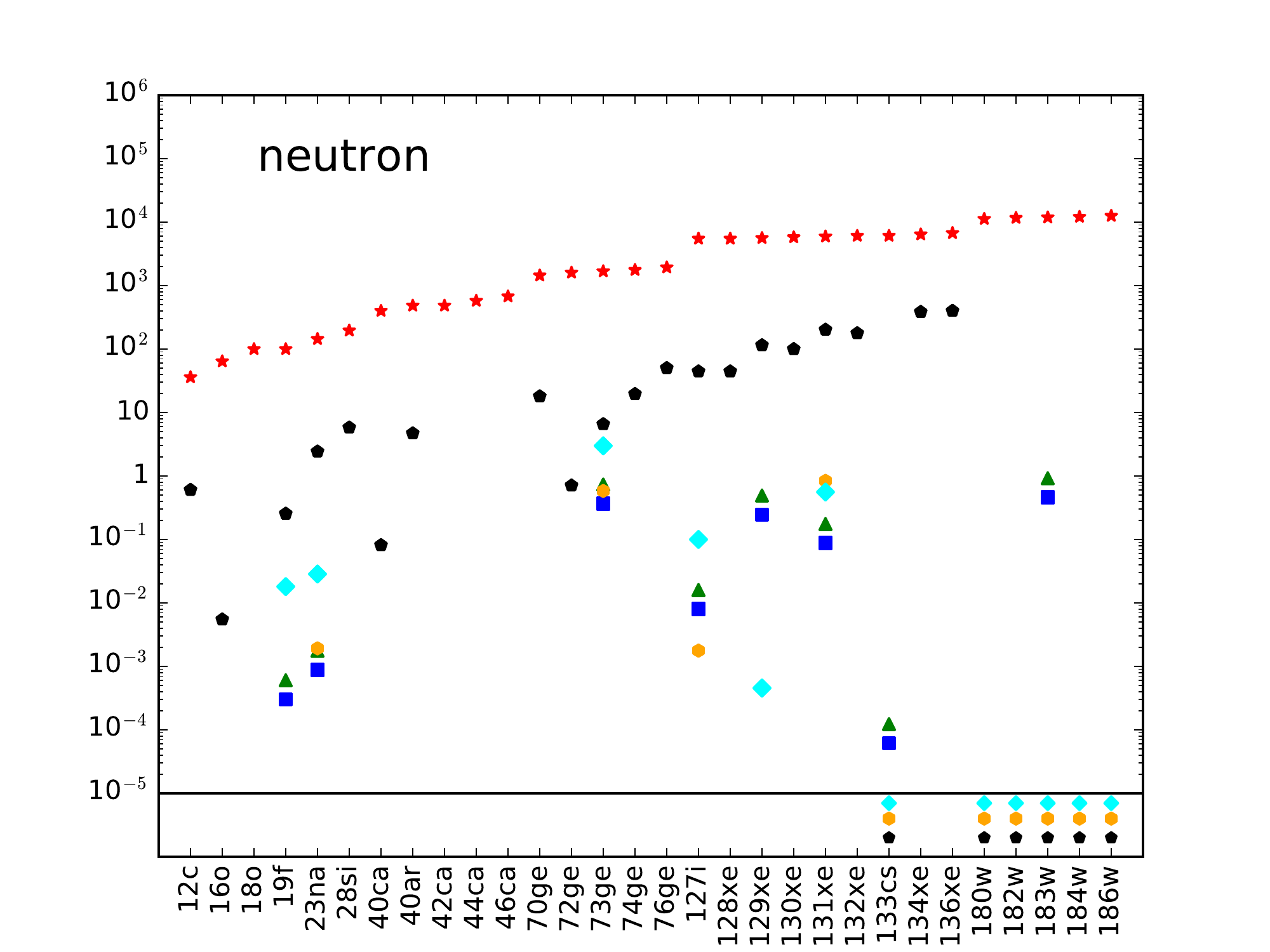}
  \includegraphics[width=0.10\columnwidth]{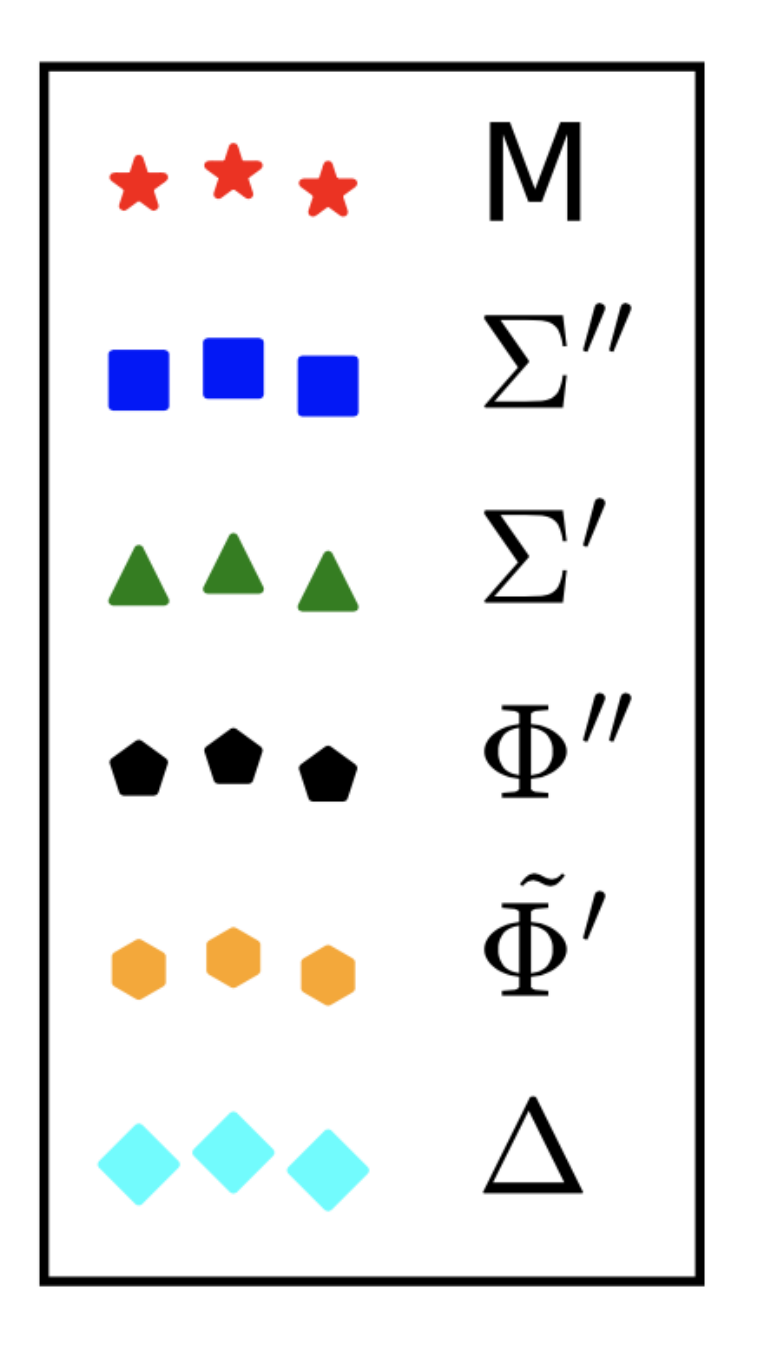}  
\end{center}
\caption{ Nuclear response functions at vanishing momentum transfer
  off protons $16\pi/(2 j_T+1)\times W^{p}_{T k}(y=0)$ (left--hand plot)
  and off neutrons $16\pi/(2 j_T+1)\times W^{n}_{T k}(y=0)$ (right--hand
  plot), for $k$=$M$, $\Phi^{\prime\prime}$, $\tilde{\Phi}^{\prime}$,
  $\Sigma^{\prime\prime}$, $\Sigma^{\prime}$, $\Delta$ and all the
  targets $T$ used in the present analysis.  The normalization factor
  is chosen so that $16\pi/(2 j_T+1)\times W^{p}_{T M}(y=0)$=$Z_T^2$ and
  $16\pi/(2 j_T+1)\times W^{n}_{T M}(y=0)$=$(A_T-Z_T)^2$ with $A_T$,
  $Z_T$ the mass and atomic numbers for target $T$. Markers below
  the horizontal solid line represent nuclear response functions that are
  missing in the literature. In our analysis we have set them to
  zero.}
\label{fig:scaling_law}
\end{figure}

With the exception of very low masses, where the effect of
$v_{min}^{th}$ is dominant, the relative sensitivity of different
detectors is determined by the scaling law of the WIMP--nucleus cross
section with different targets, which is the focus of our analysis.
In particular the SI interaction (corresponding to the $M$ effective
nuclear operator) favors heavy nuclei, so that the most stringent
bounds in Fig. \ref{fig:c1_summary} correspond to xenon experiments
(XENON1T, PANDAX-II). However the interaction terms in the Hamiltonian
of Eq.(\ref{eq:H}) lead to expected rates that depend on the full set
of possible nuclear operators ($M$, $\Phi^{\prime\prime}$,
$\tilde{\Phi}^{\prime}$, $\Sigma^{\prime\prime}$, $\Sigma^{\prime}$,
$\Delta$) leading to different scaling laws of the WIMP--nucleus cross
section on different targets. The correspondence between models and
nuclear response functions can be directly read off from the WIMP
response functions $R^{\tau\tau^{\prime}}_{k}$ (see
Eq.\ref{eq:wimp_response_functions}). In particular, using the
decomposition:

\be
R_k^{\tau\tau^{\prime}}=R_{0k}^{\tau\tau^{\prime}}+R_{1k}^{\tau\tau^{\prime}} (v^{\perp}_T)^2=R_{0k}^{\tau\tau^{\prime}}+R_{1k}^{\tau\tau^{\prime}}\left (v_T^2-v_{min}^2\right ),
\label{eq:r_decomposition}
\ee

\noindent such correspondence is summarized in Table
\ref{table:eft_summary}. In Fig. \ref{fig:scaling_law} we provide for
completeness the nuclear response functions at vanishing momentum
transfer off protons $16\pi/(2 j_T+1)\times W^{p}_{T k}(y=0)$
(left--hand plot) and off neutrons $16\pi/(2 j_T+1)\times W^{n}_{T
  k}(y=0)$ (right--hand plot), with $W^{p,n}_{T k}\equiv 1/4\times
(W^{00}_{T k}\pm W^{01}_{T k} \pm W^{10}_{T k}+ W^{11}_{T k})$ for
$k$=$M$, $\Phi^{\prime\prime}$, $\tilde{\Phi}^{\prime}$,
$\Sigma^{\prime\prime}$, $\Sigma^{\prime}$, $\Delta$ and all the
targets $T$ used in the present analysis, as calculated in
\cite{haxton2,catena}. The normalization factor is chosen so that
$16\pi/(2 j_T+1)\times W^{p}_{T M}(y=0)$=$Z_T^2$ and
$16\pi/(2 j_T+1)\times W^{n}_{T M}(y=0)$=$(A_T-Z_T)^2$ with $A_T$, $Z_T$
the mass and atomic numbers for target $T$. In the same figure values
below the horizontal line at $1\times 10^{-4}$ represent nuclear
response functions that are missing in the literature. They enter in
the calculation of expected rates in KIMS (caesium, using CsI) and
CRESST-II (tungsten, using CaWO$_4$). In both cases we have calculated
the expected rate on the targets with known nuclear response functions
and set to zero the missing ones, so that the corresponding
constraints must be considered as conservative estimates. For the
targets for which Refs.~\cite{haxton2,catena} do not provide the
nuclear response functions we evaluate the standard SI and SD
interactions following the procedure of
\ref{app:nuclear_response_functions}.

\begin{table}[t]
\begin{center}
{\begin{tabular}{@{}|c|c|c|c|c|c|@{}}
\hline
coupling  &  $R^{\tau \tau^{\prime}}_{0k}$  & $R^{\tau \tau^{\prime}}_{1k}$ & coupling  &  $R^{\tau \tau^{\prime}}_{0k}$  & $R^{\tau \tau^{\prime}}_{1k}$ \\
\hline
$1$  &   $M(q^0)$ & - & $3$  &   $\Phi^{\prime\prime}(q^4)$  & $\Sigma^{\prime}(q^2)$\\
$4$  & $\Sigma^{\prime\prime}(q^0)$,$\Sigma^{\prime}(q^0)$   & - & $5$  &   $\Delta(q^4)$  & $M(q^2)$\\
$6$  & $\Sigma^{\prime\prime}(q^4)$ & - & $7$  &  -  & $\Sigma^{\prime}(q^0)$\\
$8$  & $\Delta(q^2)$ & $M(q^0)$ & $9$  &  $\Sigma^{\prime}(q^2)$  & - \\
$10$  & $\Sigma^{\prime\prime}(q^2)$ & - & $11$  &  $M(q^2)$  & - \\
$12$  & $\Phi^{\prime\prime}(q^2)$,$\tilde{\Phi}^{\prime}(q^2)$ & $\Sigma^{\prime\prime}(q^0)$,$\Sigma^{\prime}(q^0)$ & $13$  & $\tilde{\Phi}^{\prime}(q^4)$  & $\Sigma^{\prime\prime}(q^2)$ \\
$14$  & - & $\Sigma^{\prime}(q^2)$ & $15$  & $\Phi^{\prime\prime}(q^6)$  & $\Sigma^{\prime}(q^4)$ \\
\hline
\end{tabular}}
\caption{Nuclear response functions corresponding to each coupling, for the velocity--independent and the velocity--dependent components parts of the WIMP response function, decomposed as in Eq.(\ref{eq:r_decomposition}).
  In parenthesis the power of $q$ in the WIMP response function.
  \label{table:eft_summary}}
\end{center}
\end{table}

\begin{figure}
\begin{center}
\includegraphics[width=0.80\columnwidth]{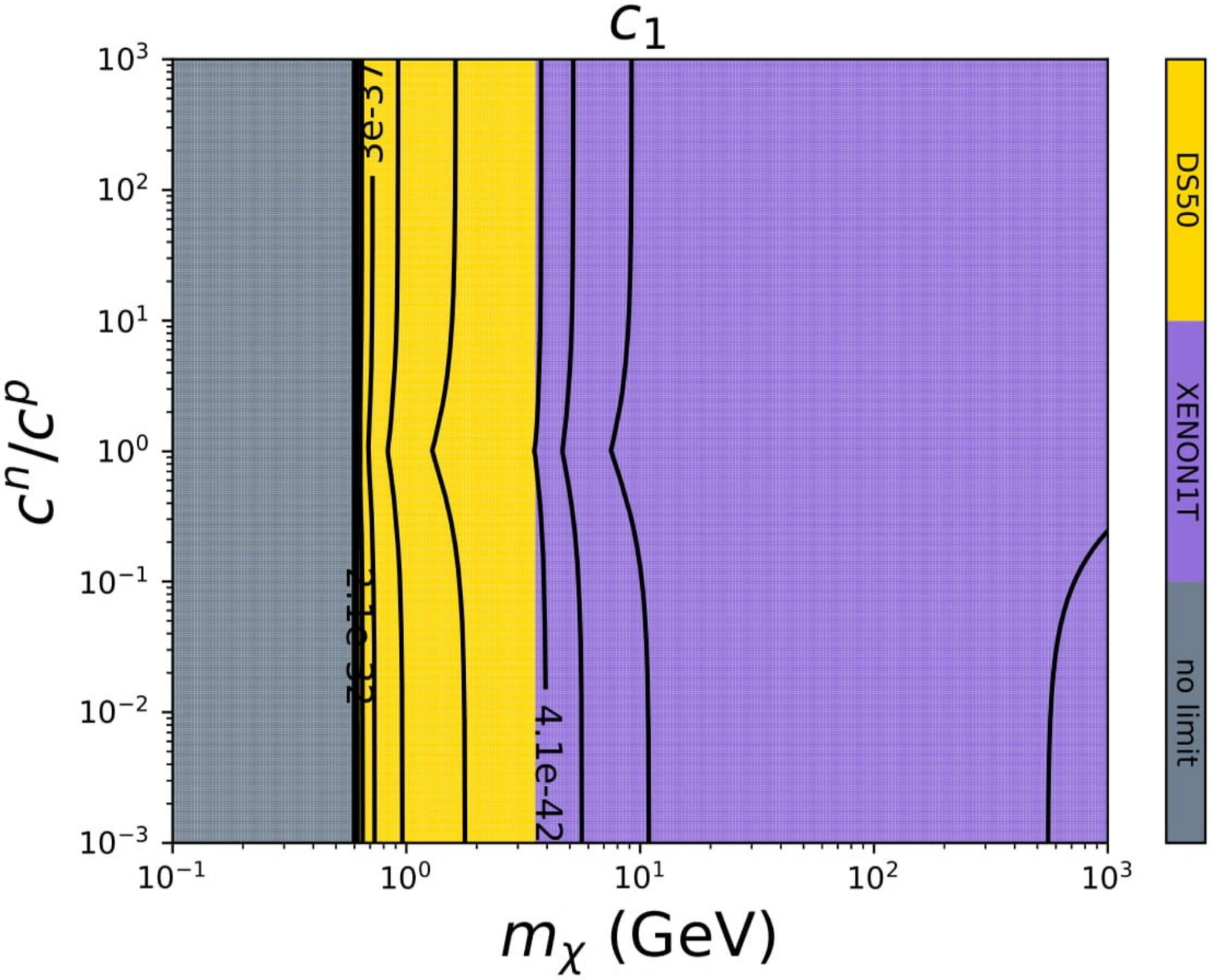}
\includegraphics[width=0.80\columnwidth]{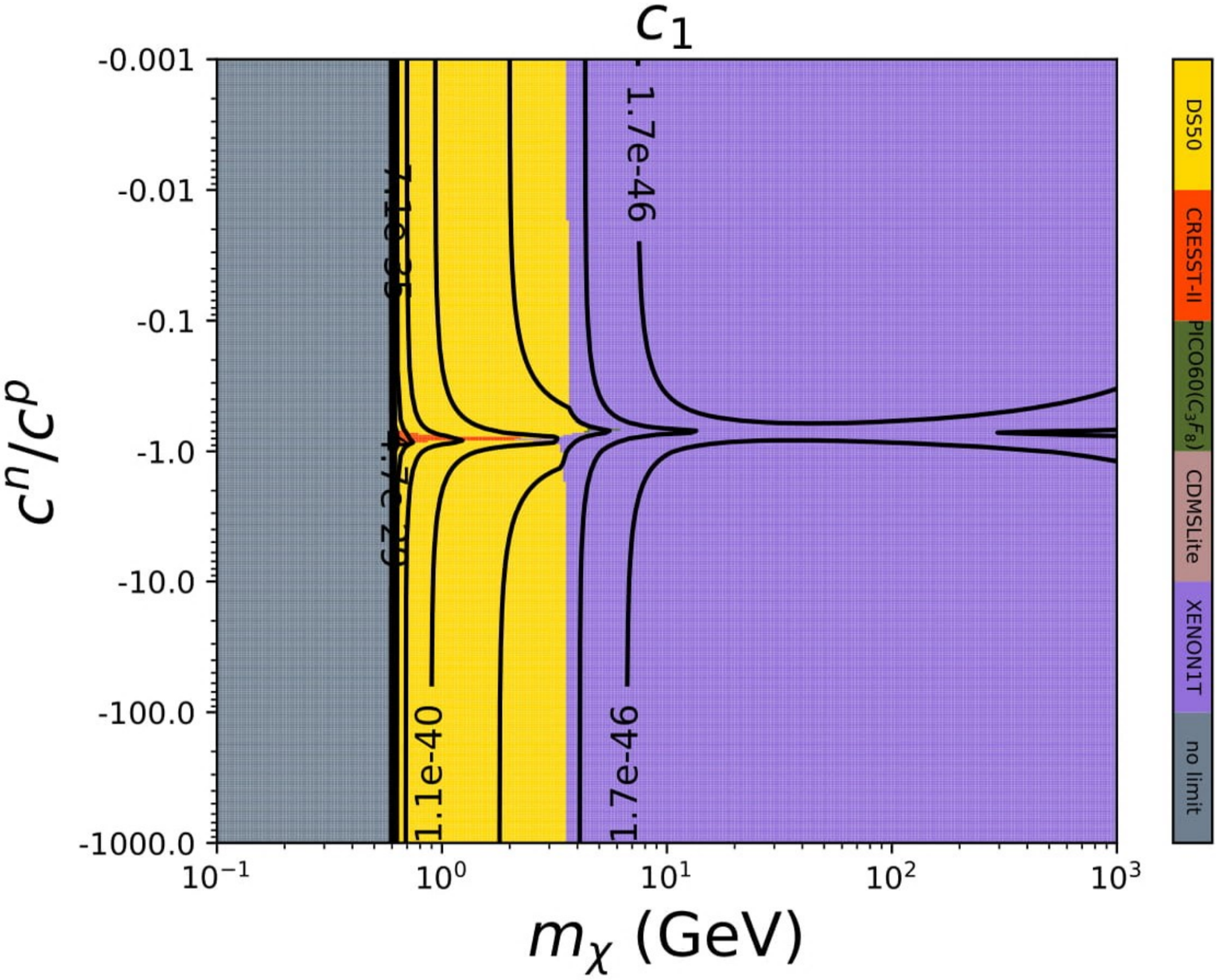}
\end{center}
\caption{Contour plots of the most stringent bound on the effective
  cross section $\sigma_{{\cal N},lim}$ introduced in
  Eq.(\ref{eq:conventional_sigma_nucleon}) as a function of the WIMP
  mass $m_{\chi}$ and of the ratio $c^n/c^p$ between the WIMP--neutron
  and the WIMP--proton couplings assuming that the operator ${\cal O}_1$
  dominates in the effective Hamiltonian of
  Eq.(\ref{eq:H}). Numerical values of the limit on $\sigma_{\cal N}$ are in
  cm$^2$. Different shadings indicate the experiment providing the
  most constraining bound, as indicated in the legend.}
\label{fig:c1_plane}
\end{figure}

\begin{figure}
\begin{center}
\includegraphics[width=0.85\columnwidth]{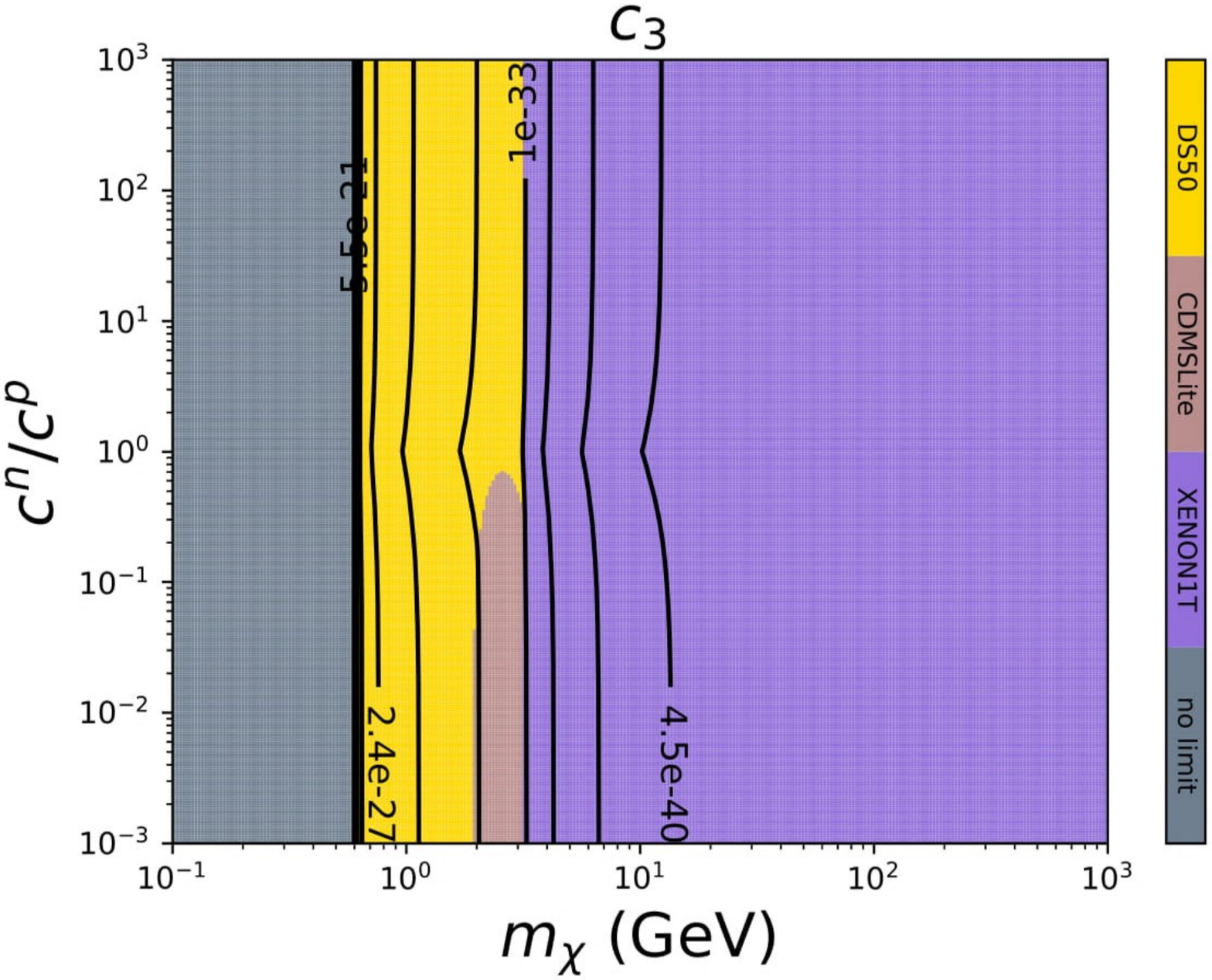}
\includegraphics[width=0.85\columnwidth]{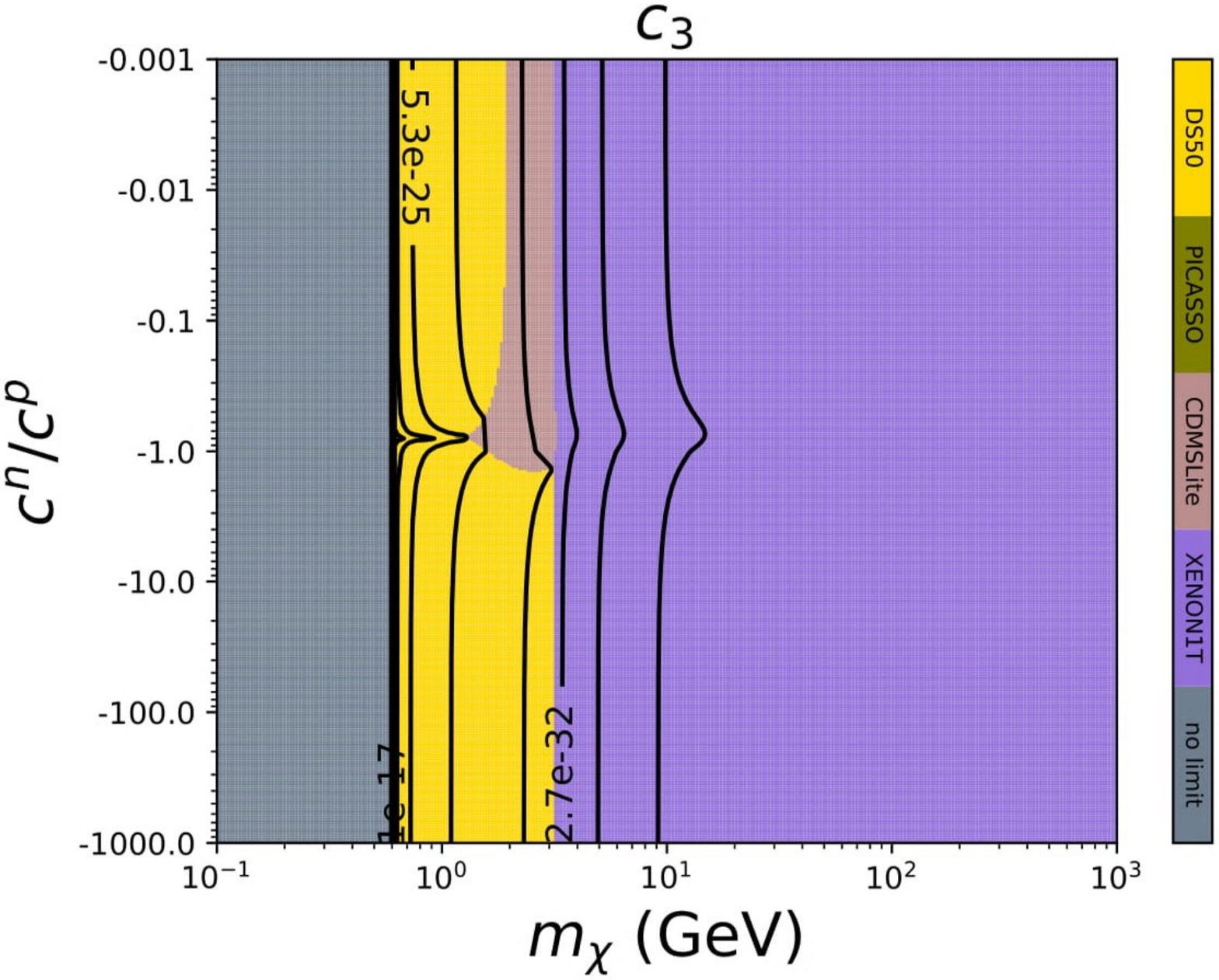}
\end{center}
\caption{The same as in Fig.\ref{fig:c1_plane} for the operator ${\cal
    O}_3$.}
\label{fig:c3_plane}
\end{figure}

\begin{figure}
\begin{center}
\includegraphics[width=0.85\columnwidth]{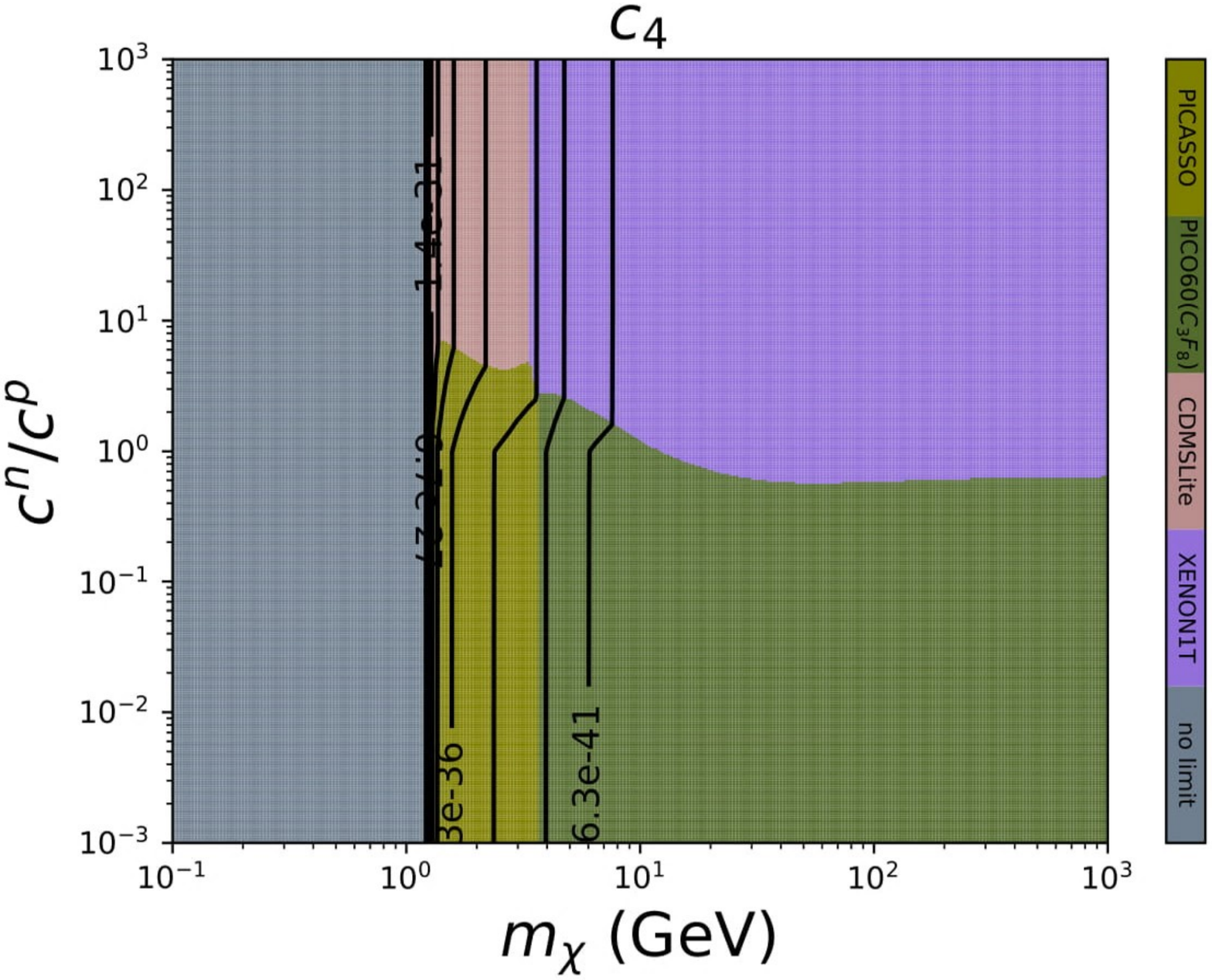}
\includegraphics[width=0.85\columnwidth]{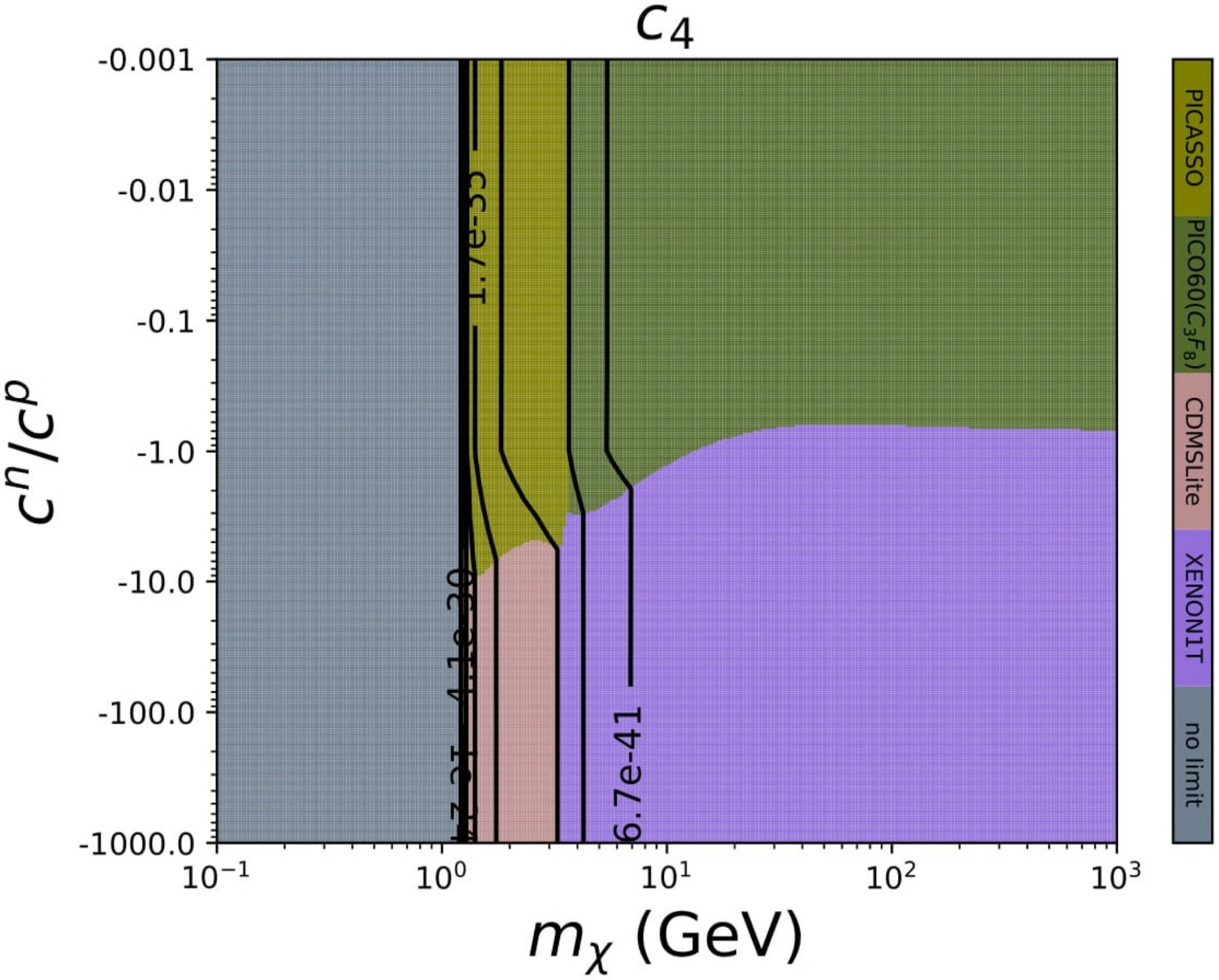}
\end{center}
\caption{The same as in Fig.\ref{fig:c1_plane} for the operator ${\cal
    O}_4$. This operator corresponds to the standard spin--dependent
  interaction of Eq. (\ref{eq:sd}).}
\label{fig:c4_plane}
\end{figure}

\begin{figure}
\begin{center}
\includegraphics[width=0.85\columnwidth]{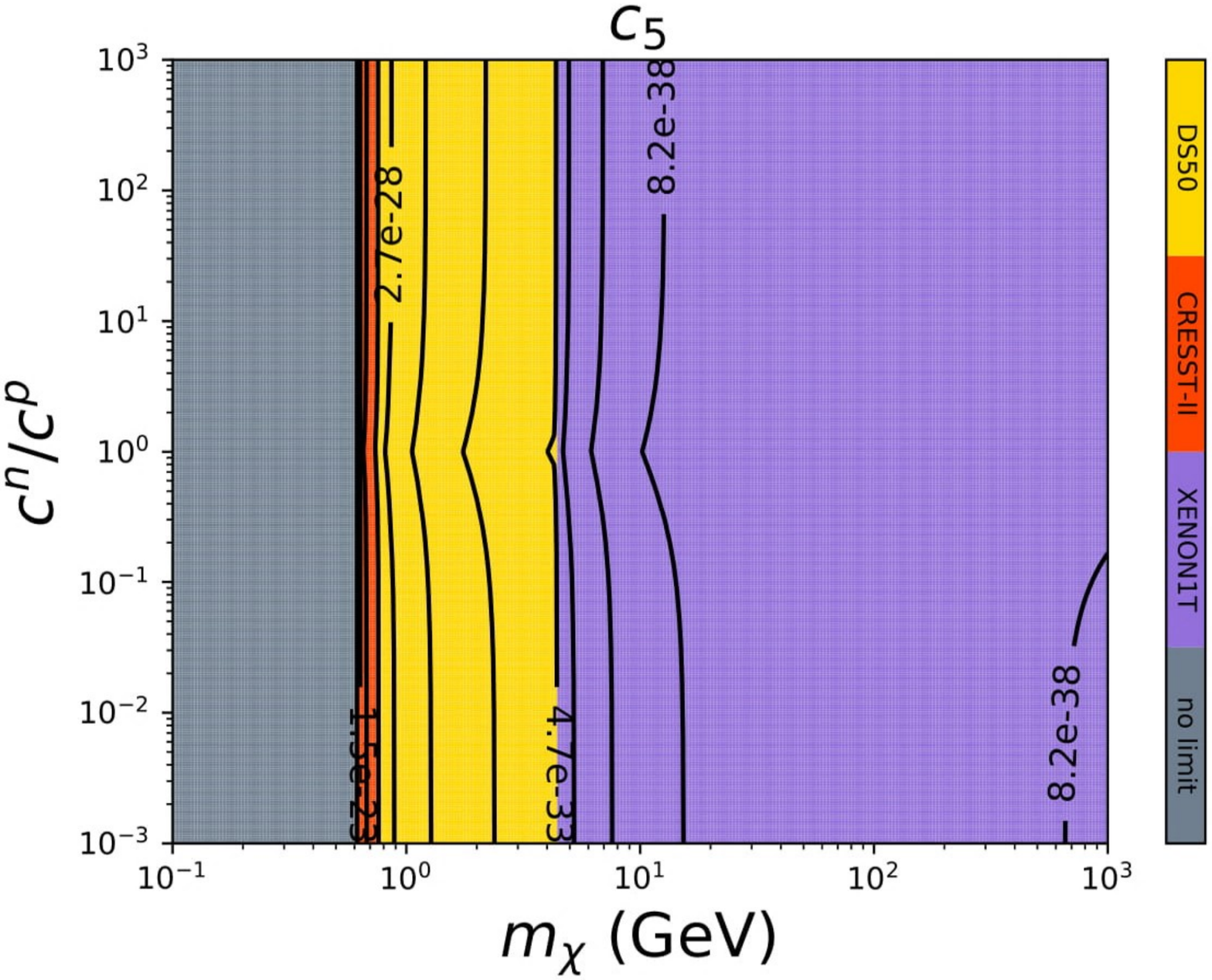}
\includegraphics[width=0.85\columnwidth]{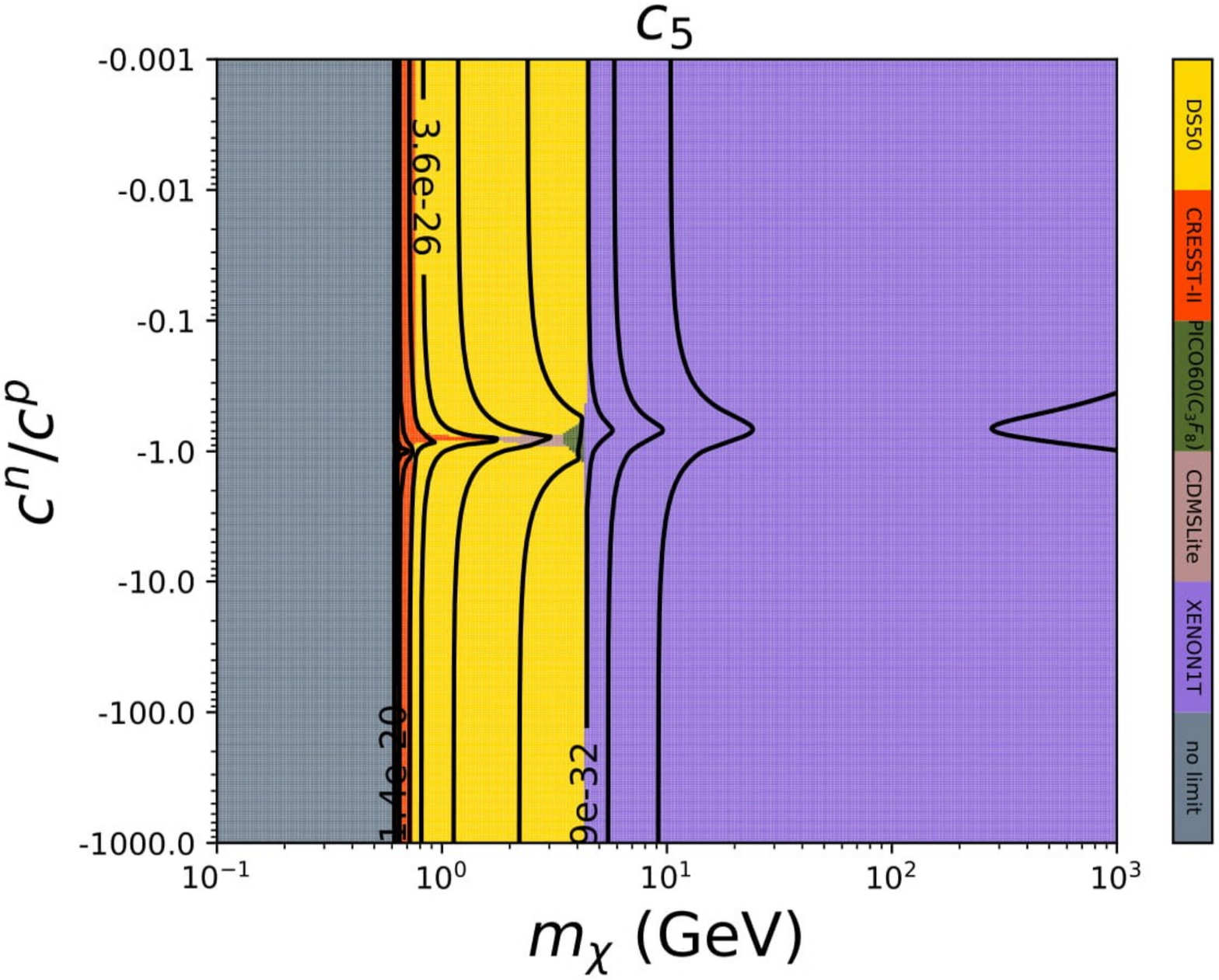}
\end{center}
\caption{The same as in Fig.\ref{fig:c1_plane} for the operator ${\cal O}_5$.}
\label{fig:c5_plane}
\end{figure}

\begin{figure}
\begin{center}
\includegraphics[width=0.85\columnwidth]{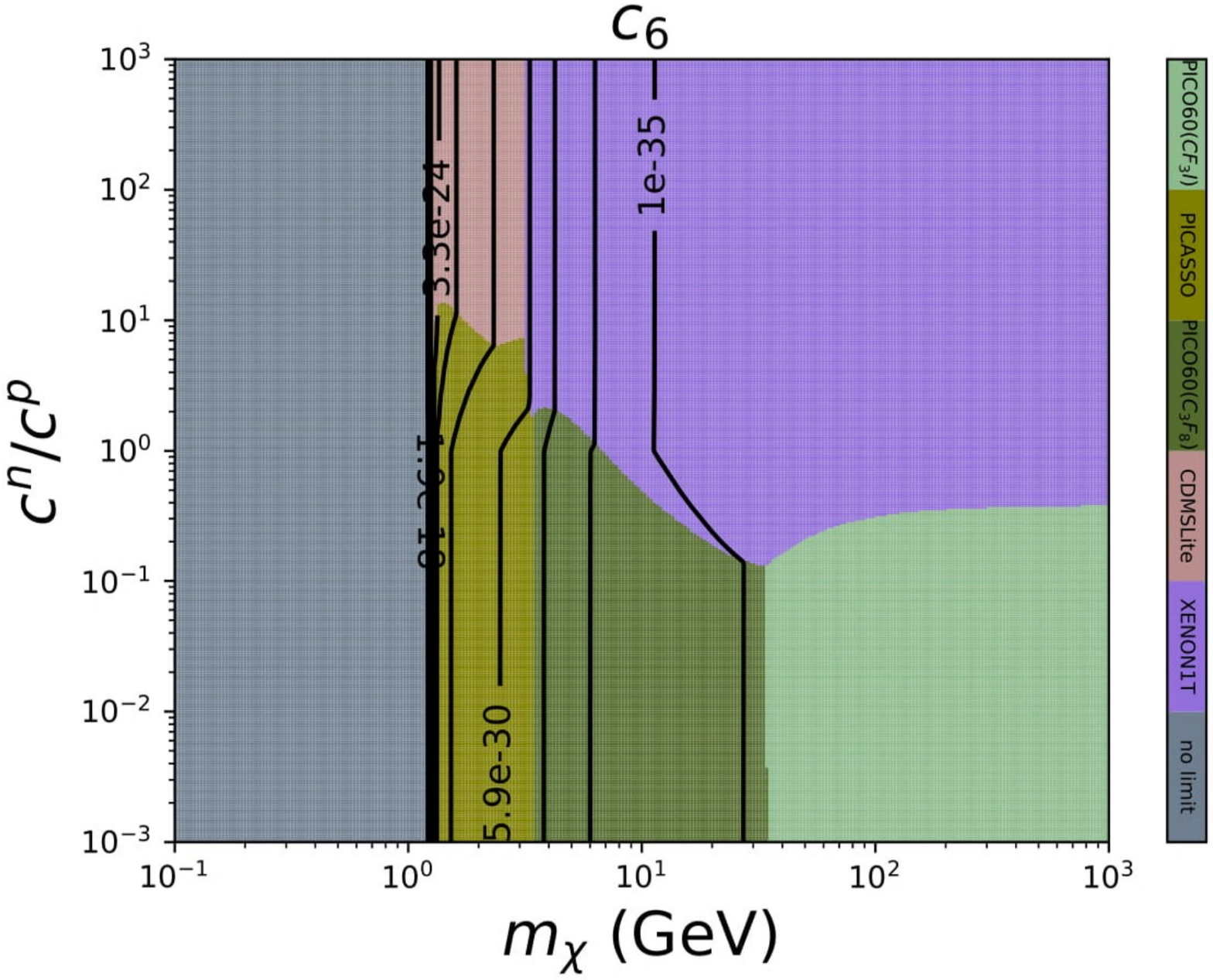}
\includegraphics[width=0.85\columnwidth]{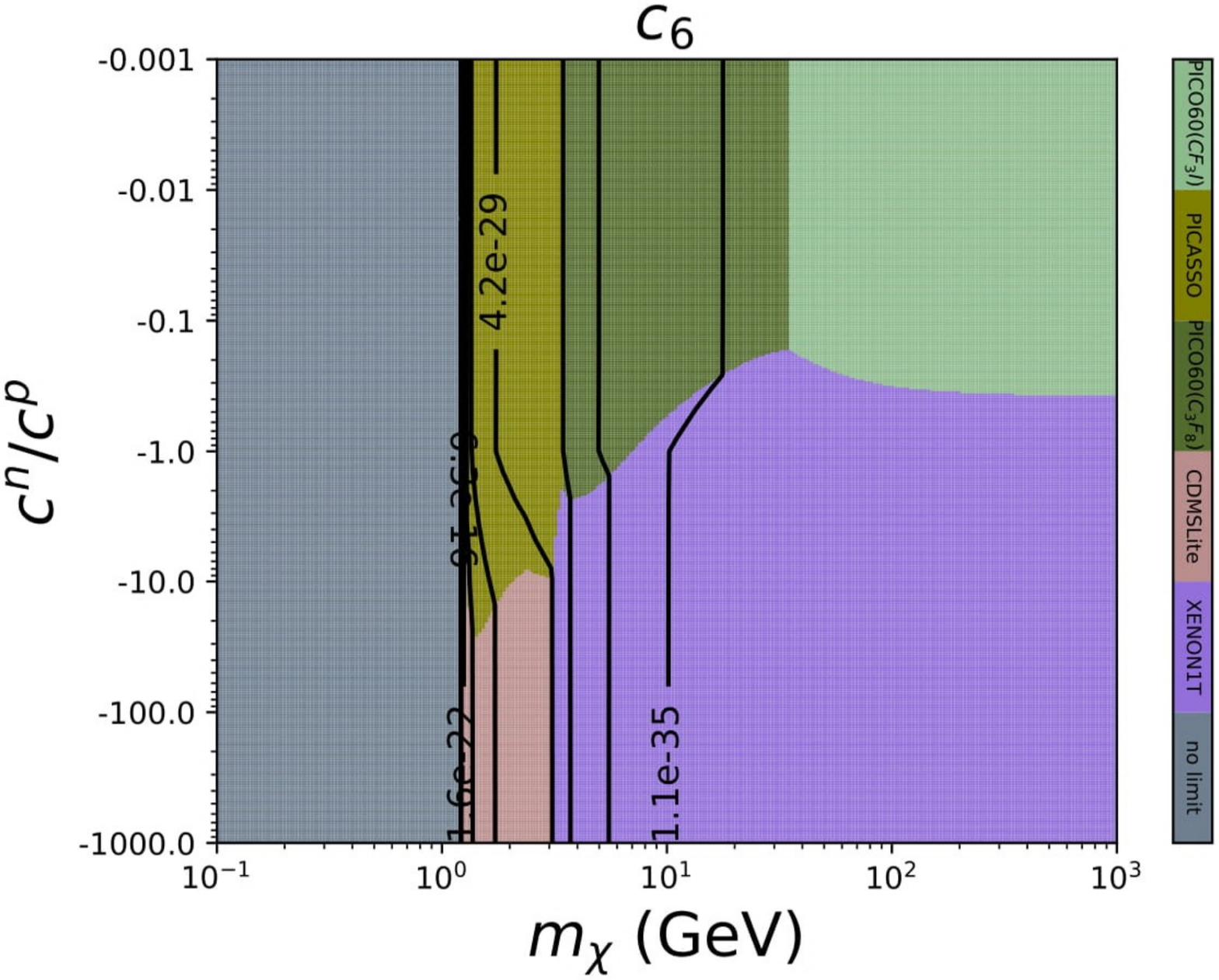}
\end{center}
\caption{The same as in Fig.\ref{fig:c1_plane} for the operator ${\cal O}_6$.}
\label{fig:c6_plane}
\end{figure}

\begin{figure}
\begin{center}
\includegraphics[width=0.85\columnwidth]{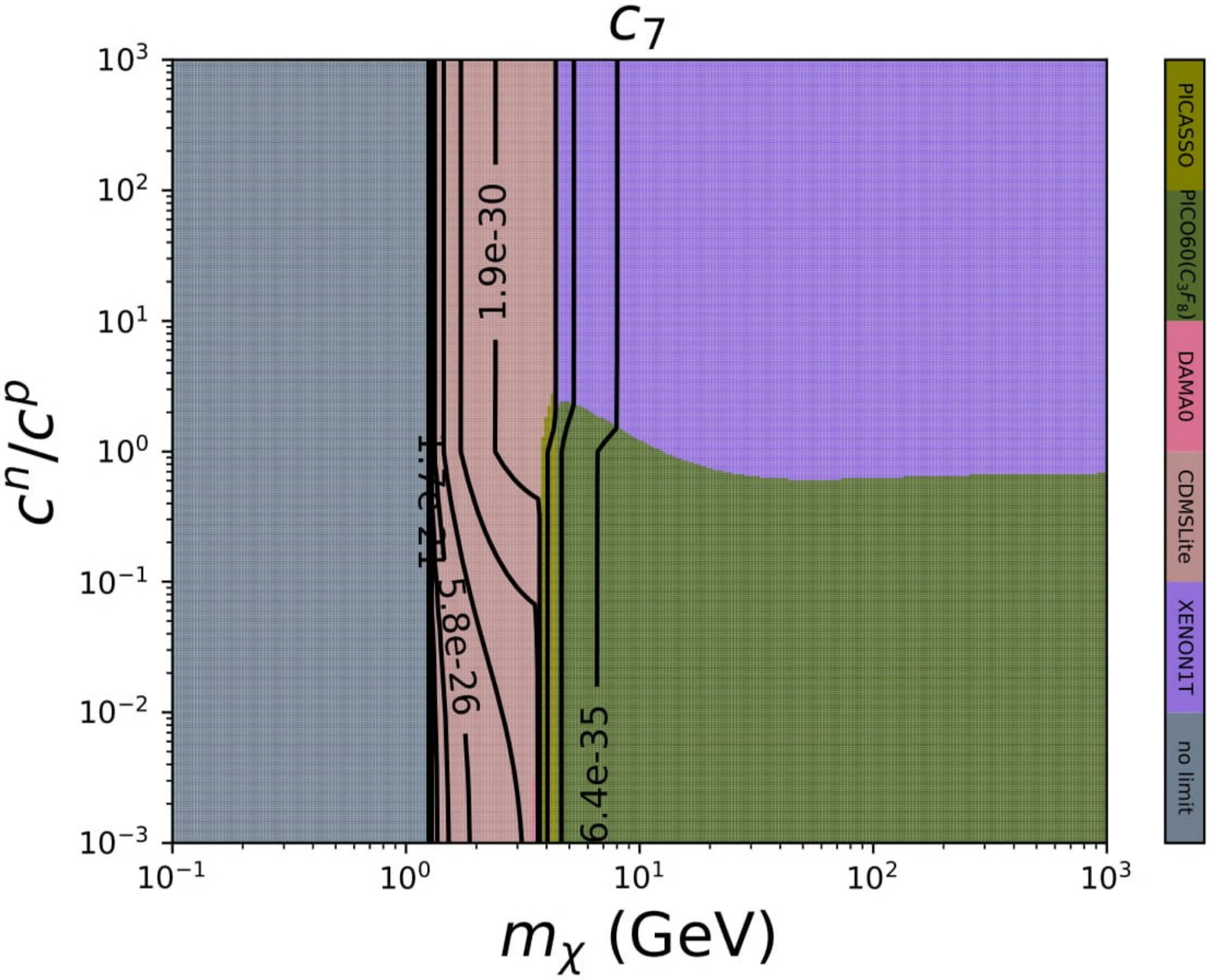}
\includegraphics[width=0.85\columnwidth]{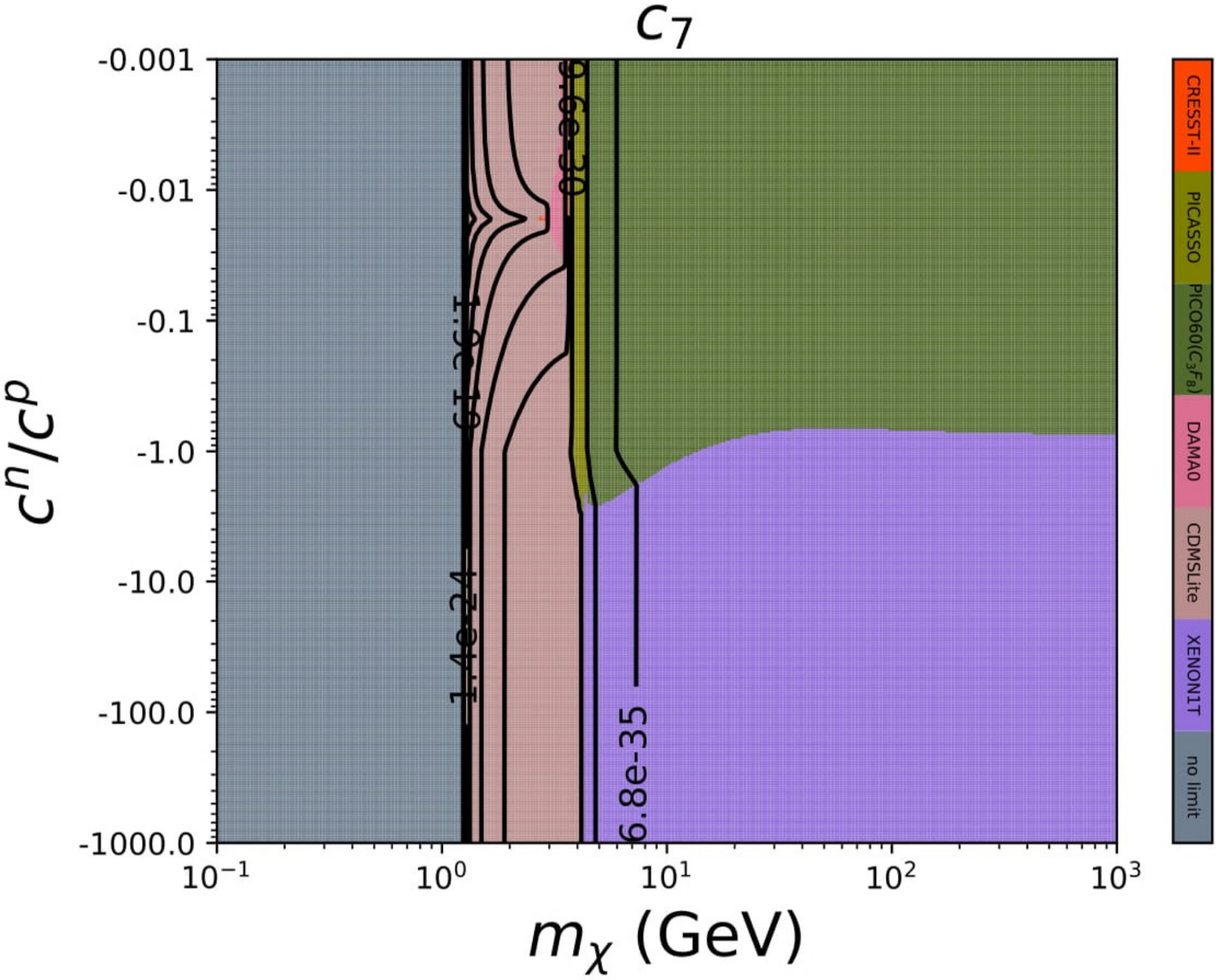}
\end{center}
\caption{The same as in Fig.\ref{fig:c1_plane} for the operator ${\cal O}_7$.}
\label{fig:c7_plane}
\end{figure}

\begin{figure}
\begin{center}
\includegraphics[width=0.85\columnwidth]{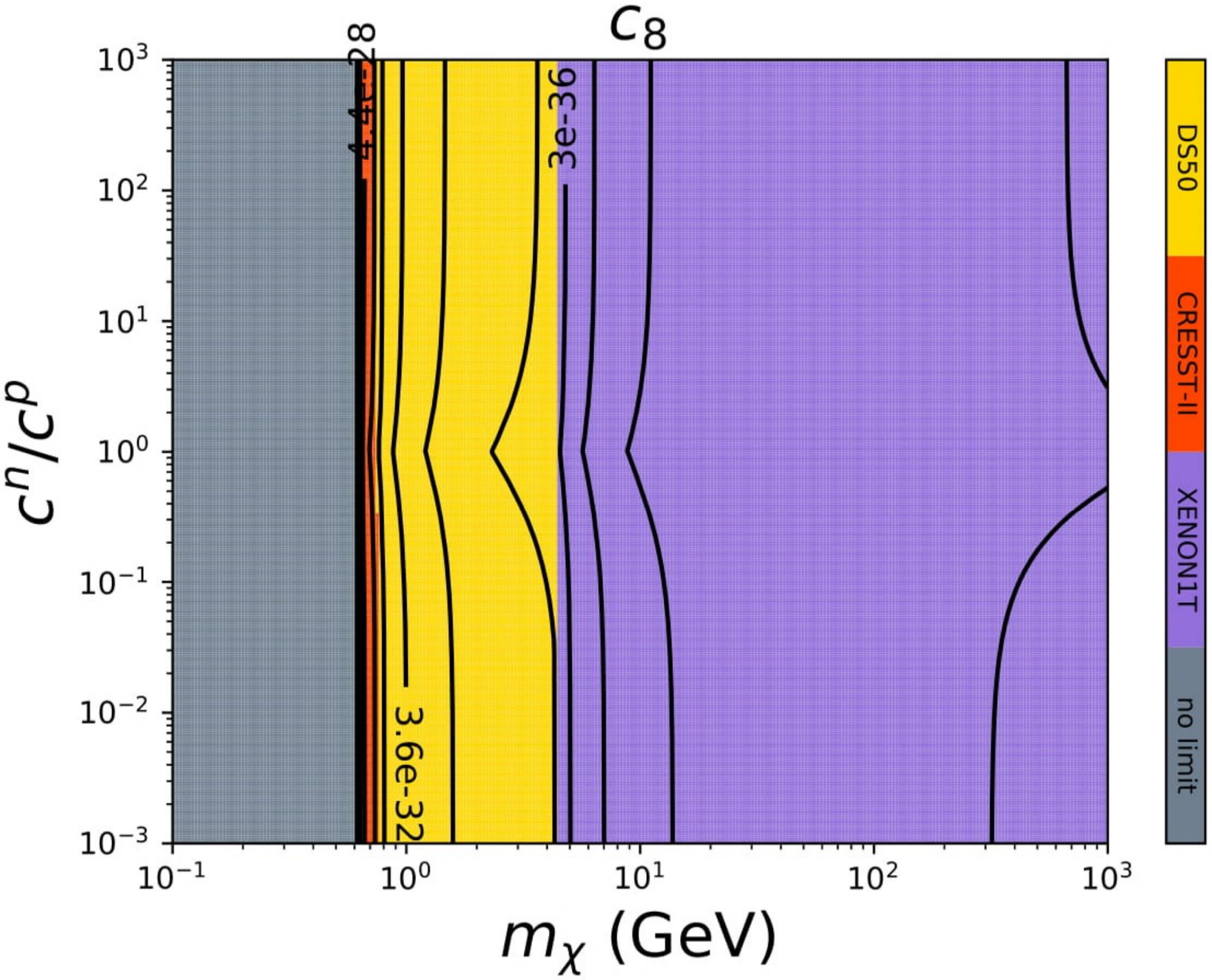}
\includegraphics[width=0.85\columnwidth]{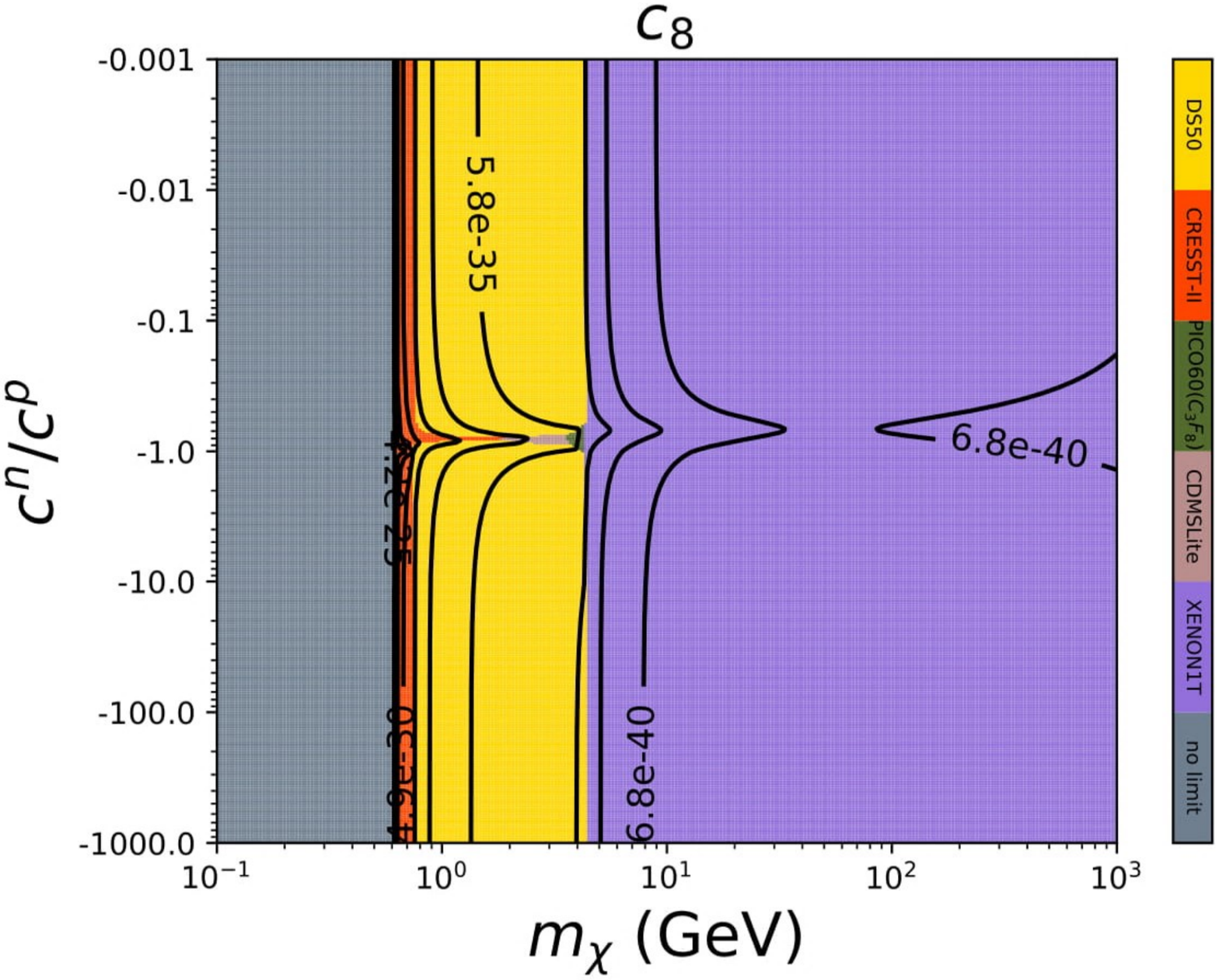}
\end{center}
\caption{The same as in Fig.\ref{fig:c1_plane} for the operator ${\cal O}_8$.}
\label{fig:c8_plane}
\end{figure}

\begin{figure}
\begin{center}
\includegraphics[width=0.85\columnwidth]{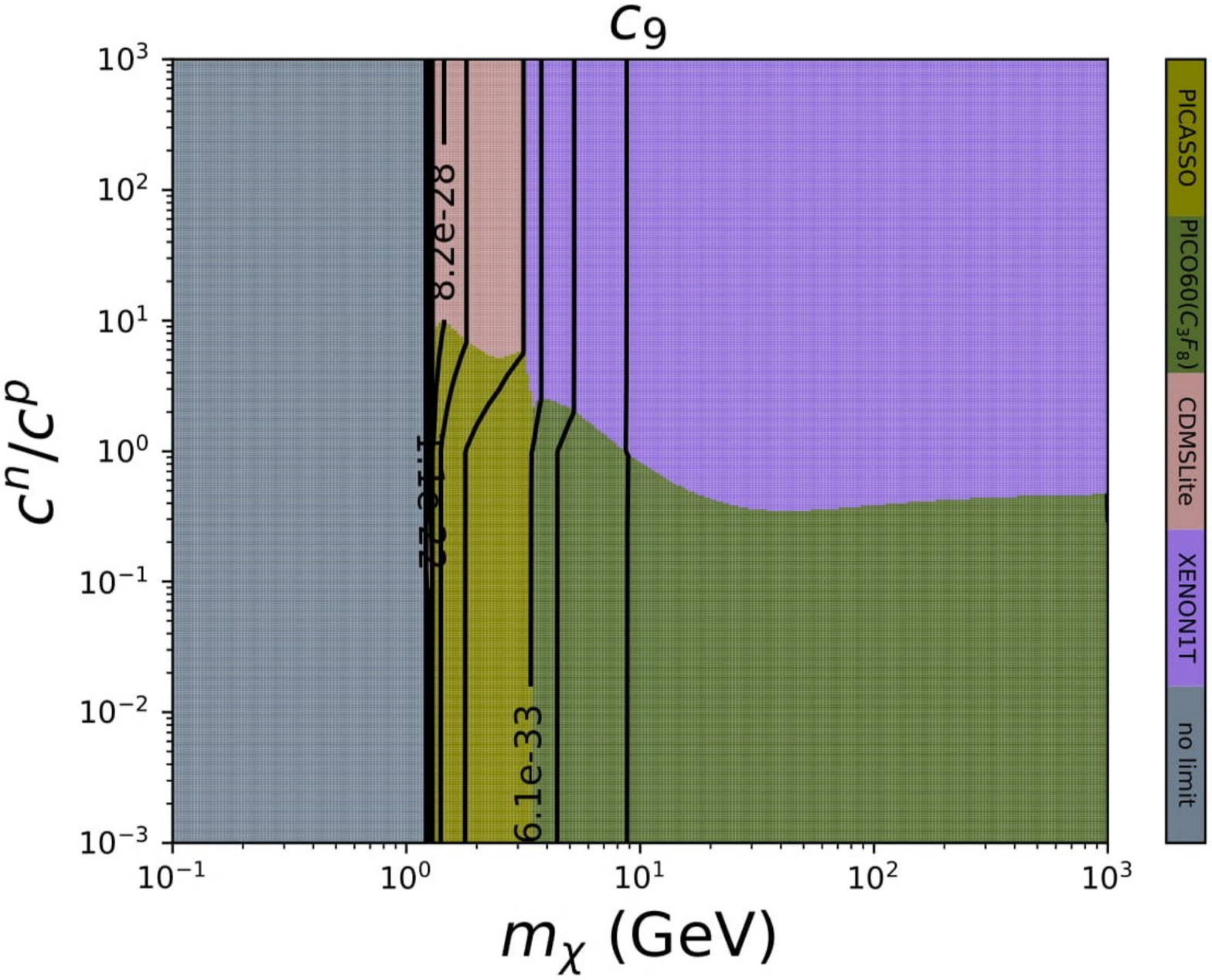}
\includegraphics[width=0.85\columnwidth]{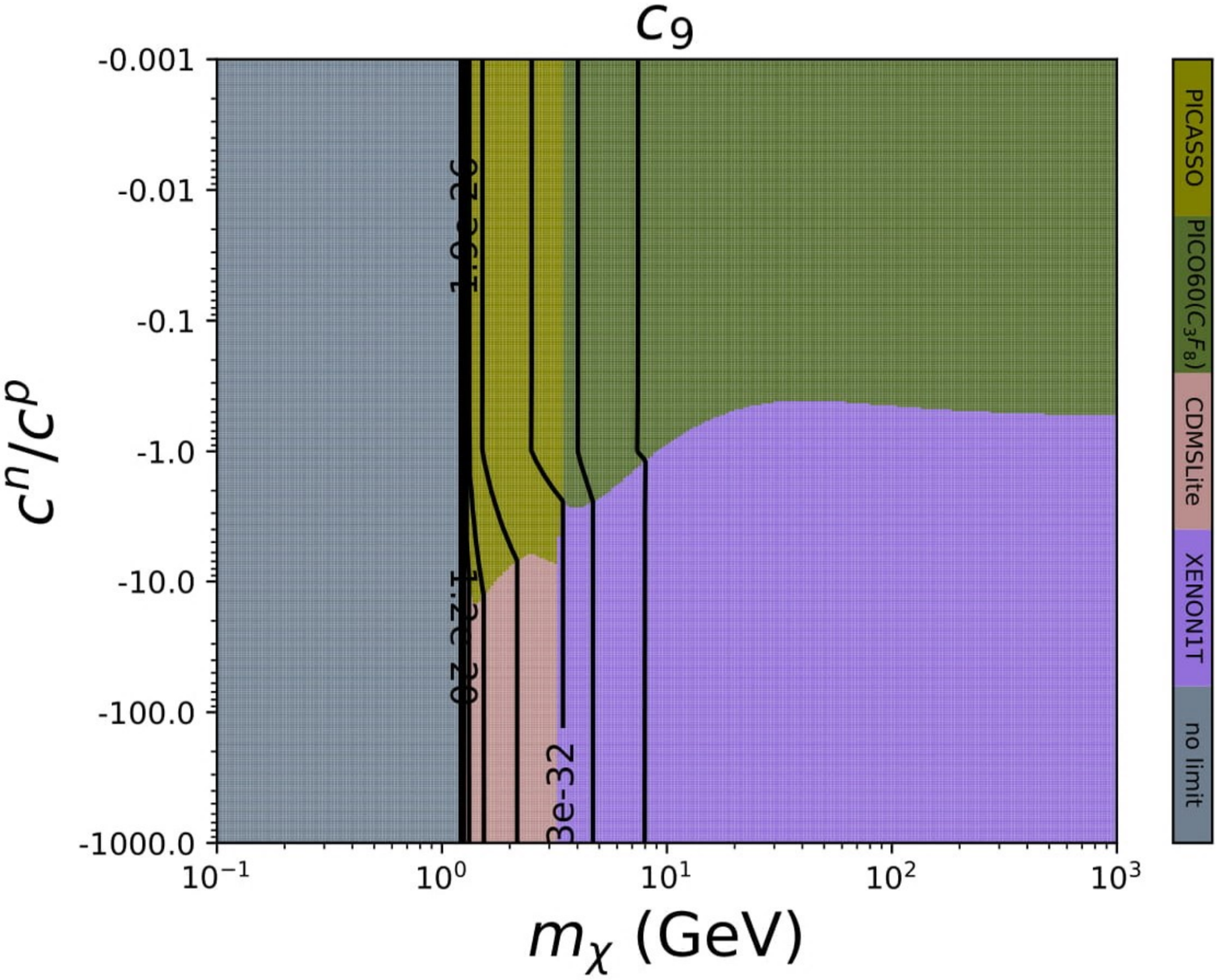}
\end{center}
\caption{The same as in Fig.\ref{fig:c1_plane} for the operator ${\cal O}_9$.}
\label{fig:c9_plane}
\end{figure}

\begin{figure}
\begin{center}
\includegraphics[width=0.85\columnwidth]{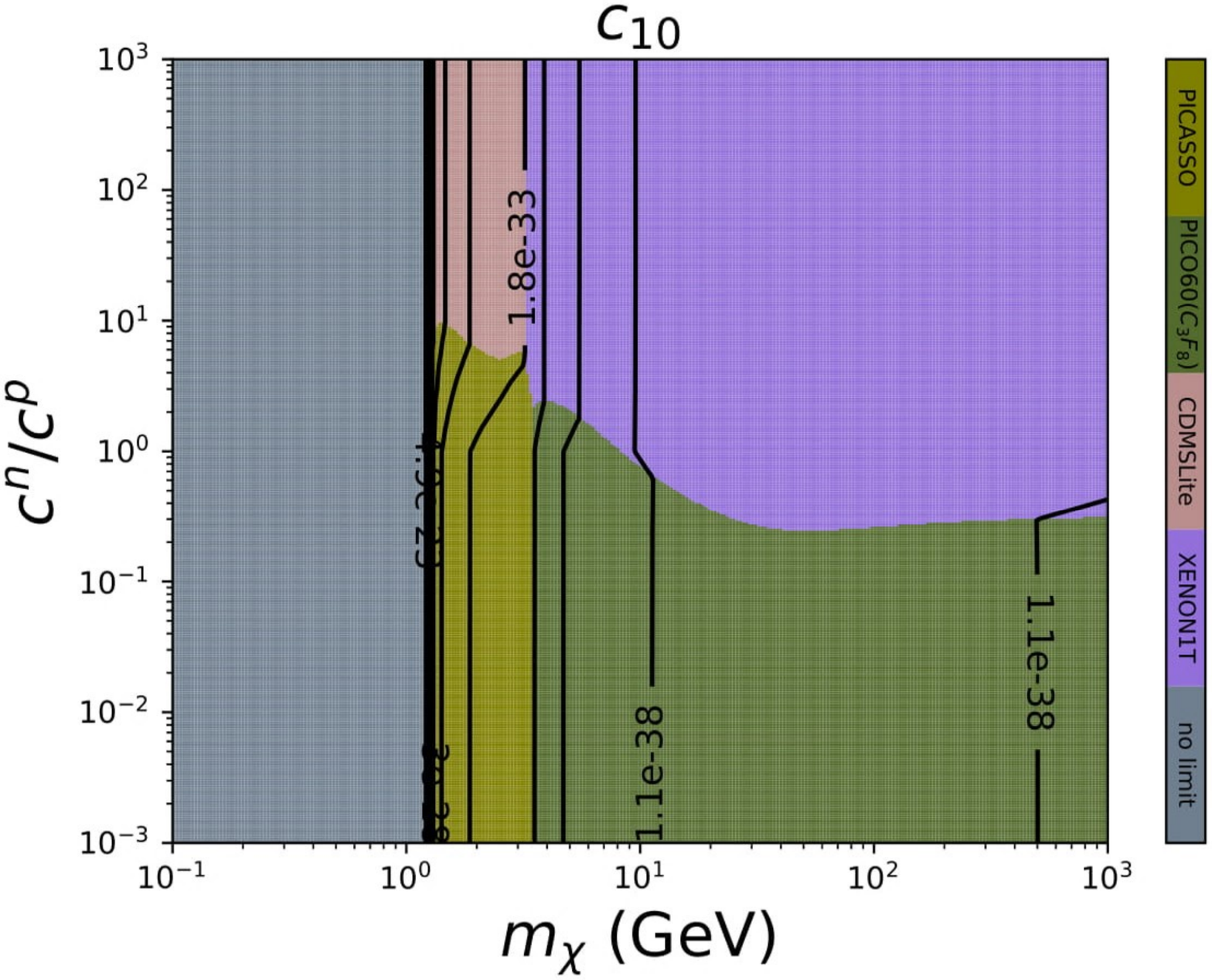}
\includegraphics[width=0.85\columnwidth]{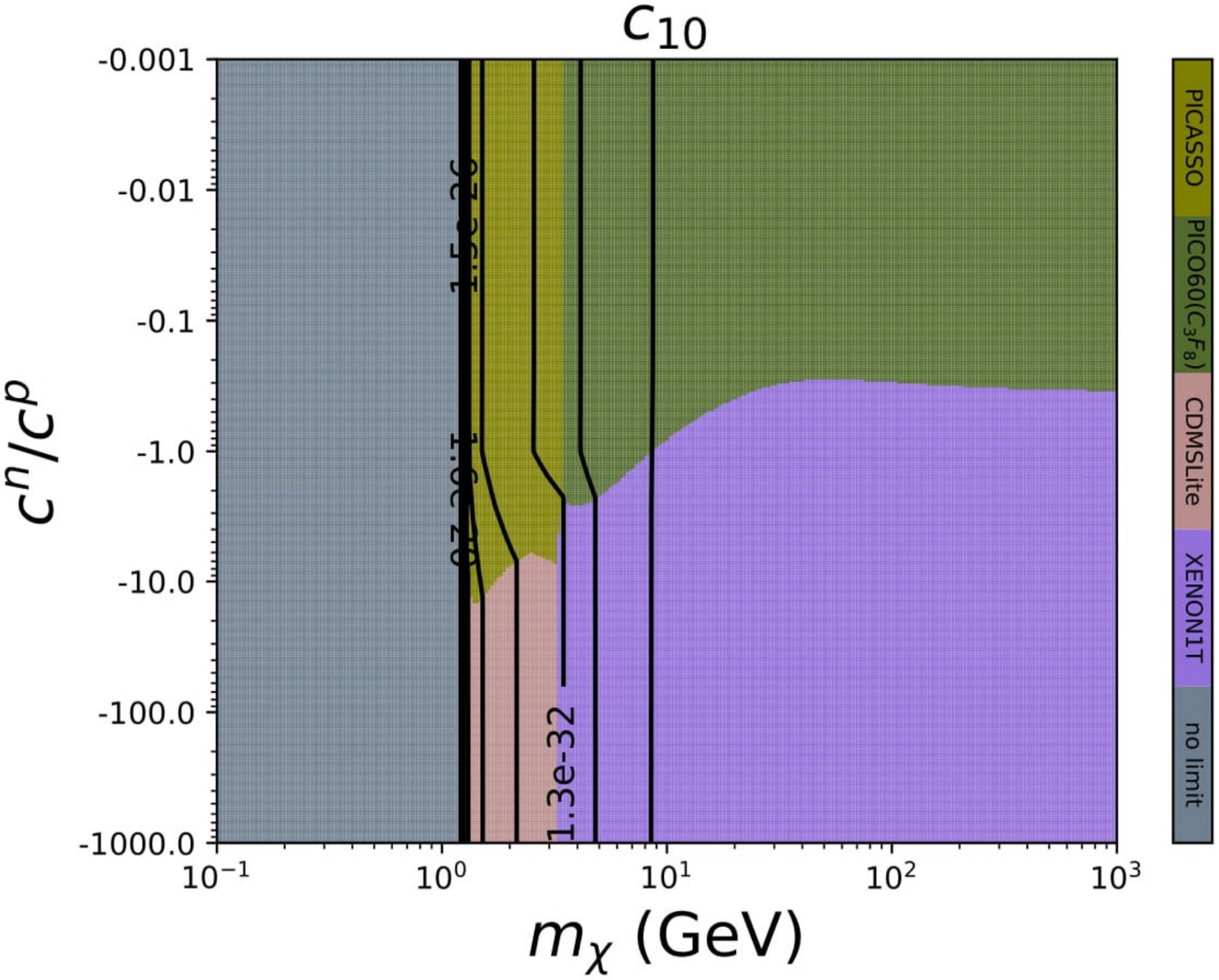}
\end{center}
\caption{The same as in Fig.\ref{fig:c1_plane} for the operator ${\cal O}_{10}$.}
\label{fig:c10_plane}
\end{figure}

\begin{figure}
\begin{center}
\includegraphics[width=0.85\columnwidth]{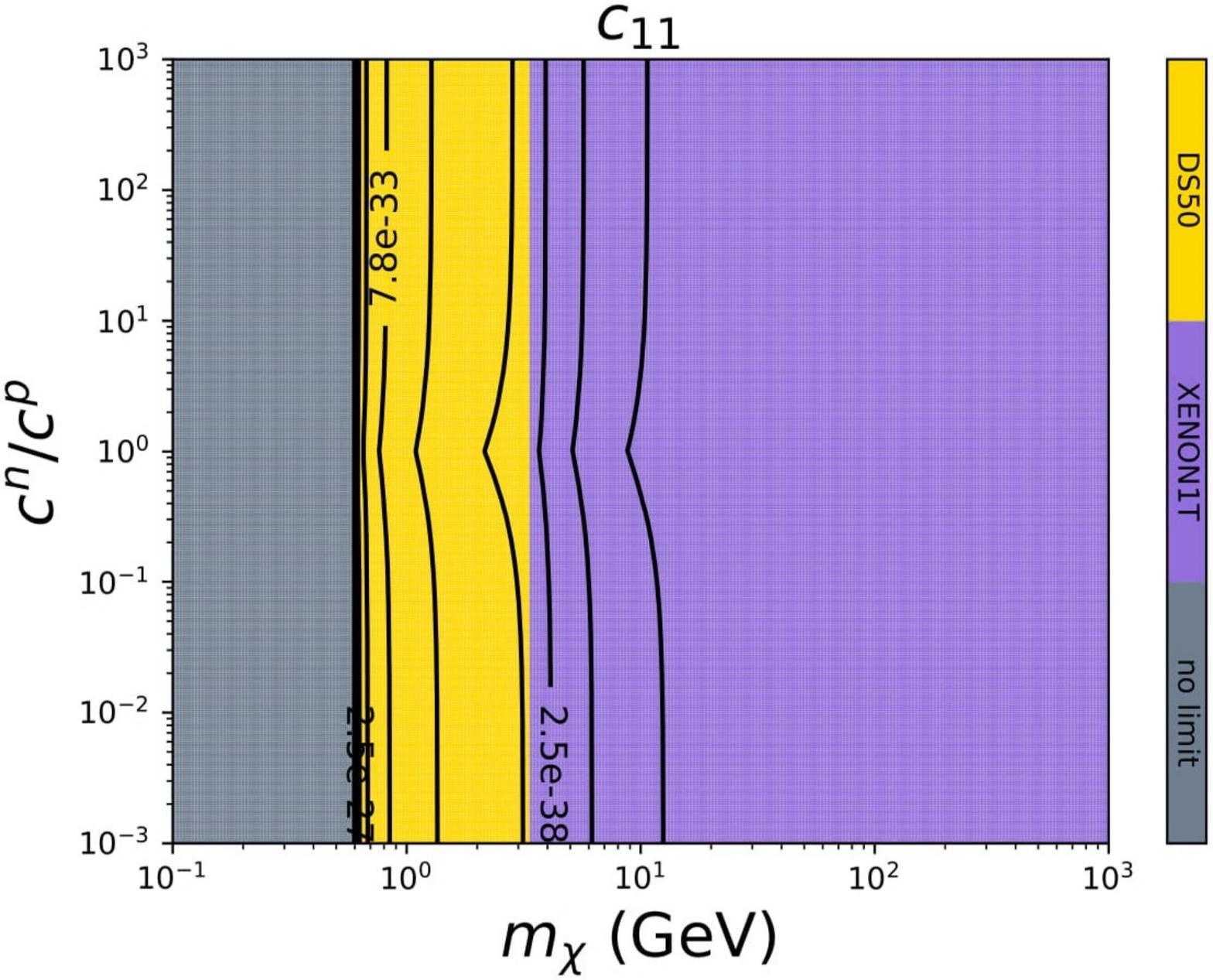}
\includegraphics[width=0.85\columnwidth]{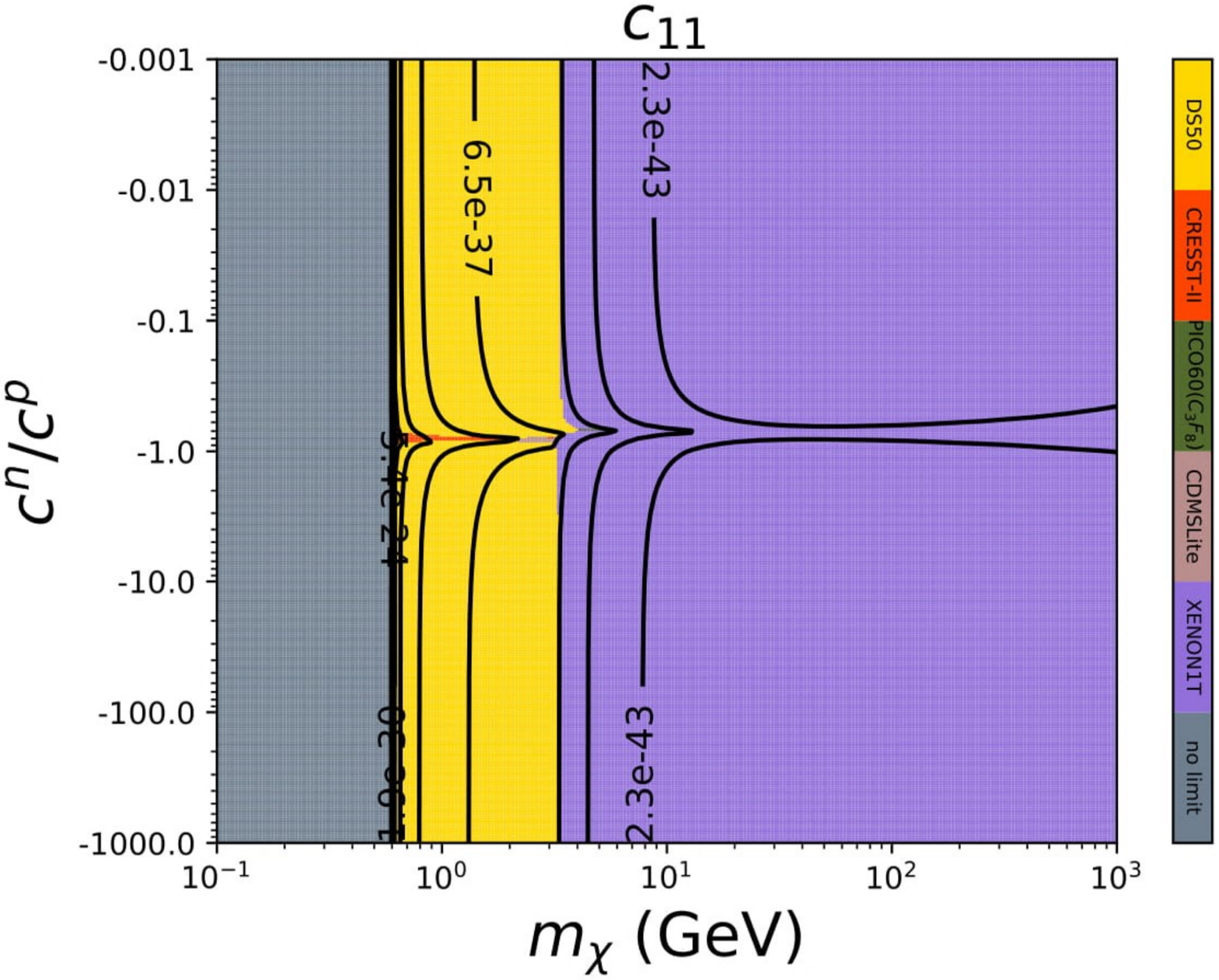}
\end{center}
\caption{The same as in Fig.\ref{fig:c1_plane} for the operator ${\cal O}_{11}$.}
\label{fig:c11_plane}
\end{figure}

\begin{figure}
\begin{center}
\includegraphics[width=0.85\columnwidth]{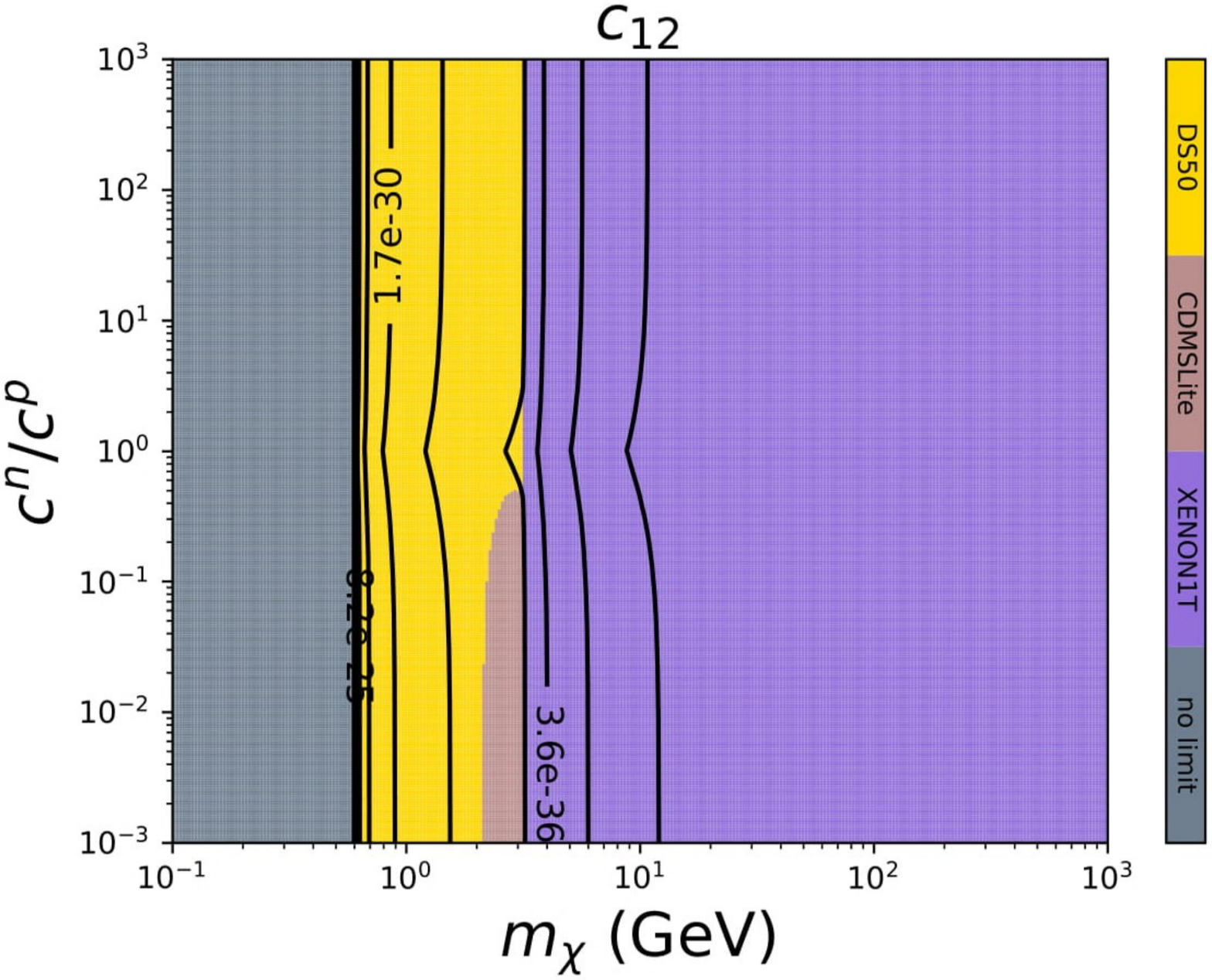}
\includegraphics[width=0.85\columnwidth]{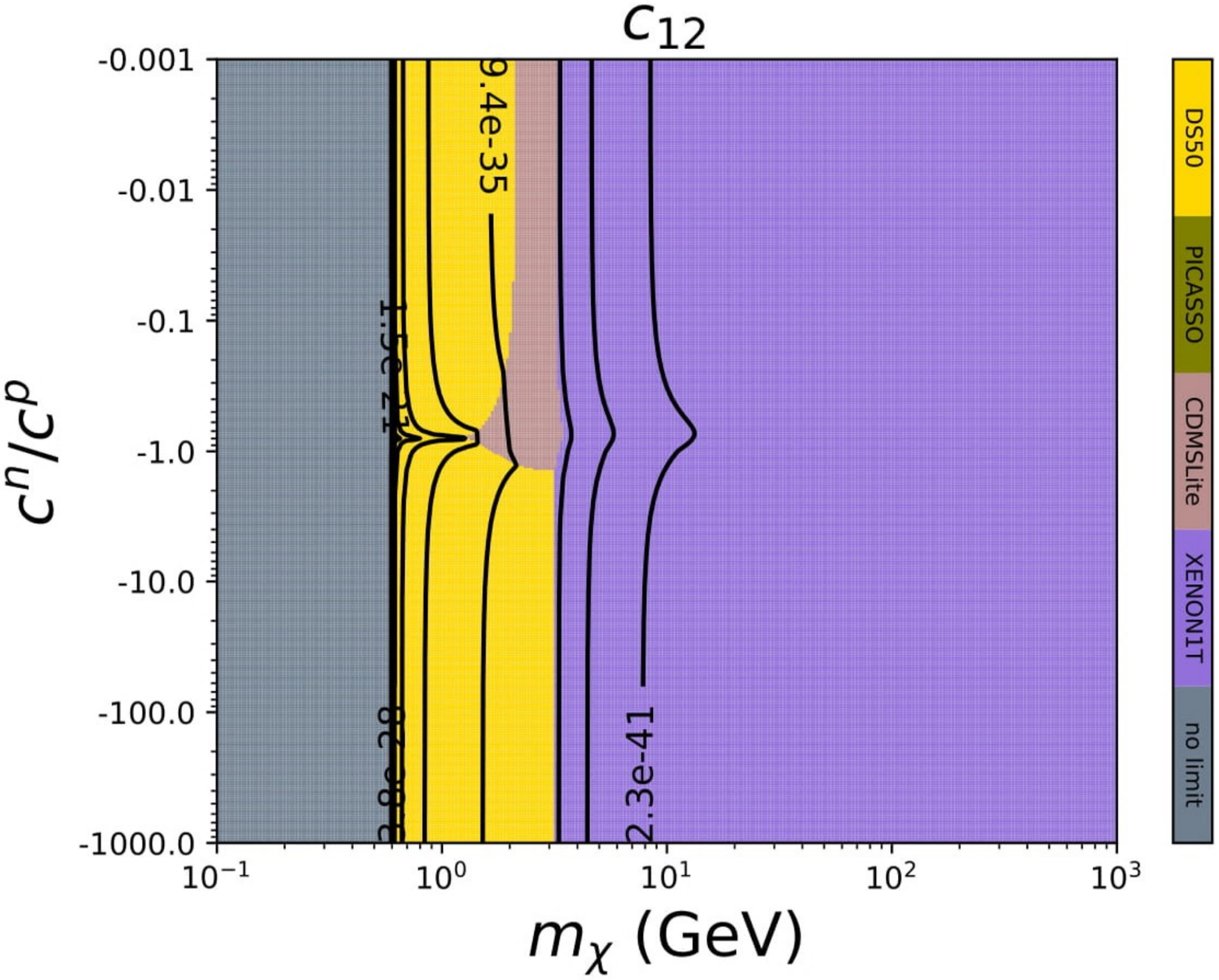}
\end{center}
\caption{The same as in Fig.\ref{fig:c1_plane} for the operator ${\cal O}_{12}$.}
\label{fig:c12_plane}
\end{figure}

\begin{figure}
\begin{center}
\includegraphics[width=0.85\columnwidth]{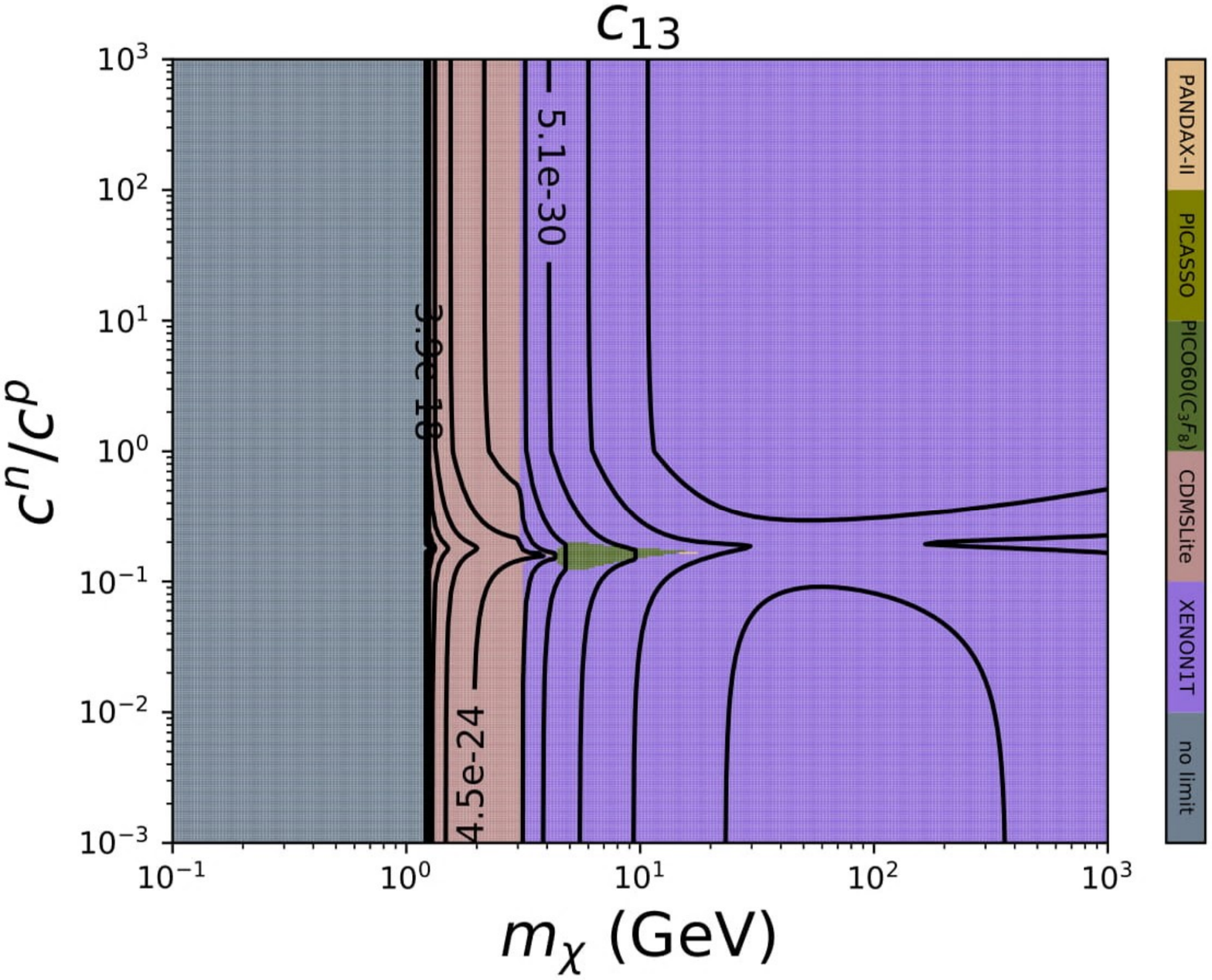}
\includegraphics[width=0.85\columnwidth]{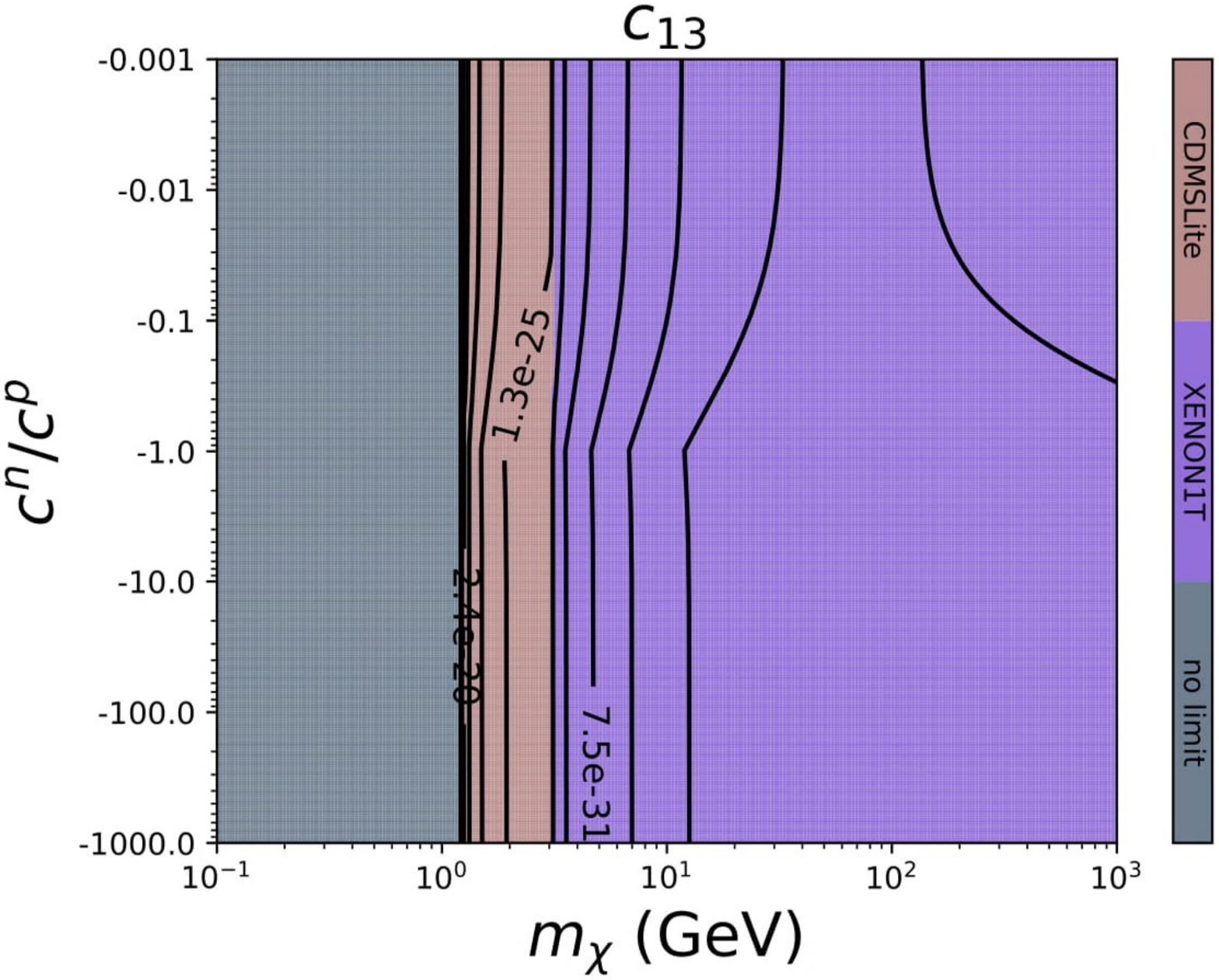}
\end{center}
\caption{The same as in Fig.\ref{fig:c1_plane} for the operator ${\cal O}_{13}$.}
\label{fig:c13_plane}
\end{figure}

\begin{figure}
\begin{center}
\includegraphics[width=0.85\columnwidth]{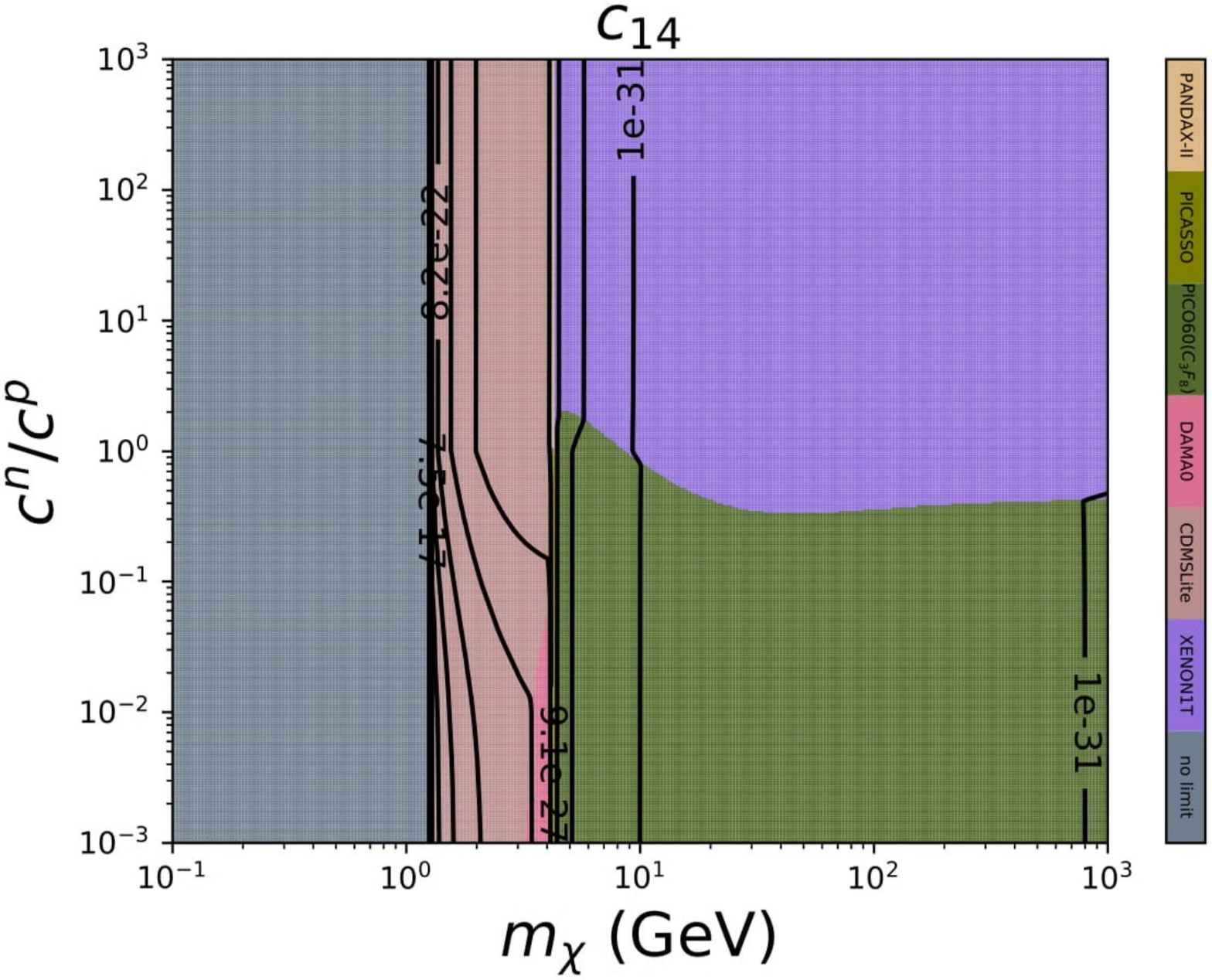}
\includegraphics[width=0.85\columnwidth]{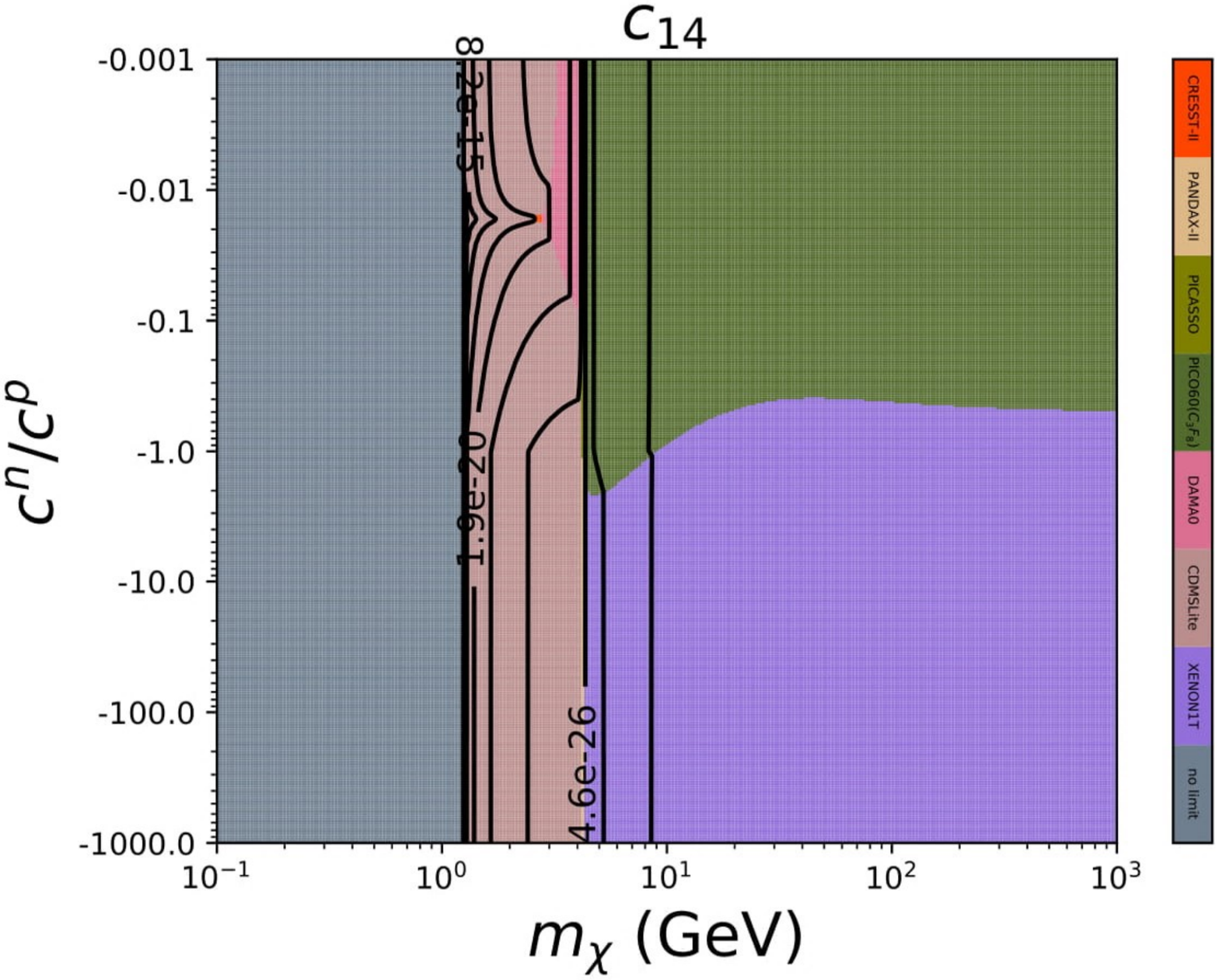}
\end{center}
\caption{The same as in Fig.\ref{fig:c1_plane} for the operator ${\cal O}_{14}$.}
\label{fig:c14_plane}
\end{figure}

\begin{figure}
\begin{center}
\includegraphics[width=0.85\columnwidth]{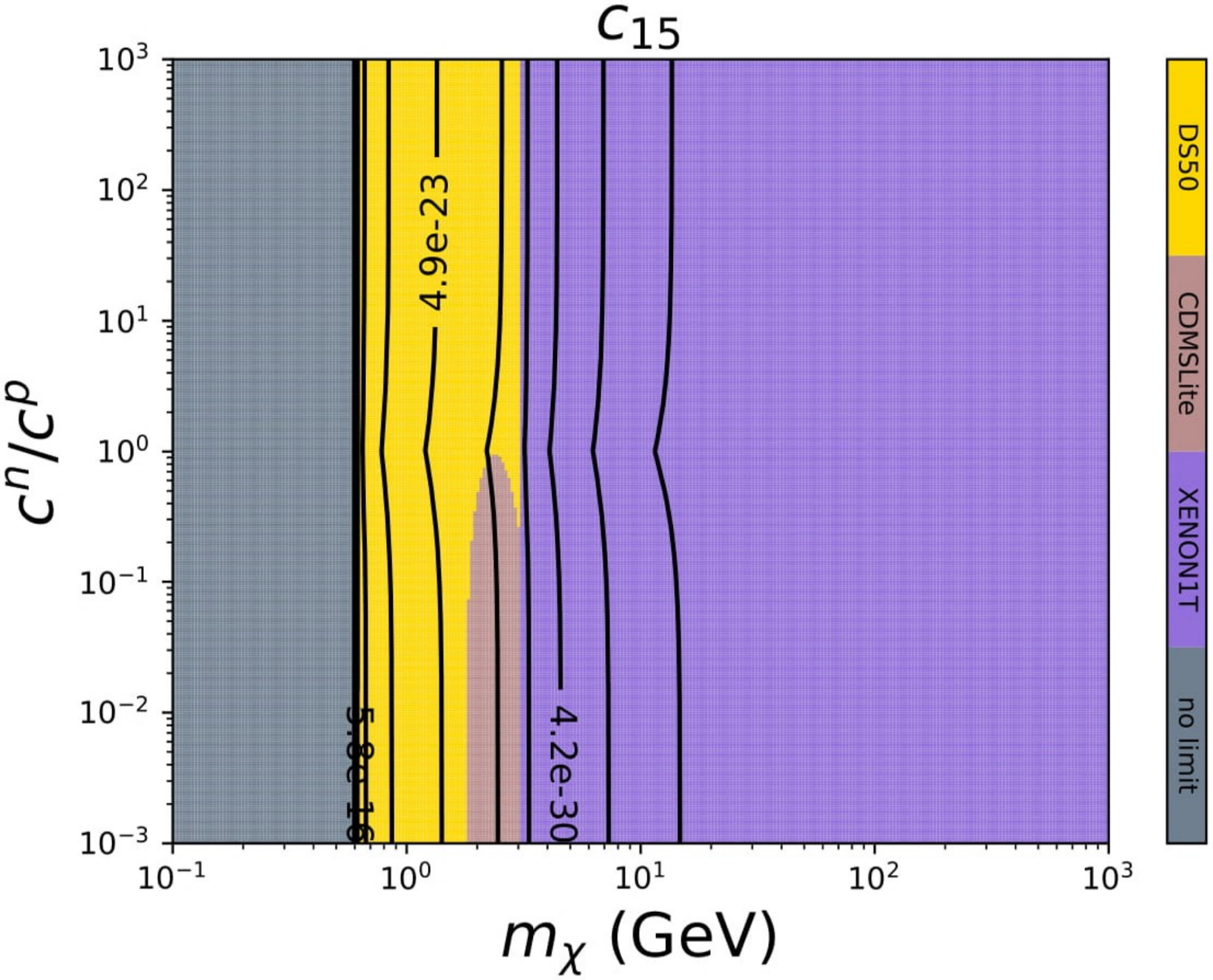}
\includegraphics[width=0.85\columnwidth]{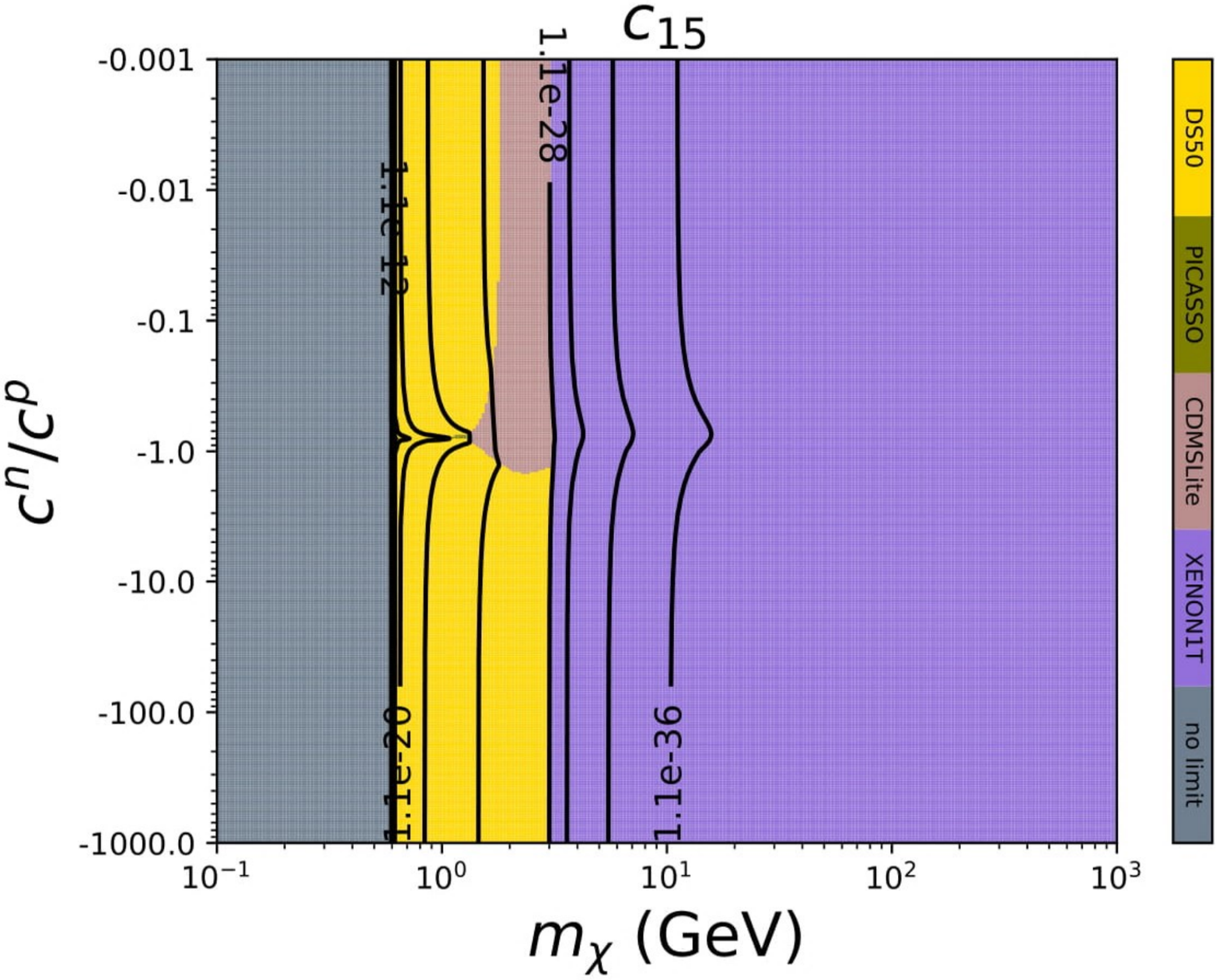}
\end{center}
\caption{The same as in Fig.\ref{fig:c1_plane} for the operator ${\cal O}_{15}$.}
\label{fig:c15_plane}
\end{figure}

The sensitivity of present experiments to each of the couplings of the
effective Hamiltonian of Eq.(\ref{eq:H}) is discussed in
Figs. \ref{fig:c1_plane}--\ref{fig:c15_plane}, which show the contour
plots of the most stringent 90\% C.L. bound on the effective
WIMP--nucleon cross section $\sigma_{\cal N}$, defined as:

\begin{equation}
\sigma_{\cal N}=\max(\sigma_{p},\sigma_{n}),
  \label{eq:conventional_sigma_nucleon}
\end{equation}

\noindent as a function of the WIMP mass $m_{\chi}$ and of the ratio
$c^n/c^p$ between the WIMP--neutron and the WIMP--proton
couplings. The numerical values in the figures indicate the most
stringent bound on $\sigma_{\cal N}$ in cm$^2$.  In each plot the
different shadings (colors) indicate the experiment providing the most
constraining bound, as indicated in the corresponding legend.  To make
such plots of practical use, in~\ref{app:program}
we introduce {\verb NRDD_constraints }, a simple interpolating code written in Python
that allows to extract the numerical values of $\sigma_{\cal N}$ from
Figs.~\ref{fig:c1_plane}--\ref{fig:c15_plane}. In the calculation of
all the plots the lower part of the 2--sigma DAMA modulation amplitude
region in the $m_{\chi}$--$\sigma_p$ plane is included as if it were
an additional constraint, in order to locate possible regions of
compatibility between the DAMA excess and other constraints in the
parameter space. As can be seen from Figs.
\ref{fig:c1_plane}--\ref{fig:c15_plane} DAMA never appears as the most
constraining bound, indicating that an explanation of its annual
modulation excess in terms of a WIMP signal is in tension with the
constraints of other experiments no matter which of the effective
operators among those in Eq.(\ref{eq:H}) is assumed to dominate in the
WIMP--nucleus interaction and for all $(m_{\chi}$--$\sigma_{\cal N}$)
combinations. This result is in agreement with the findings of
Ref.~\cite{Catena_dama,dama_2018_sogang}.

The different patterns of the regions appearing in
Figs. \ref{fig:c1_plane}--\ref{fig:c15_plane} can be understood with
the help of Table \ref{table:eft_summary} and
Fig.\ref{fig:scaling_law}. In particular, the velocity--dependent
contribution proportional to $(v^{\perp}_T)^2$ is negligible or absent
in all cases with 5 exceptions: $c_7$ and $c_{14}$ (where the
velocity--independent term is not present), $c_5$ and $c_8$ (where it
is enhanced by the $M$ coherent response functions) and for the
coupling $c_{13}$ (see below). As a consequence of this, the
interaction terms $c_1$, $c_5$, $c_8$ and $c_{11}$ depend on the $M$
coherent response function with a consequent strong sensitivity of
xenon detectors (XENON1T and PANDAX--II) except for $c^n/c^p\simeq$
-0.7 corresponding to a suppression on xenon targets. By the same
token, the interactions terms $c_4$, $c_6$, $c_7$, $c_9$, $c_{10}$ and
$c_{14}$ depend on the response functions $\Sigma^{\prime\prime}$
and/or $\Sigma^{\prime}$, that are related to the spin--dependent
coupling of Eq.(\ref{eq:sd}): in particular, $\Sigma^{\prime\prime}$
corresponds to the coupling of the WIMP to the component of the
nucleon spin along the direction of the transferred momentum $\vec{q}$
while $\Sigma^{\prime}$ to that perpendicular to it, with
$W^{\tau\tau^{\prime}}_{\Sigma^{\prime}}(q^2)\simeq 2
W^{\tau\tau^{\prime}}_{\Sigma^{\prime\prime}}(q^2)$ when
$q^2\rightarrow 0$. Since inside nuclei the nucleons spins tend to
cancel each other the contribution from even--numbered nucleons to the
response functions $\Sigma^{\prime\prime}$ and $\Sigma^{\prime}$ is
strongly suppressed. As a consequence of this for such interactions
neutron--odd targets (such as xenon and germanium) are mostly
sensitive to the regime $|c^n/c^p|\gsim$ 1 while proton--odd ones
(such as fluorine and iodine) mainly constrain the opposite case
$|c^n/c^p|\lsim$ 1. This reflects in the pattern of the shaded areas
of Figs.  \ref{fig:c4_plane}, \ref{fig:c6_plane}, \ref{fig:c7_plane},
\ref{fig:c9_plane}, \ref{fig:c10_plane} and \ref{fig:c14_plane}, where
for $m_{\chi}\gsim$ 1 GeV the PICASSO and PICO(C$_3$F$_8$) bounds
(using proton--odd fluorine) are the most constraining limits for
$|c^n/c^p|\lsim$ 1 (with the exception of $c_6$, where also
PICO(CF$_3$I) becomes competitive in spite of the relatively large
energy threshold ($E_R$=13.6 keV) due to the $q^4$ momentum dependence
that enhances the Iodine nuclear response function), while XENON1T and
PANDAX-II (using neutron--odd xenon) drive the constraints for
$|c^n/c^p|\gsim$ 1.  On the other hand, at lower masses the constraint
is driven by CDMSlite, which is the experiment with a non--vanishing
spin target (germanium) which has the lowest energy threshold (in
particular DS50 and CRESST-II do not put any constraint in this regime
since argon and oxygen are spinless nuclei). One can also notice in
Figs. \ref{fig:c7_plane} and \ref{fig:c14_plane} the loss of
sensitivity of PICASSO at low WIMP mass ($m_{\chi}\lsim$ 5 GeV) for
the velocity--dependent couplings $c_7$ and $c_{14}$ compared to the
case of other spin--dependent couplings: this is due to the fact that
the constraint from PICASSO is driven by its low energy threshold
($E_{th}\simeq$ 1 keV) and consequent low $v_{min}$: however, for a
velocity--dependent cross section the contribution to the rate of the
part of the velocity integral close to $v_{min}$ is suppressed by the
term $v^2-v_{min}^2$, weakening the corresponding bound.

Another set of plots with a similar color pattern is given by
Figs.\ref{fig:c3_plane}, \ref{fig:c12_plane} and \ref{fig:c15_plane},
corresponding to the couplings $c_3$, $c_{12}$ and $c_{15}$. As can be
seen from Table \ref{table:eft_summary} in all these cases expected
rates are driven by the $\Phi^{\prime\prime}$ nuclear response
function, which is related to spin-orbit coupling $\vec{\sigma}\cdot
\vec{l}$~\cite{haxton1}. Such response function is non vanishing for
all nuclei and favors heavier elements with large nuclear shell model
orbitals not fully occupied and, as can be seen from
Fig.\ref{fig:scaling_law}, its scaling with the nuclear target is
similar to the SI interaction (albeit the corresponding nuclear
response functions are about two orders of magnitude smaller), with
XENON1T the most constraining experiment at large--enough WIMP masses
and DS50 constraining the low--mass range due to its low velocity
threshold. In all the three figures one can also observe a region for
2 GeV $\lsim m_{\chi}\lsim$ 4 GeV and $|c^n| < |c^p|$ where the CDMSlite
constraint becomes competitive with DS50. Indeed, in this range of
masses the two constraints are quite close in all the range of
$c^n/c^p$, with a slight weakening of the CDMSlite bound for
$|c^n|>|c^p|$ due to the suppressed response off neutrons in the
semi--magic isotope $^{72}Ge$ (the dot product $\vec{\sigma}\cdot
\vec{l}$ vanishes for completely filled angular momentum
orbitals~\cite{haxton1}), as can be seen in the right--hand plot of
Fig.\ref{fig:scaling_law}.

The case of the coupling $c_{13}$ shown in Fig.\ref{fig:c13_plane} is
the only one that depends in a sizable way on the nuclear response
function $\tilde{\Phi}^{\prime}$, which is related to a
vector-longitudinal operator that transforms as a tensor under
rotations~\cite{haxton1,haxton2}. From the phenomenological point of
view, such operator requires a nuclear spin $j>$1/2, so that, among
the isotopes used in DM searches, it is non--vanishing only for the
four isotopes $^{23}$Na, $^{73}$Ge, $^{127}$I and $^{131}$Xe. Indeed,
the most stringent constraints arise in this case from CDMSlite at
low WIMP masses and from XENON1T at larger values. Nevertheless, in
Fig. \ref{fig:c13_plane} two fluorine detectors (PICASSO and PICO-60)
yield the stronger constraints in the mass range 5 GeV $\lsim m_{\chi}
\lsim$ 7 GeV and for $|c^n/c^p|<$1. This is explained by the
spin--dependent term with an explicit velocity dependence in the
decomposition of Eq.(\ref{eq:r_decomposition}), that, in spite of the
suppression due to the slow incoming WIMP speeds, can become as
constraining as the velocity--independent coupling off xenon in
XENON1T.

We conclude our discussion with a comment on the dependence of the
direct detection signal on the recoil energy.  Besides a different
scaling of the cross section with the target, the non--standard
interactions listed in Table~\ref{table:eft_summary} involve cases
where the cross section depends explicitly on the momentum transfer
$q=\sqrt{2 m_T E_R}$, implying a harder energy spectrum of the
expected signal compared to the usual exponentially decaying case
observed for the standard SI and SD cases. This may lead to a
weakening of the constraints compared to the standard case when, as
for DS50 and KIMS (see \ref{app:exp} and
Eq.(\ref{eq:bck_chi2})), a background estimation growing with energy
is subtracted from the data. Indeed, we observe this effect in our
analysis, but it is significant only for WIMP masses large enough for
the expected rate to be insensitive to the high--speed tail of the
velocity distribution. On the other hand large count rates requiring
background subtraction are typically present only in experiments that,
such as DS50 and KIMS, focus on a low energy threshold to reach a
competitive sensitivity at low WIMP masses at the expense of the
efficiency of their background discrimination. For such low values of
$m_{\chi}$ the signal spectrum has a steep decay with energy also in
presence of a $q^n$ term in the cross section and the effectiveness of
background subtraction is similar to the standard case. A dependence
of the constraint on the expected signal spectral shape enters also in
the optimal--interval method \cite{yellin} that we have applied in the
case of SuperCDMS.

\begin{figure}
\begin{center}
\includegraphics[width=0.49\columnwidth]{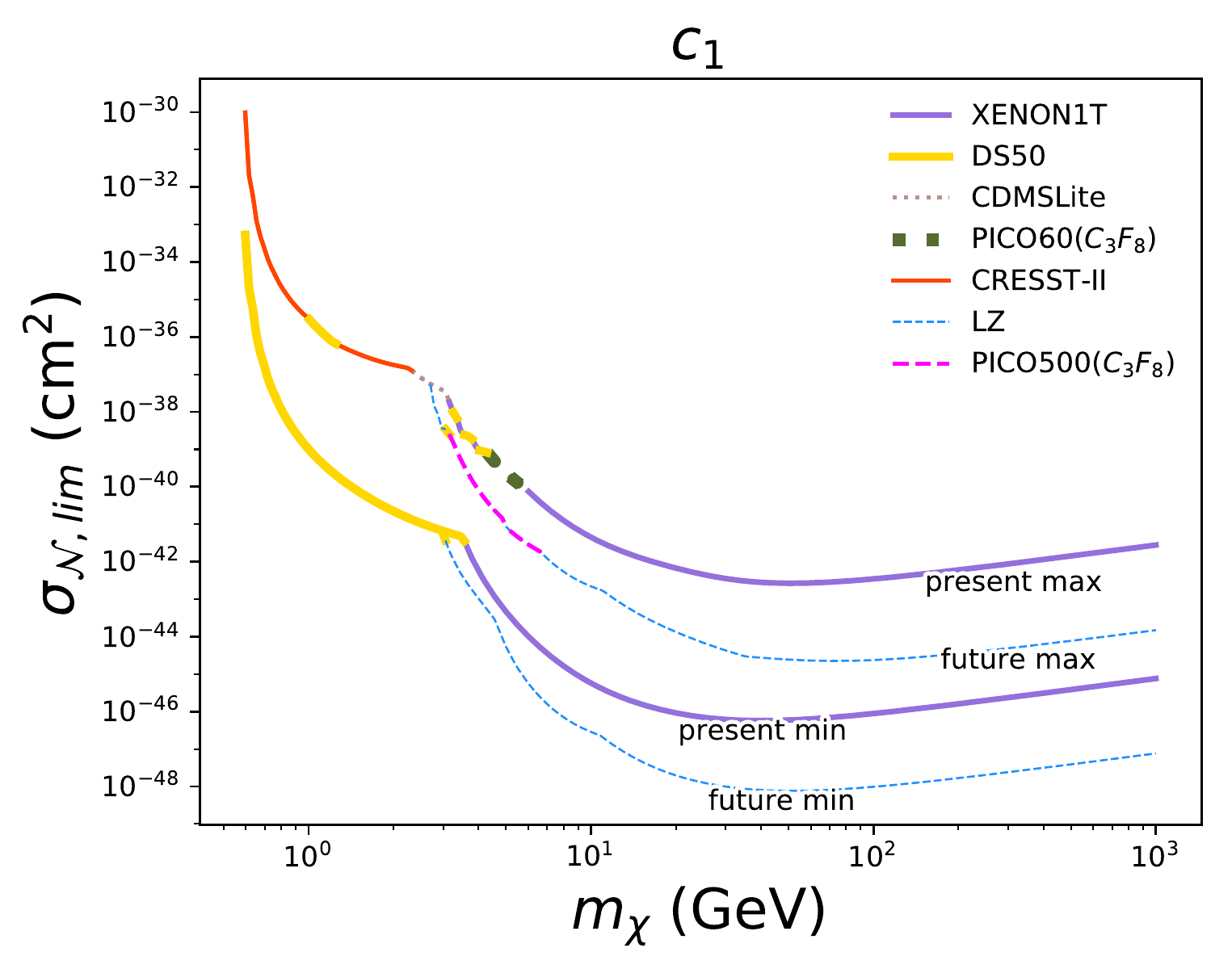}
\includegraphics[width=0.49\columnwidth]{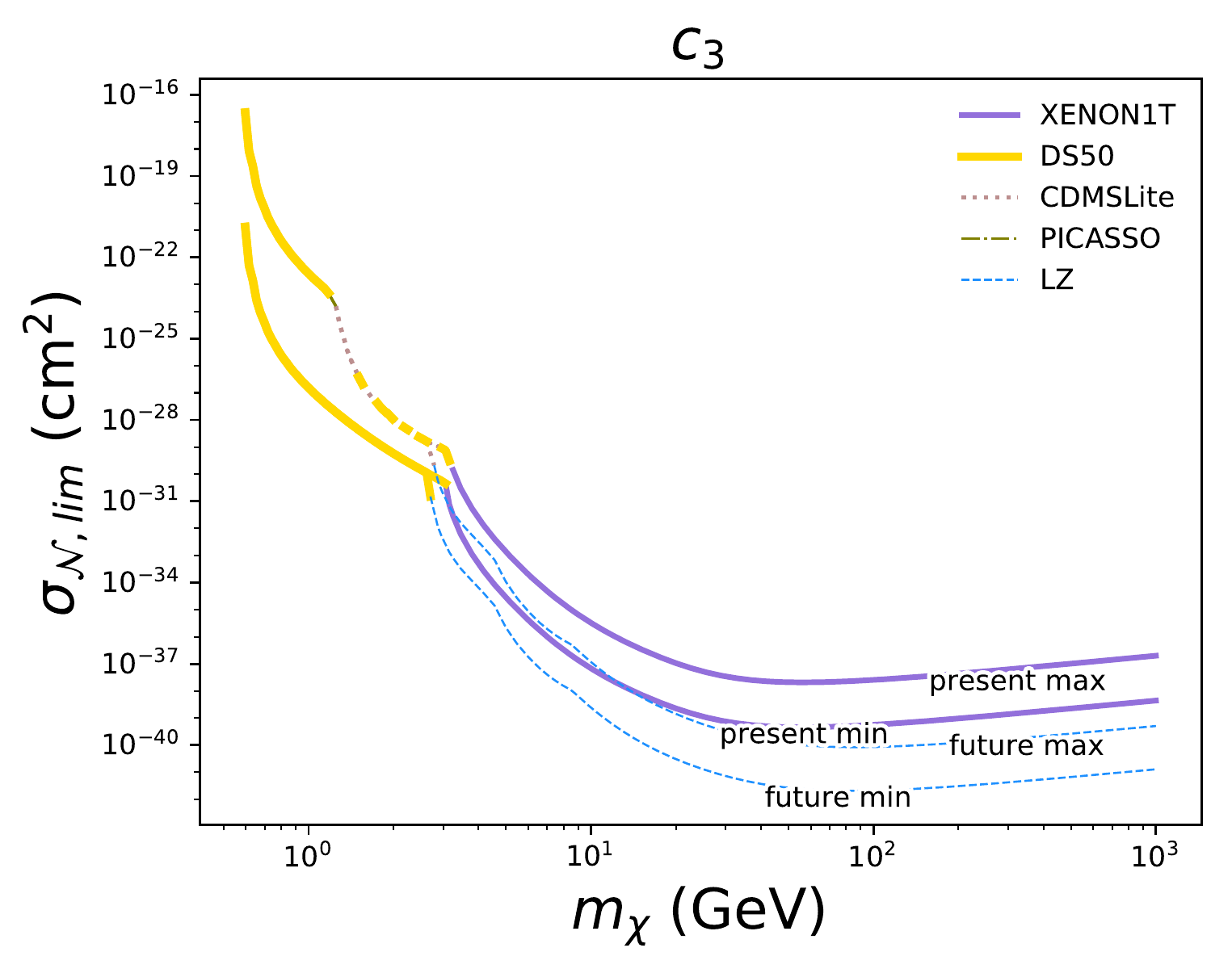}
\includegraphics[width=0.49\columnwidth]{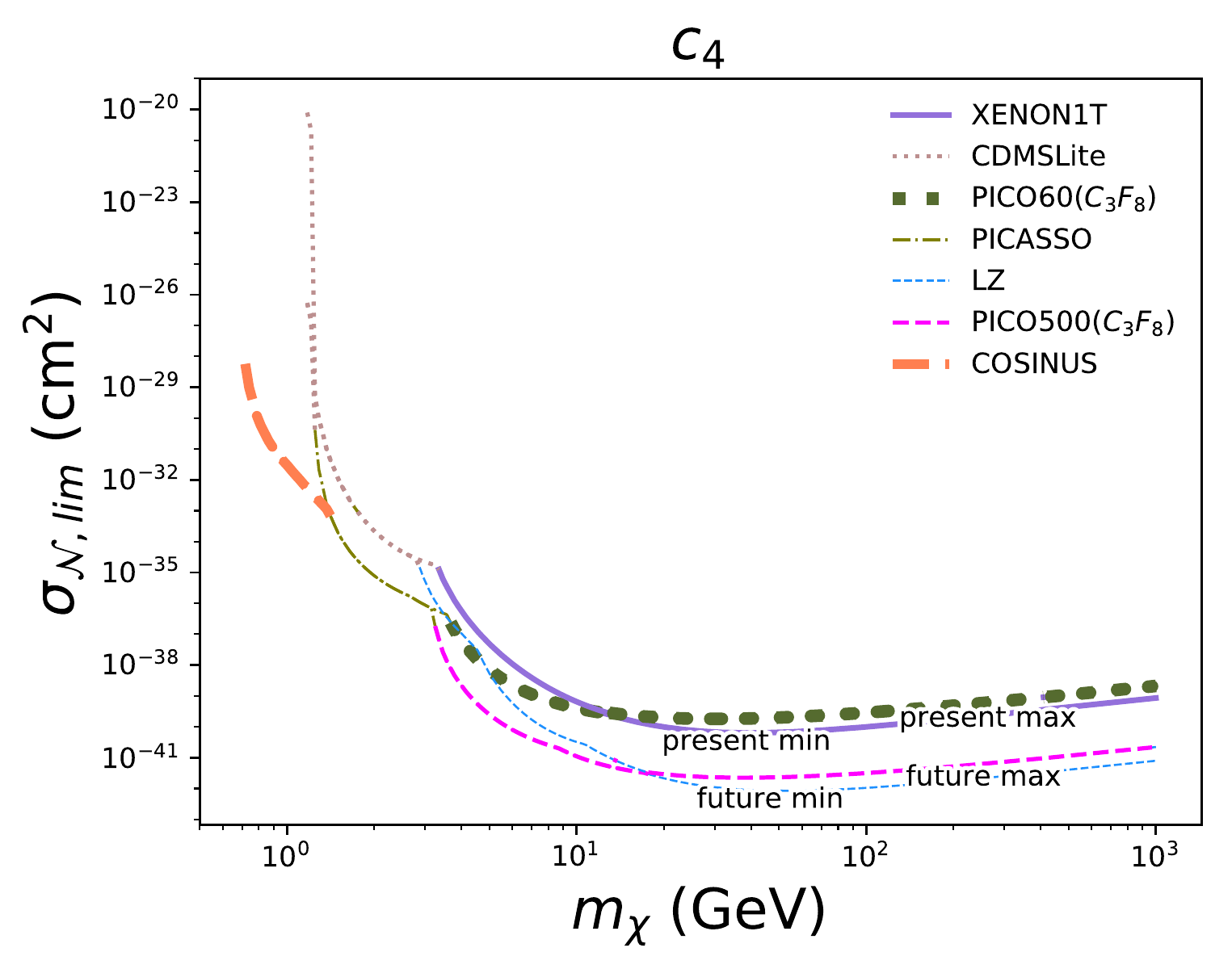}
\includegraphics[width=0.49\columnwidth]{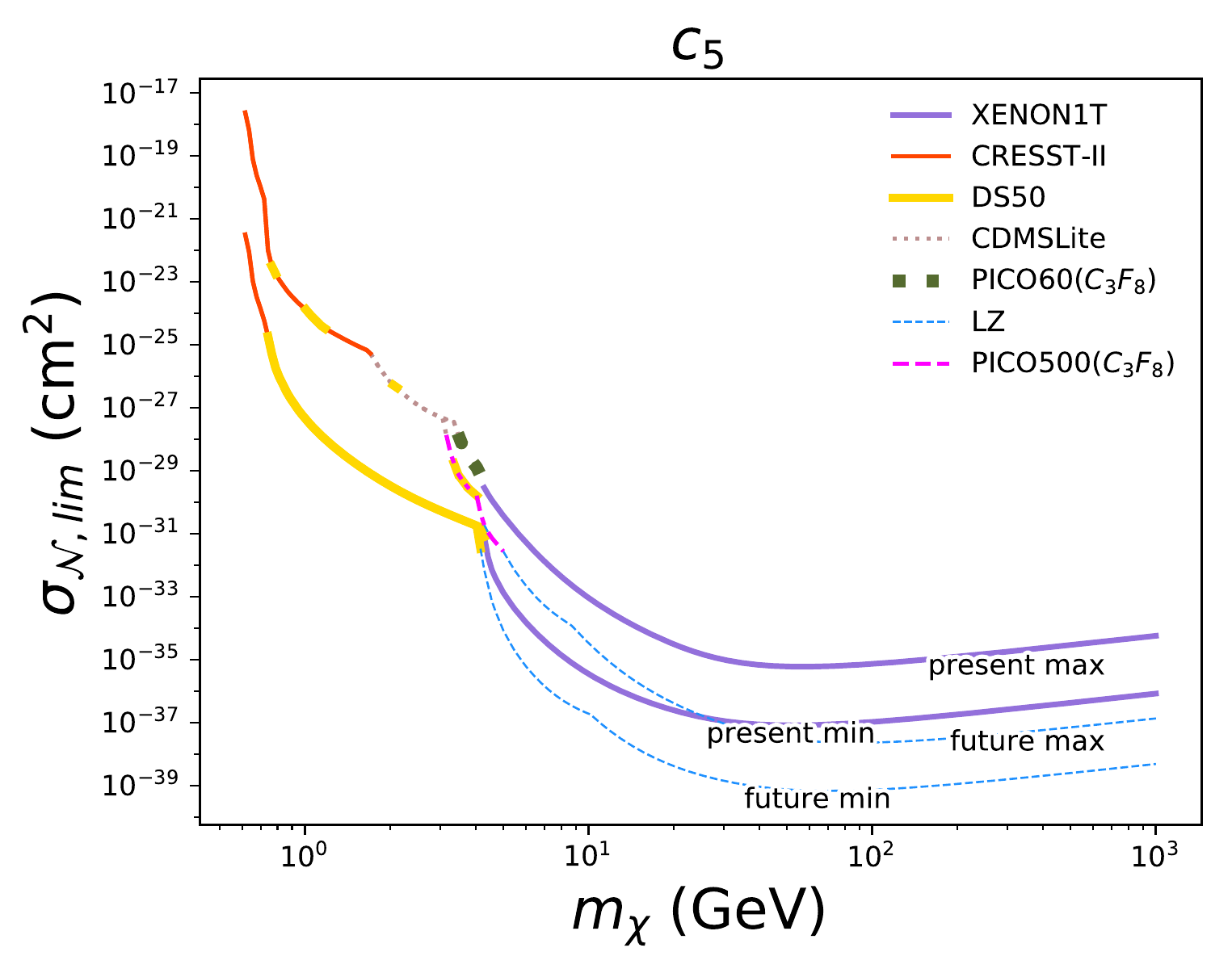}
\includegraphics[width=0.49\columnwidth]{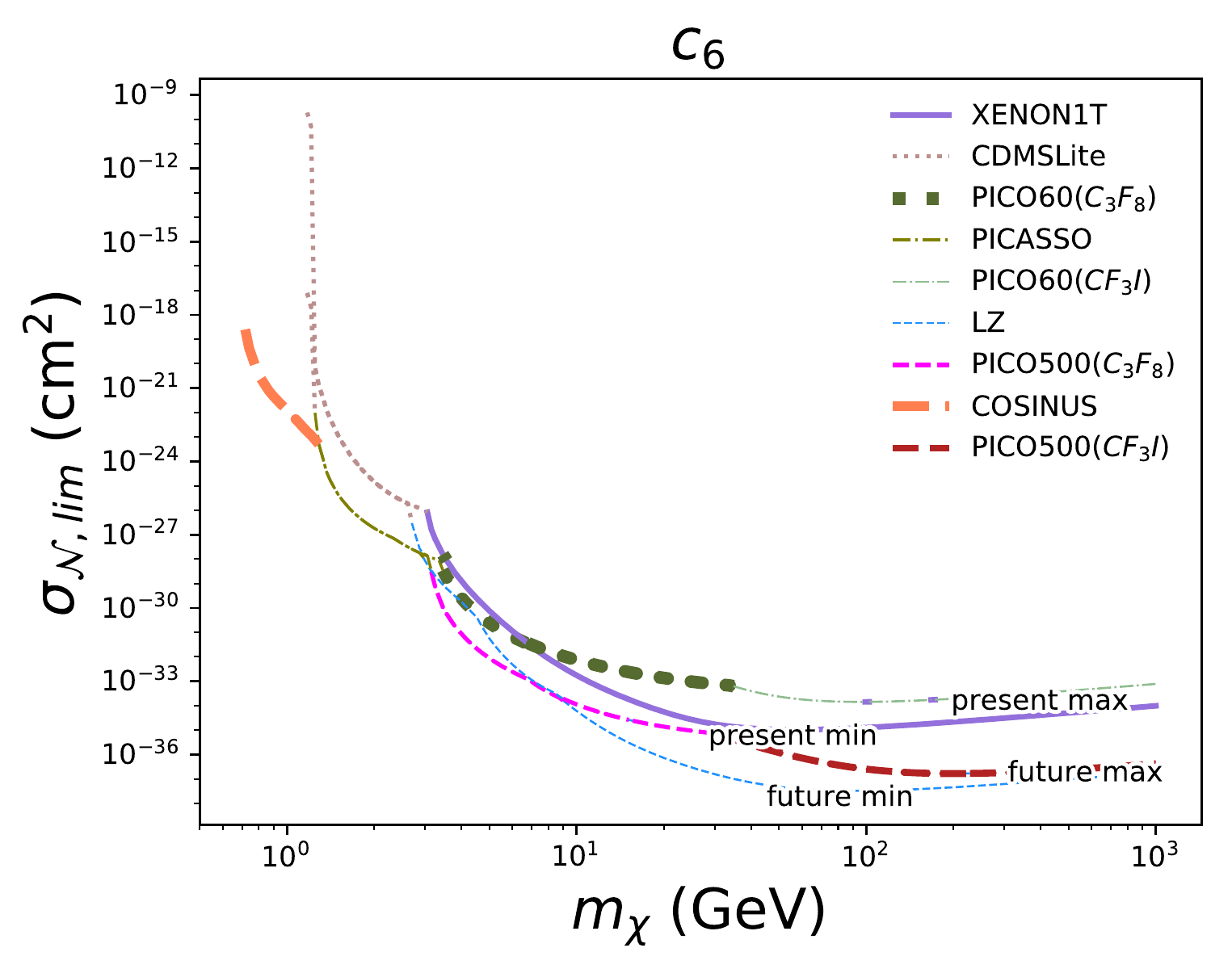}
\includegraphics[width=0.49\columnwidth]{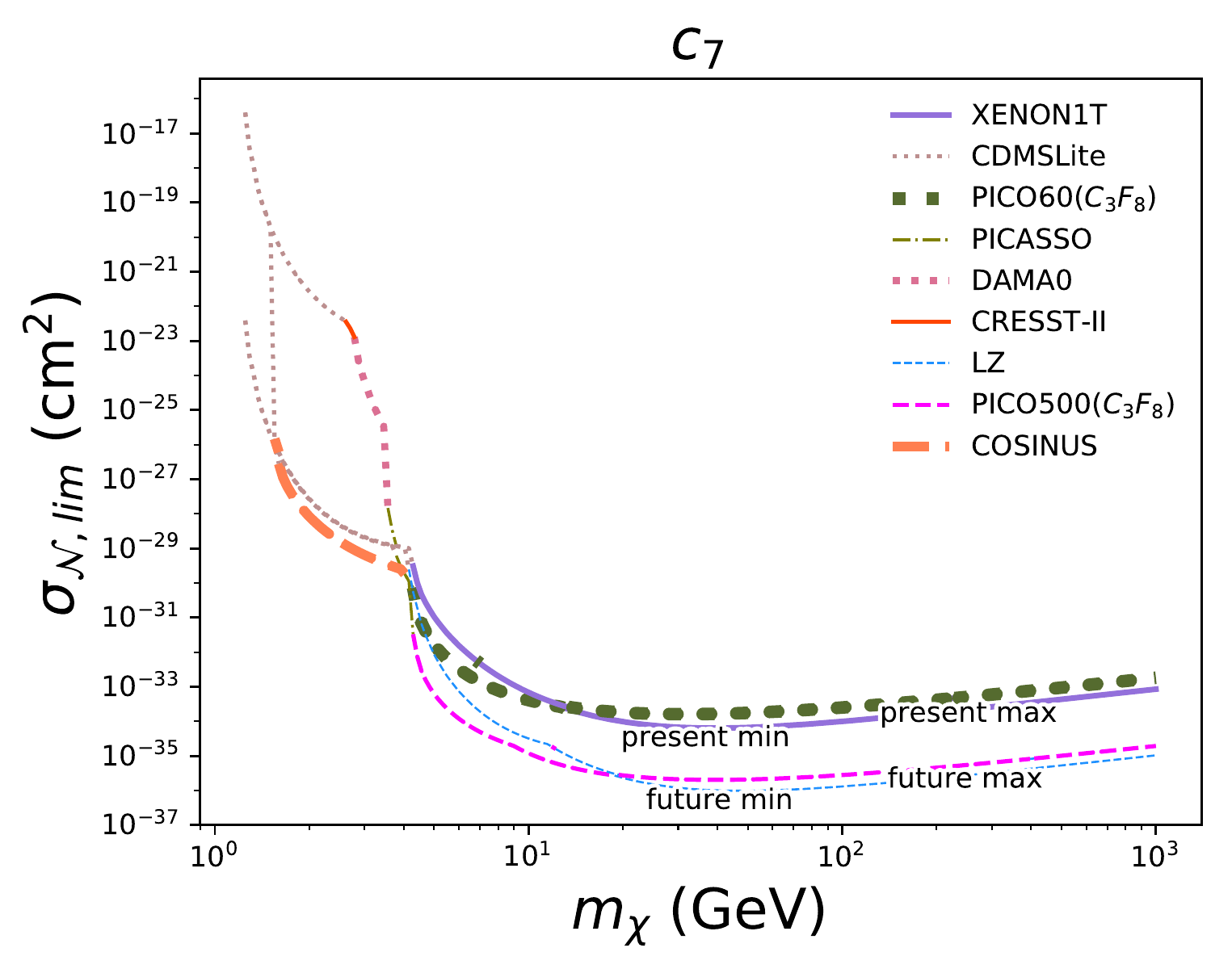}

\end{center}
\caption{Most stringent bound among those from the experiments listed
  in Fig.\ref{fig:c1_summary} on the effective WIMP--nucleon cross
  section $\sigma_{{\cal N},lim}$ defined in
  Eq.(\ref{eq:conventional_sigma_nucleon}) as a function of the WIMP
  mass $m_{\chi}$ for operators $c_1$, $c_3$, $c_4$, $c_5$, $c_6$ and
  $c_7$. In each plot the two curves indicated by ``present min'' and
  ``present max'' show the range of the limit from present experiments
  on $\sigma_{{\cal N},lim}$ when $c^n/c^p$ is varied, while the
  curves indicated by ``future min'' and ``future max'' show the same
  range when the limits from projected experiments are included. In
  each curve the different styles indicate the experiment providing
  the most stringent bound, as shown by the legend.}
\label{fig:summary_c1_c7}
\end{figure}

\begin{figure}
\begin{center}
\includegraphics[width=0.49\columnwidth]{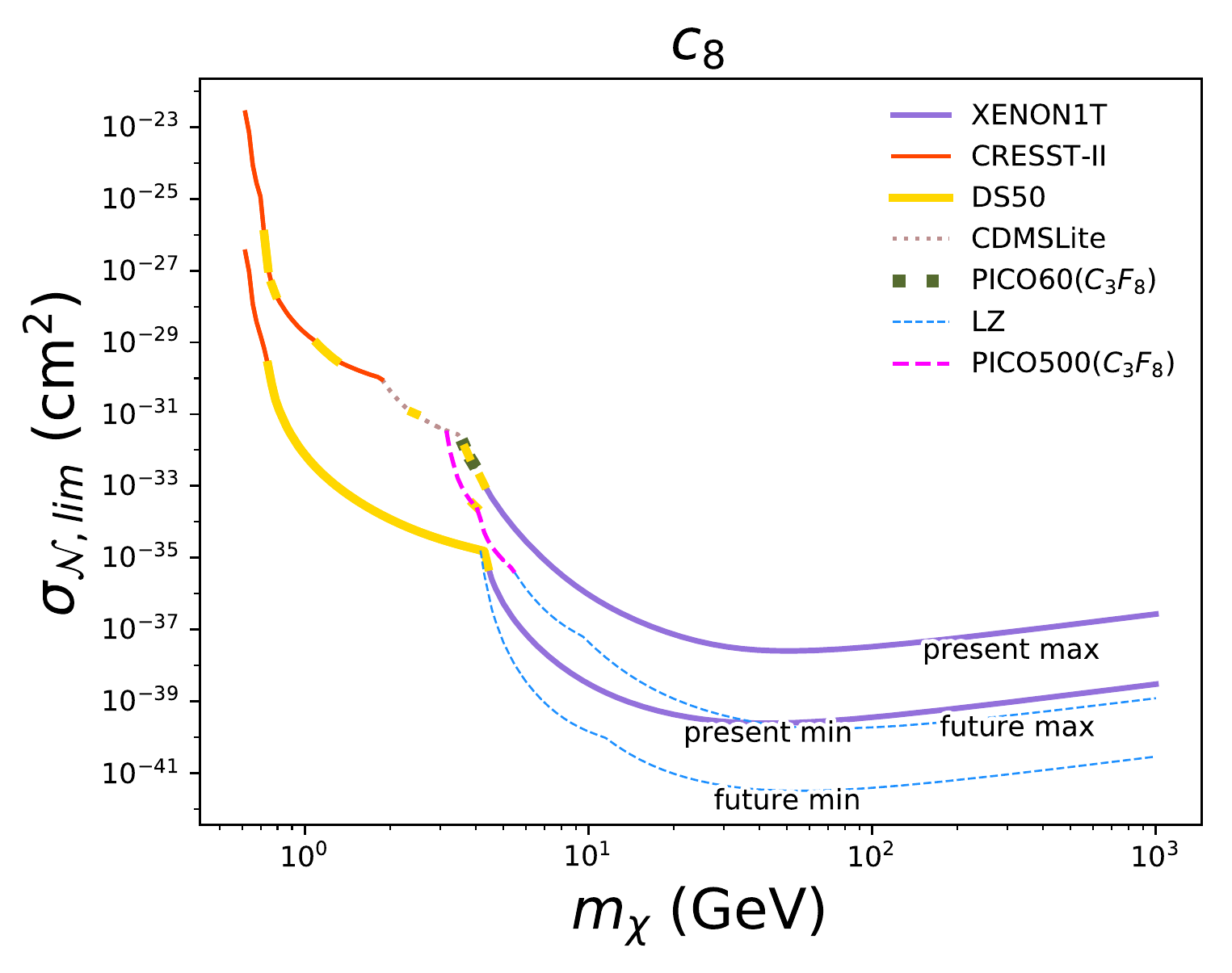}
\includegraphics[width=0.49\columnwidth]{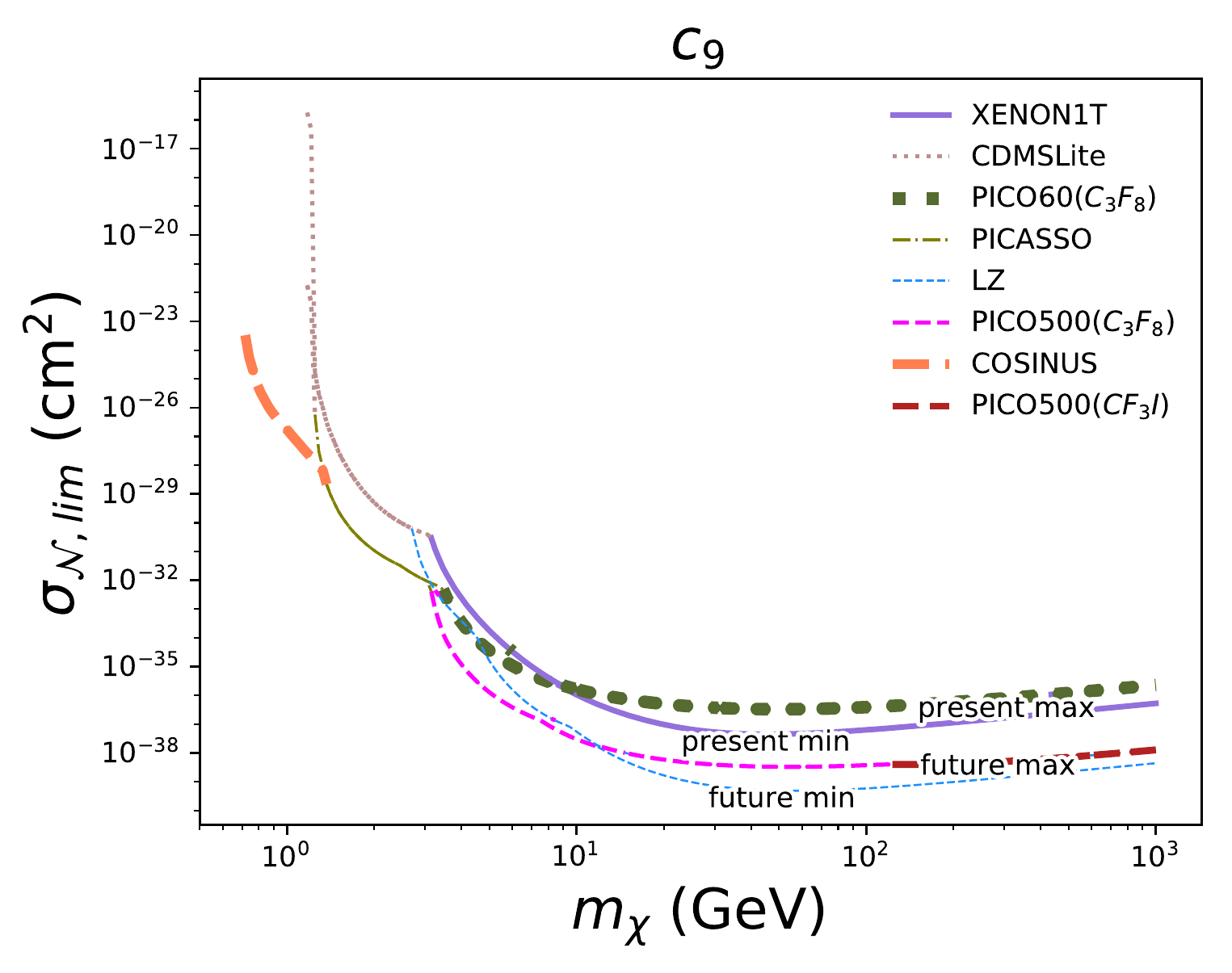}
\includegraphics[width=0.49\columnwidth]{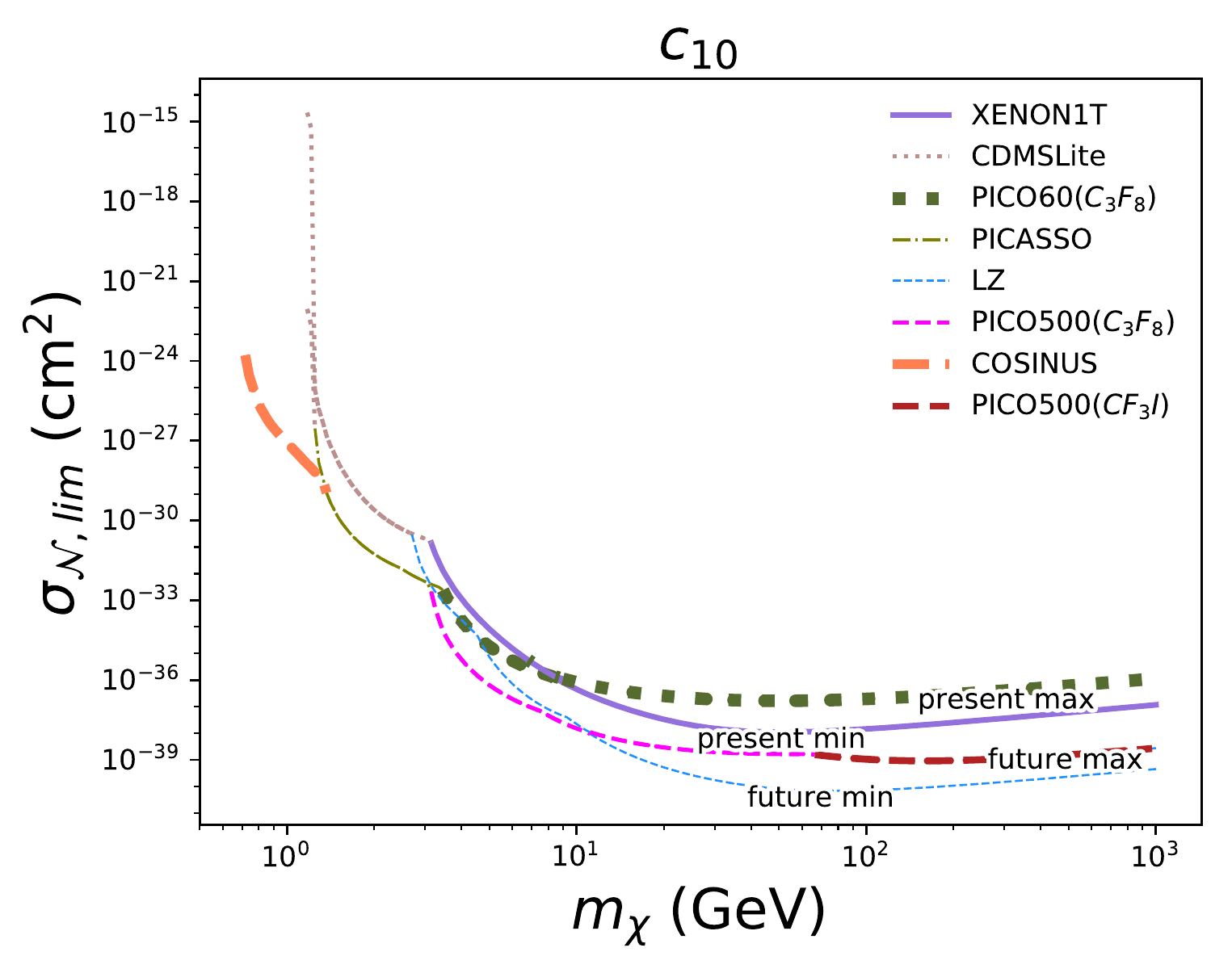}
\includegraphics[width=0.49\columnwidth]{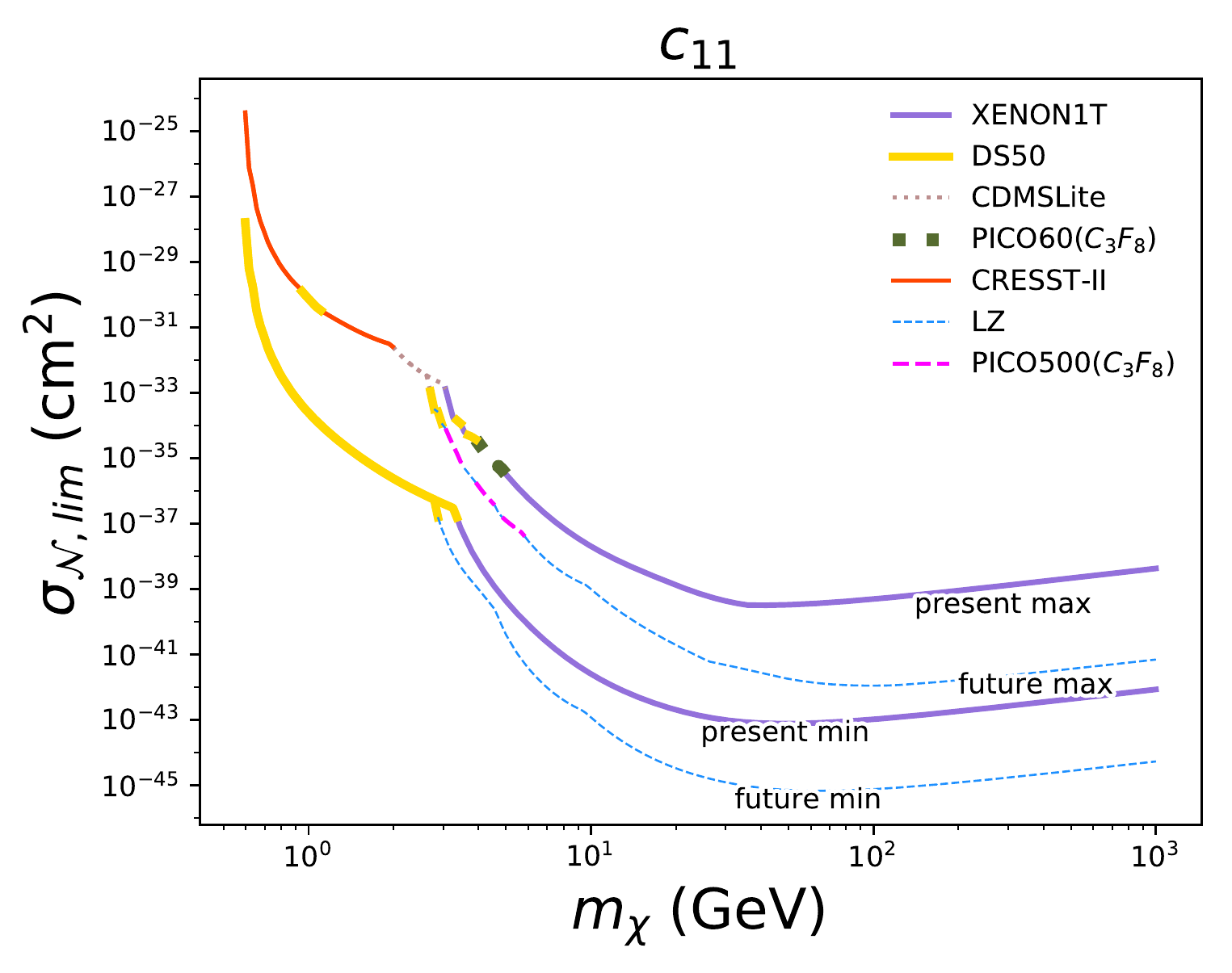}
\includegraphics[width=0.49\columnwidth]{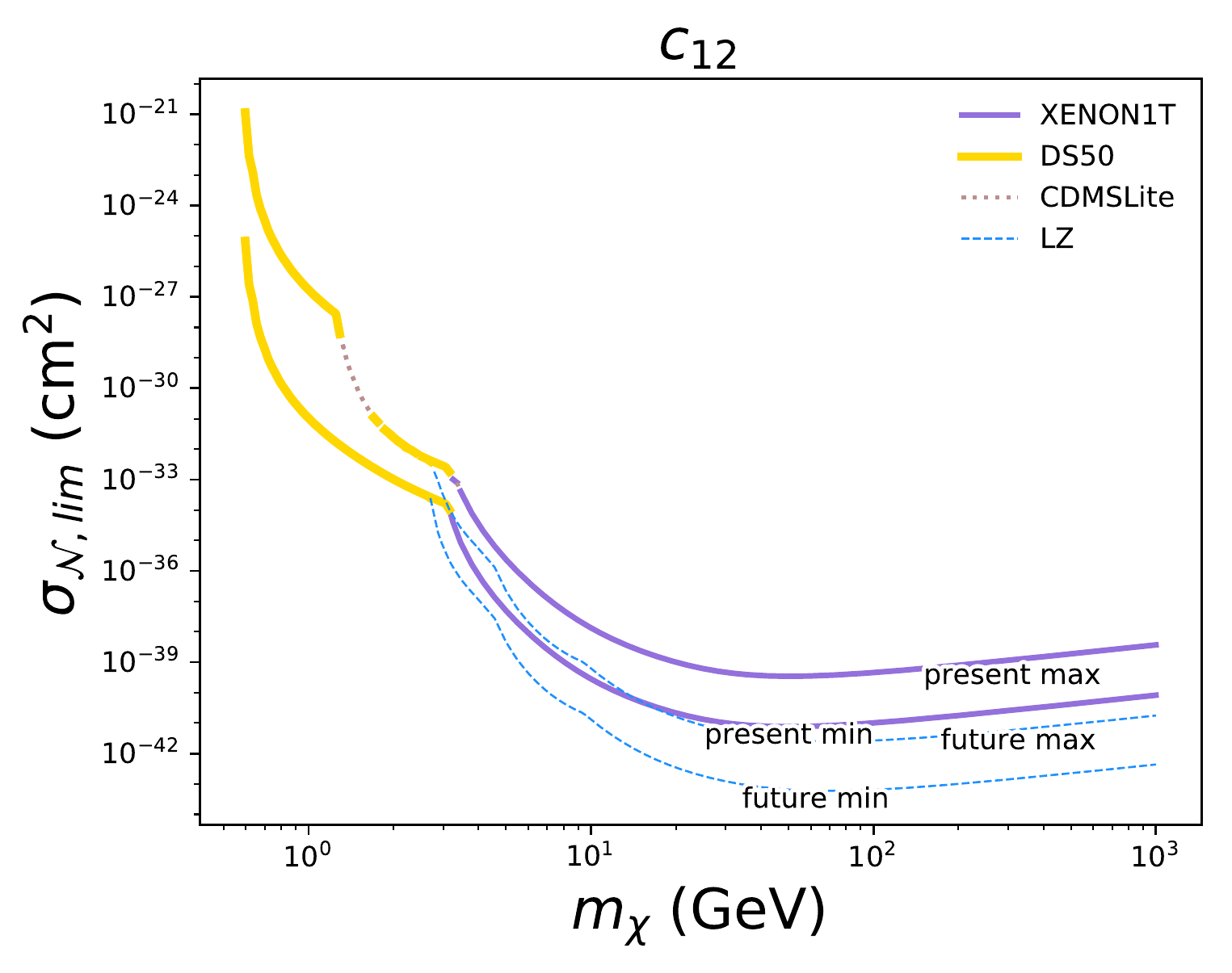}
\includegraphics[width=0.49\columnwidth]{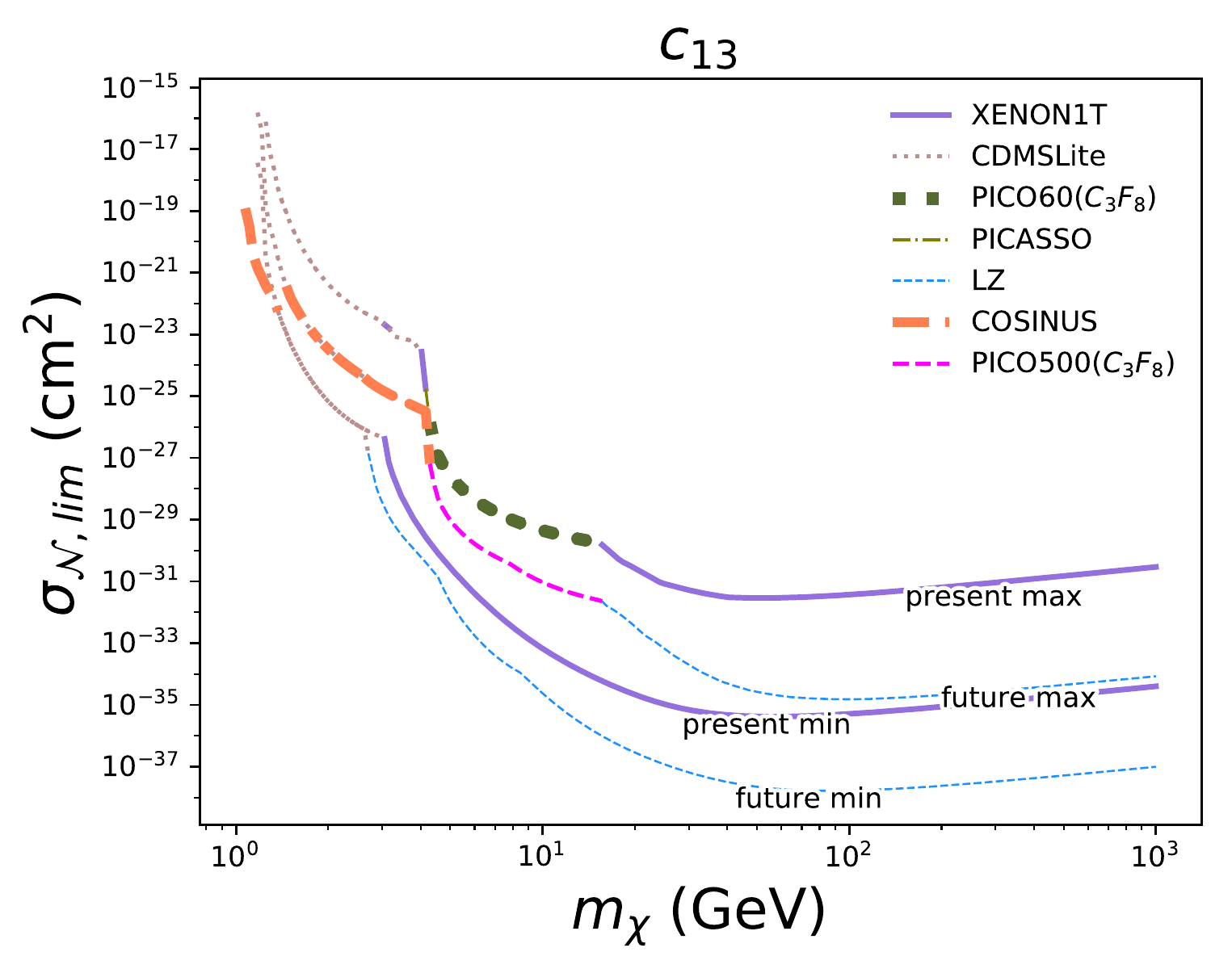}

\end{center}
\caption{The same as in Fig. \ref{fig:summary_c1_c7} for operators
  $c_8$, $c_9$, $c_{10}$, $c_{11}$, $c_{12}$ and $c_{13}$.}
\label{fig:summary_c8_c13}
\end{figure}

\begin{figure}
\begin{center}
\includegraphics[width=0.49\columnwidth]{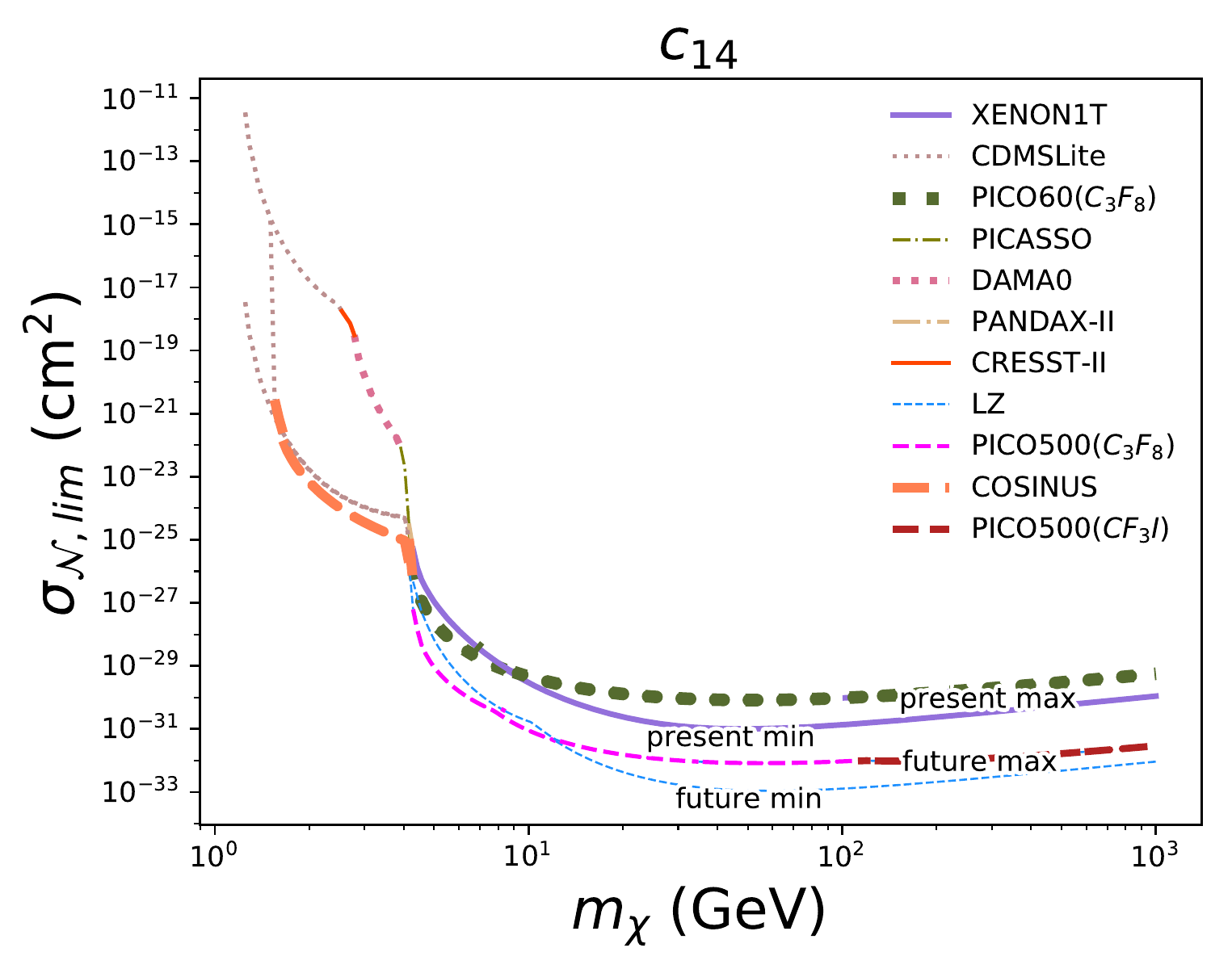}
\includegraphics[width=0.49\columnwidth]{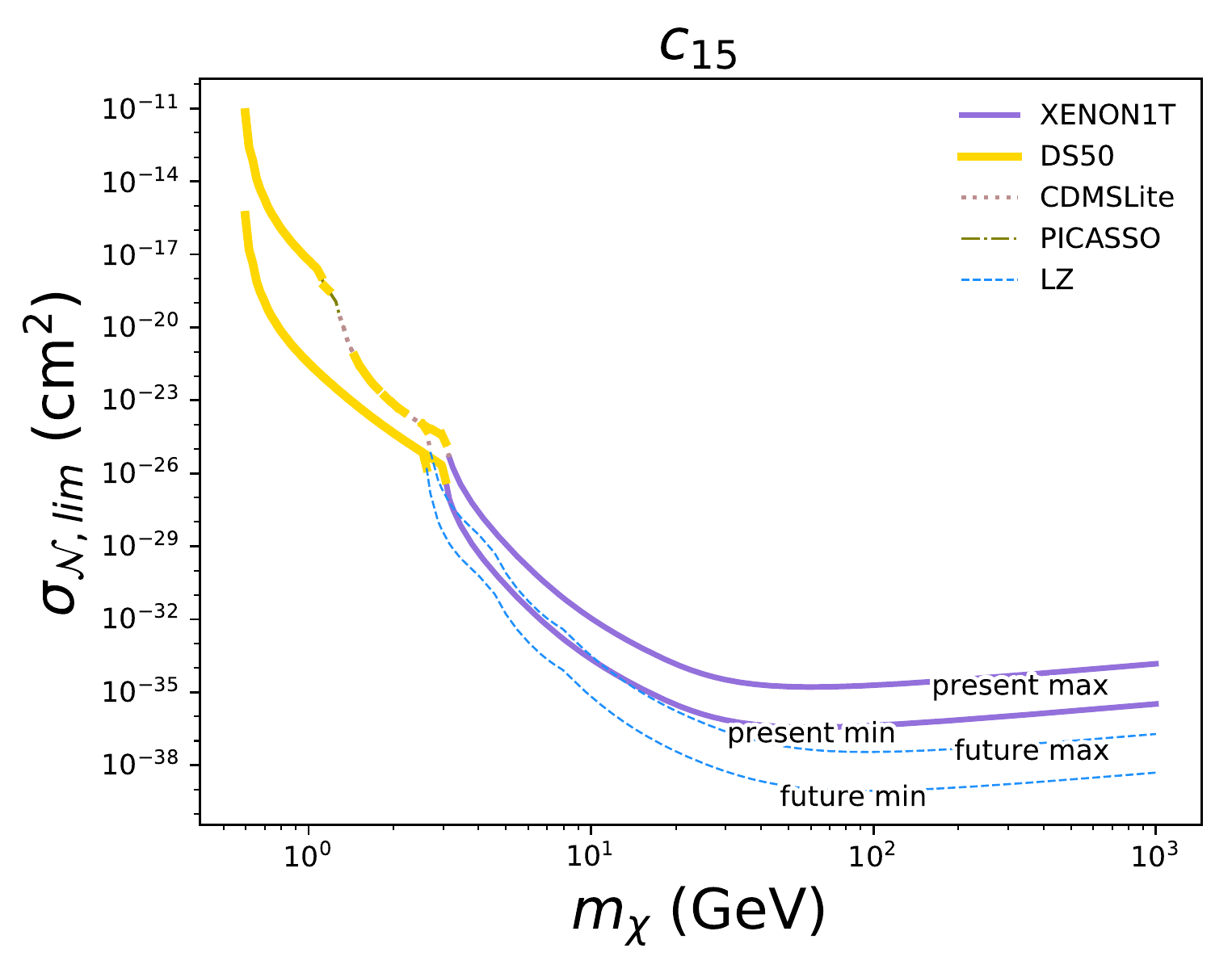}

\end{center}
\caption{The same as in Fig. \ref{fig:summary_c8_c13} for operators
  $c_{14}$ and $c_{15}$.}
\label{fig:summary_c14_c15}
\end{figure}

The full set of bounds is summarized in
Fig. \ref{fig:summary_c1_c7}--\ref{fig:summary_c8_c13}, where for each
of the couplings of the Hamiltonian of Eq.(\ref{eq:H}) the most
stringent constraint on $\sigma_{{\cal N},lim}$ is plotted as a
function of the WIMP mass $m_{\chi}$. In each plot the two curves
indicated by ``present min'' and ``present max'' show the range of the
most stringent limit at fixed $m_{\chi}$ on $\sigma_{{\cal N},lim}$
from present experiments when the ratio $c^n/c^p$ is varied in the
same interval of Figs.\ref{fig:c1_plane}--\ref{fig:c15_plane}, while
the curves indicated by ``future min'' and ``future max'' show the
same range when the expected bound from some projected experiments
(LUX--ZEPLIN (LZ), PICO-500 (C$_3$F$_8$), PICO-500 (CF$_3$I) and
COSINUS) are included (see \ref{app:exp} for details). The styles of
each curve indicate the experiment providing the most stringent bound,
as shown by the corresponding legend.  Notice that at a given value of
the WIMP mass an upper bound on the effective cross section
$\sigma_{\cal N}$ corresponds to a lower bound on the effective
coupling $c^{n,p}$ which, in case of a spin--1/2 particle, has
dimension GeV$^{-2}$. Writing $c^{n,p}=1/M_{EFT}^2$ with $M_{EFT}$
a cut--off scale, the validity of the NR effective theory requires
$M_{EFT}\gsim 1$ GeV, which implies $\sigma_{\cal N}\lsim 10^{-30}
\mbox{cm}^2$. Some of the bounds on $\sigma_{\cal N}$ if
Figs.\ref{fig:summary_c1_c7}, \ref{fig:summary_c8_c13} and
\ref{fig:summary_c14_c15} are not compatible to such condition,
especially for low WIMP masses where expected rates are suppressed by
the velocity distribution. This can be simply interpreted as the fact
that in such regimes present sensitivities do not pose any sensible
bound on the corresponding coupling.

For each of the couplings of the
effective Hamiltonian the most stringent bounds from present and
future experiments on $\sigma_{{\cal N},lim}$ are tabulated in Table
\ref{tab:max_reach_r_strong}, where the ratio $c^n/c^p$ is fixed in
each case to the value that corresponds to the stronger constraint,
and in Table \ref{tab:max_reach_r_weak} when $c^n/c^p$ corresponds to
the weaker constraint. One can see that the expected reach on
$\sigma_{{\cal N},lim}$ varies by many orders of magnitude with the
effective coupling. For all the NR couplings we consider XENON1T
yields the most constraining bound among existing experiments for WIMP
masses below 1 TeV. Among future experiments LZ has the highest
sensitivity at all WIMP masses for $c_1$, $c_3$, $c_5$, $c_8$,
$c_{11}$, $c_{12}$, $c_{13}$ and $c_{15}$ while for other couplings
either LZ or PICO-500 (C$_3$F$_8$) corresponds to the most
constraining limit depending on $m_{\chi}$. On the other hand as shown
in Figs. \ref{fig:c1_plane}--\ref{fig:c15_plane}, 9 present
experiments out of the total of 14 considered in our analysis provide
the most stringent bound on some of the effective couplings for a
given choice of $(m_{\chi},c^n/c^p)$: XENON1T, PANDAX-II, CDMSlite,
PICASSO, PICO-60 (CF$_3$I), PICO-60 (C$_3$F$_8$), CRESST-II, DAMA0
(average count rate) and DarkSide--50. This is evidence of the
complementarity of different target nuclei and/or different
combinations of count--rates and energy thresholds when the search of
a DM particle is extended to a wide range of possible
interactions. The variation of the best reach on $\sigma_{{\cal
    N},lim}$ with $c^n/c^p$ is about 3 orders of magnitude for $c_1$,
$c_{11}$ and $c_{13}$, about 2 orders of magnitude for $c_8$, between
1 and 2 orders of magnitude for $c_{3}$, $c_{5}$, $c_{12}$ and
$c_{15}$, about one order of magnitude for $c_{6}$, $c_{10}$ and less
than one order of magnitude for $c_{4}$, $c_{7}$, $c_{9}$ and
$c_{14}$. For all couplings future experiments could improve the
present best reach between two and three orders of magnitude.



\clearpage

\clearpage
\clearpage
\clearpage
\clearpage
\clearpage
\clearpage
\clearpage
\clearpage
\clearpage
\clearpage
\clearpage
\clearpage
\clearpage


\clearpage


\clearpage


 \begin{table}
 \centering
 \caption{Most stringent constraints on the effective cross section
   $\sigma_{{\cal N},lim}$ for each of the couplings in the effective
   Hamiltonian of Eq.(\ref{eq:H}) among the present and future
   experiments included in our analysis. In each case the ratio
   $c^n/c^p$ is fixed to the value that corresponds to the most
   stringent bound.}
 \begin{tabular}{*5c}
 \toprule
 Coupling &  \multicolumn{2}{c}{Present} & \multicolumn{2}{c}{Future}\\
 \midrule
 {}   & $m_{\chi}$ (GeV)   & $\sigma_{{\cal N},lim}$(cm$^2$)   & $m_{\chi}$ (GeV)   & $\sigma_{{\cal N},lim}$(cm$^2$)\\
 $c_1$   &  40.6 & 5.7$\times 10^{-47}$   & 52.0 & 7.6$\times 10^{-49}$\\
 $c_3$   &  57.0 & 4.5$\times 10^{-40}$  & 82.5  & 2.1$\times 10^{-42}$\\
 $c_4$   &  39.4  &  6.3$\times 10^{-41}$   & 52.0  & 8.4$\times 10^{-43}$\\
 $c_5$   &  52.0  &  8.2$\times 10^{-38}$   & 70.7  & 6.9$\times 10^{-40}$\\
 $c_6$   &  57.0  &  1.0$\times 10^{-35}$   & 90.5  & 3.3$\times 10^{-38}$\\
 $c_7$   &  39.4  &  6.4$\times 10^{-35}$   & 47.4  & 9.8$\times 10^{-37}$\\
 $c_8$   &  43.2  &  2.4$\times 10^{-40}$   & 55.3  & 3.2$\times 10^{-42}$\\
 $c_9$   &  45.9  &  4.5$\times 10^{-38}$   & 57.0  & 4.9$\times 10^{-40}$\\
 $c_{10}$   &  52.0  &  1.1$\times 10^{-38}$   & 77.6  & 6.9$\times 10^{-41}$\\
 $c_{11}$   &  48.9  &  7.8$\times 10^{-44}$   & 64.5  & 6.9$\times 10^{-46}$\\
 $c_{12}$   &  50.4  &  7.6$\times 10^{-42}$   & 68.6  & 5.9$\times 10^{-44}$\\
 $c_{13}$   &  57.0  &  4.2$\times 10^{-36}$   & 85.1  & 1.6$\times 10^{-38}$\\
 $c_{14}$   &  47.4  &  1.0$\times 10^{-31}$   & 58.8  & 1.1$\times 10^{-33}$\\
 $c_{15}$   &  60.6  &  3.6$\times 10^{-37}$   & 93.3  & 8.8$\times 10^{-40}$\\
 \bottomrule
 \end{tabular}
 \label{tab:max_reach_r_strong}
 \end{table}

 \begin{table}
 \centering
 \caption{Most stringent constraints on the effective cross section $\sigma_{{\cal N},lim}$
   for each of the couplings in the effective Hamiltonian of
   Eq.(\ref{eq:H}) among the present and future experiments included in
   our analysis. In each case the ratio $c^n/c^p$ is fixed to the value
   that corresponds to the less stringent bound.}
 \begin{tabular}{*5c}
 \toprule
 Coupling &  \multicolumn{2}{c}{Present} & \multicolumn{2}{c}{Future}\\
 \midrule
 {}   & $m_{\chi}$ (GeV)   & $\sigma_{{\cal N},lim}$(cm$^2$)   & $m_{\chi}$ (GeV)   & $\sigma_{{\cal N},lim}$(cm$^2$)\\
 $c_1$   &  52.0 & 2.7$\times 10^{-43}$   & 72.9 & 2.2$\times 10^{-45}$\\
 $c_3$   &  57.0 & 2.1$\times 10^{-38}$  & 85.1  & 8.4$\times 10^{-41}$\\
 $c_4$   &  29.8  &  1.8$\times 10^{-40}$   & 37.0  & 2.2$\times 10^{-42}$\\
 $c_5$   &  57.0  &  6.0$\times 10^{-36}$   & 87.7  & 2.3$\times 10^{-38}$\\
 $c_6$   &  96.2  &  1.4$\times 10^{-34}$   & 207.8  & 1.7$\times 10^{-37}$\\
 $c_7$   &  31.7  &  1.6$\times 10^{-34}$   & 39.4  & 2.0$\times 10^{-36}$\\
 $c_8$   &  50.4  &  2.5$\times 10^{-38}$   & 72.9  & 1.8$\times 10^{-40}$\\
 $c_9$   &  50.4  &  3.3$\times 10^{-37}$   & 53.6  & 3.3$\times 10^{-39}$\\
 $c_{10}$   &  50.4  &  1.7$\times 10^{-37}$   & 162.4  & 9.2$\times 10^{-40}$\\
 $c_{11}$   &  39.4  &  3.2$\times 10^{-40}$   & 99.2  & 1.1$\times 10^{-42}$\\
 $c_{12}$   &  50.4  &  3.4$\times 10^{-40}$   & 72.9  & 2.5$\times 10^{-42}$\\
 $c_{13}$   &  52.0  &  2.9$\times 10^{-32}$   & 99.2  & 1.5$\times 10^{-35}$\\
 $c_{14}$   &  52.0  &  8.2$\times 10^{-31}$   & 55.3  & 8.2$\times 10^{-33}$\\
 $c_{15}$   &  60.6  &  1.6$\times 10^{-35}$   & 93.3  & 3.5$\times 10^{-38}$\\
 \bottomrule
 \end{tabular}
 \label{tab:max_reach_r_weak}
 \end{table}


 \section{Operator mixing}
 \label{sec:mixing}
In this Section we show how the results of the previous Section can be
used to estimate approximate limits also in the case of more than one
NR operator. To be quantitative, we consider here the specific
examples of two interaction Lagrangians between the DM
particle and quarks, valid at the scale $\Lambda$=2 GeV. In the notation of~\cite{bishara}:

\begin{equation}
  {\cal Q}^{(6)}_{2}=\sum_q \hat{{\cal C}}^{(6)}_{2,q}(\bar{\chi}\gamma_{\mu}\gamma_{5}\chi) (\bar{q}\gamma^{\mu}q),\,\,\,\,
  {\cal Q}^{(6)}_{3}=\sum_q \hat{{\cal C}}^{(6)}_{3,q}(\bar{\chi}\gamma_{\mu}\chi) (\bar{q}\gamma^{\mu}\gamma_5 q).
    \label{eq:rel_models}
  \end{equation}

\begin{figure}
\begin{center}
  \includegraphics[width=0.44\columnwidth]{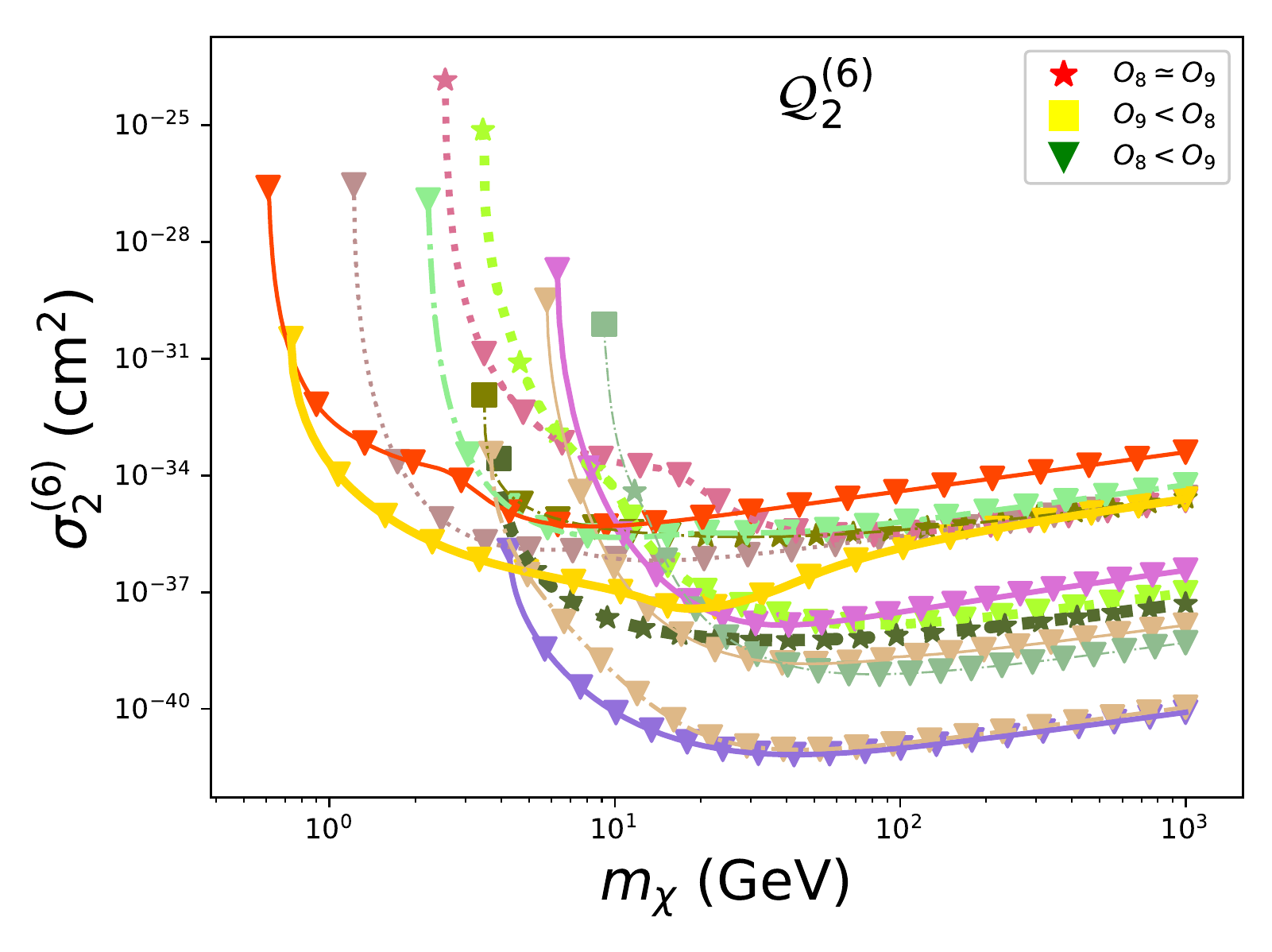}
  \includegraphics[width=0.44\columnwidth]{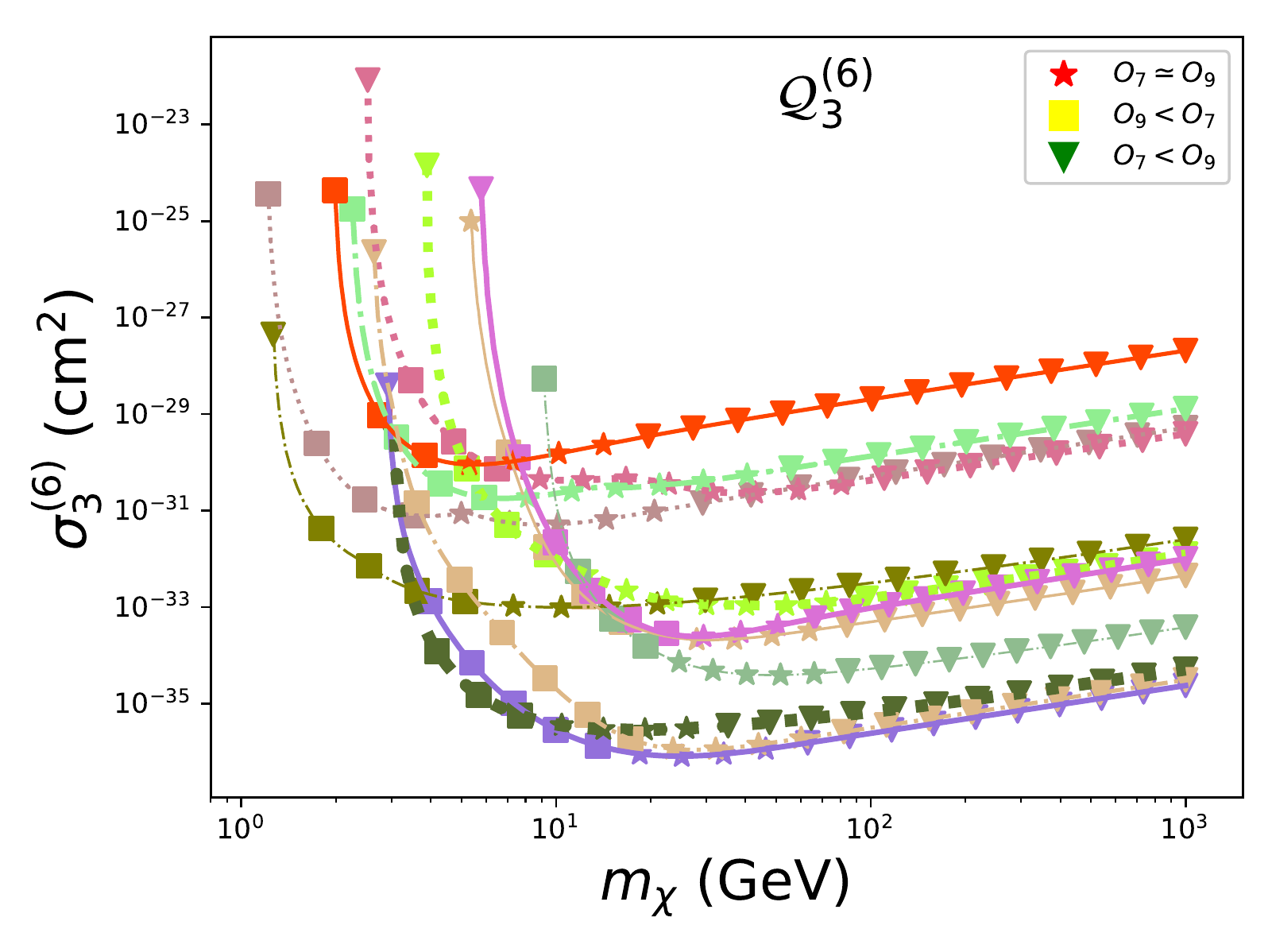}
  \includegraphics[width=0.10\columnwidth]{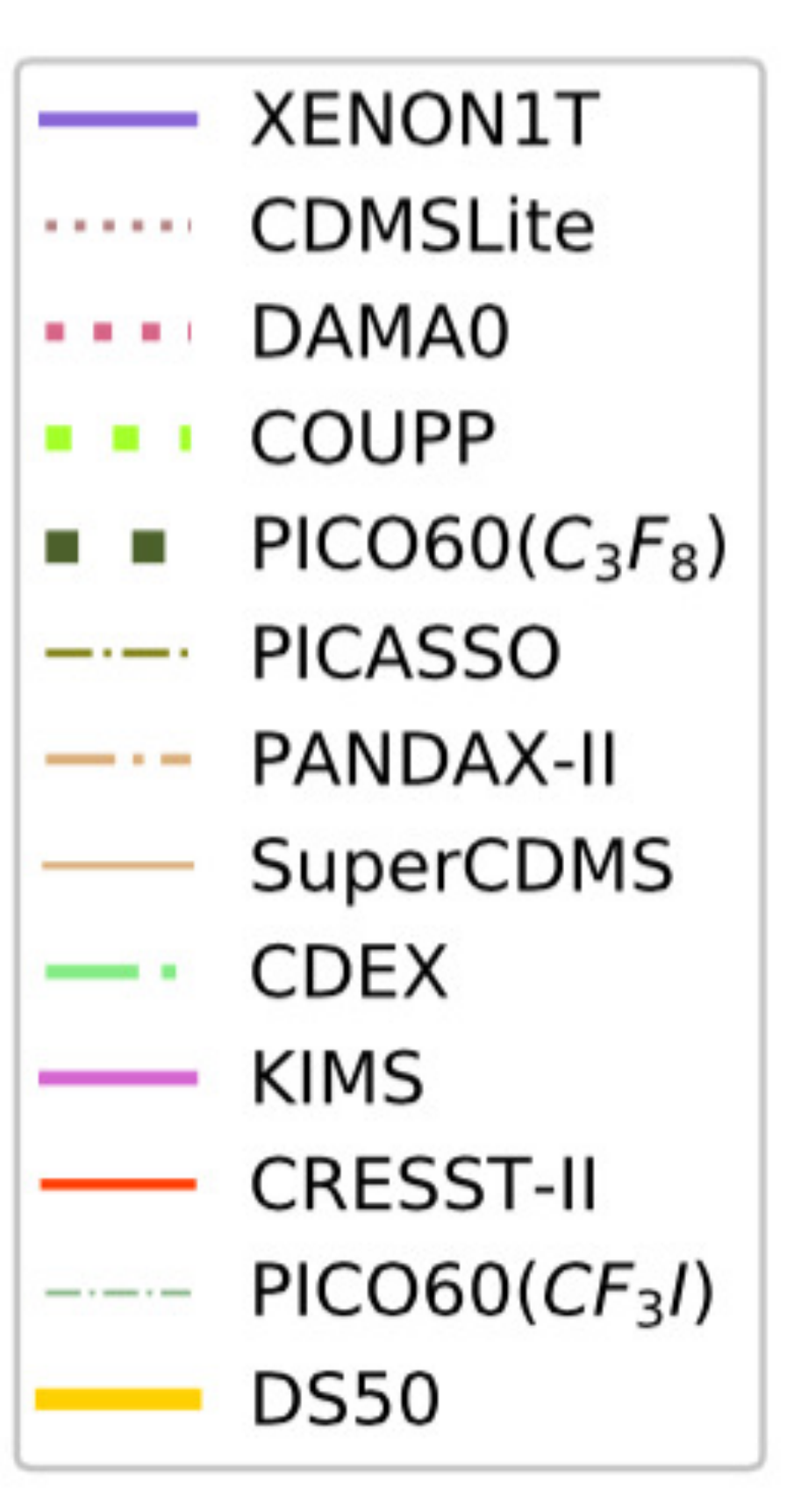}
\end{center}
\caption{Current 90\% C.L. exclusion plots to the effective cross
  sections $\sigma^{(6)}_2$=$\left [\hat{{\cal C}}^{(6)}_2\right ]^2
  \mu^2_{\chi{\cal N}}/\pi$ and $\sigma^{(6)}_3$=$\left [\hat{{\cal
        C}}^{(6)}_3\right ]^2 \mu^2_{\chi{\cal N}}/\pi$. Different markers
  show which operator drives the limit, ${\cal O}_i<{\cal O}_j$ meaning
  that the bound using only the contribution of ${\cal O}_i$ is at least
  a factor of 3 better than the one obtained including only
  ${\cal O}_j$.}
\label{fig:mixing}
\end{figure}

\noindent For simplicity, we assume in the following that the
couplings are same for all quarks, $\hat{{\cal
    C}}^{(6)}_{2,q}$=$\hat{{\cal C}}^{(6)}_{2}$, $\hat{{\cal
    C}}^{(6)}_{3,q}$=$\hat{{\cal C}}^{(6)}_{3}$. With this assumption
the non--relativistic limits of (\ref{eq:rel_models}) are given by:

\begin{eqnarray}
  {\cal Q}^{(6)}_{2}&\rightarrow& c_8{\cal O}_8+c_9 {\cal O}_9,\\
  {\cal Q}^{(6)}_{3}&\rightarrow& c_7{\cal O}_7+c_9^{\prime} {\cal O}_9,
  \label{eq:nr_limits}
  \end{eqnarray}

\noindent with $c_8=c_8^n=c_8^p=k_8\hat{{\cal C}}^{(6)}_{2}$,
$c_9=c_9^n=c_9^p= k_9\, \hat{{\cal C}}^{(6)}_{2}$,
$c_7=c_7^n=c_7^p=k_7\hat{{\cal C}}^{(6)}_{3}$, $c_9^{\prime}=c_9^{\prime
  n}=c_9^{\prime p}=k_9^{\prime}\hat{{\cal C}}^{(6)}_{3}$, with $k_8$=6,
$k_9$=4.89, $k_7$=-6 and $k_9^{\prime}$=$6
m_N/m_{\chi}$ \cite{bishara}. The corresponding calculation of the 90\% C.L.
upper bounds on the effective cross sections $\sigma^{(6)}_2$=$\left
[\hat{{\cal C}}^{(6)}_2\right ]^2 \mu^2_{\chi{\cal N}}/\pi$ and
$\sigma^{(6)}_3$=$\left [\hat{{\cal C}}^{(6)}_3\right ]^2
\mu^2_{\chi{\cal N}}/\pi$ are shown in Fig. \ref{fig:mixing}. In each
plot the different markers show which operator drives the constraint,
${\cal O}_i<{\cal O}_j$ meaning that the bound using only the
contribution of ${\cal O}_i$ is at least a factor of 3 better than the
one obtained including only ${\cal O}_j$. From such Figure one can see
that, depending on the value of the WIMP mass, the constraints of
different experiments can either be driven by the same NR operator or
by different ones. In any case the result of the full calculation in
Fig. \ref{fig:mixing} can be obtained with sufficient accuracy by
using the NR results of Section \ref{sec:analysis}. Let's assume the
value $m_{\chi}$=30 GeV (at larger masses all limits for ${\cal
  C}^{(6)}_2$ being driven by ${\cal O}_8$ and those for ${\cal
  C}^{(6)}_3$ by ${\cal O}_7$). To proceed, for a given value of the
WIMP mass one needs to obtain the corresponding bound on each NR
coupling from the contour plots in the $m_{\chi}$--$r$ planes of
Figs. \ref{fig:c1_plane}--\ref{fig:c15_plane}.  In particular the
relativistic theory predicts for each NR coupling the $r_i=c^n/c^p$
ratio (in our simple example they are both equal to one).
The rate on
experiment $exp$ can be written as:

\begin{equation}
  R_{exp}=\left(\hat{{\cal C}}^{(d)}_a\right)^2 \sum_i k_i^2 {\cal R}_{i,exp}(m_{\chi},r_i)
  =\sum_i c_i^2 {\cal R}_{i,exp}(m_{\chi},r_i)
  \end{equation}

\noindent where the sum over operators $i$ neglects interferences and
${\cal R}_{i,exp}$ is a response function that depends on the
experimental details. Assuming no cancellations in the sum the bound
on the coupling $\hat{{\cal C}}^{(d)}_a$ from the upper bound
$R_{exp,lim}$ on the count rate can be approximated by:

  \begin{equation}
\left(\hat{{\cal C}}^{(d)}_a\right)^2\lsim\min_{i,exp}\frac{R_{exp,lim}}{k_i^2 {\cal
          R}_{i,exp}(m_{\chi},r_i)}=\min_i \frac{1}{k_i^2}
      \min_{exp}\frac{R_{exp,lim}}{{\cal R}_{i,exp}(m_{\chi},r_i)}=
    \min_i \frac{1}{k_i^2}\left [c_i^{lim}(m_{\chi},r_i)\right]^2.
    \end{equation}

  \noindent The constraint:
  \begin{equation}
\left[c_i^{lim}(m_{\chi},r_i)\right ]^2=\min_{exp}\frac{R_{exp,lim}}{{\cal R}_{i,exp}(m_{\chi},r_i)},
    \end{equation}
\noindent can be read from the planes of
Figs.~\ref{fig:c1_plane}--\ref{fig:c15_plane}. Notice that this
procedure does not require that the same operator dominates at the
same WIMP mass in different experiments, because the $k_i$ quantities
depend on the interaction but not on the experiment.

In our specific example, for $m_{\chi} \simeq$ 30 GeV, $r$=1 the bound
for the operator ${\cal O}_7$ can be obtained from
Fig. \ref{fig:c7_plane} ($\sigma^7_{lim}\simeq$ 6.9$\times 10^{-35}$
cm$^2$), that for the operator ${\cal O}_8$ can be obtained from
Fig. \ref{fig:c8_plane} ($\sigma^8_{lim}\simeq$ 2.7$\times 10^{-40}$
cm$^2$) and that for the operator ${\cal O}_9$ can be obtained from
Fig. \ref{fig:c9_plane} ($\sigma^9_{lim}\simeq$ 5.5$\times 10^{-38}$
cm$^2$).  Then assuming dominance of one coupling at a time one gets
for model ${\cal Q}^{(6)}_{2}$ either $k_8^2
\sigma^{(6)}_2<\sigma^8_{lim}$ or $k_9^2
\sigma^{(6)}_2<\sigma^9_{lim}$. The most constraining of the two
bounds is $\sigma^{(6)}_2\simeq \sigma^8_{lim}/k_8^2\simeq 7.5\times
10^{-42}$ cm$^2$, in agreement to the result in Fig. \ref{fig:mixing}.
Proceeding in the same way, for model ${\cal Q}^{(6)}_{3}$ either
$k_7^2 \sigma^{(6)}_3<\sigma^7_{lim}$ or $(k_9^{\prime})^2
\sigma^{(6)}_3<\sigma^9_{lim}$. In this case the two bounds are
similar, $\sigma^{(6)}_3\simeq \sigma^7_{lim}/k_7^2\simeq 1.9\times
10^{-36}$ cm$^2$, $\sigma^{(6)}_3\simeq
\sigma^9_{lim}/(k_9^{\prime})^2\simeq 1.6\times 10^{-36}$
cm$^2$. Since in this case the two operators have similar
contributions this approximate bound is about a factor of 2 weaker
than the result in Fig. \ref{fig:mixing}, $\sigma^{(6)}_3\lsim$
8.4$\times$ 10$^{-37}$ cm$^2$.

Although in the present analysis we only considered the case of a
contact interaction, a similar procedure can be generalized in a
straightforward way to the case of relativistic couplings whose NR
limit leads to operators multiplied by a propagator $q^{-2}$. For
instance, the coupling
$1/q^2\bar{\chi}\sigma^{\mu\nu}\frac{q_{\nu}}{m_N}\chi
\bar{q}\gamma_{\mu}q$ (a DM particle coupling through a magnetic
dipole moment) leads to the combination $-\frac{1}{2 m_{\chi}} {\cal
  O}_1+\frac{2 m_N}{q^2} {\cal O}_5-2m_N(\frac{1}{m_N^2}{\cal O}_4-\frac{1}{q^2}{\cal
  O}_6)$. This would require to extend our systematic discussion in
Figs.\ref{fig:c1_plane}--\ref{fig:c15_plane} to the limits for the
generalized operator $q^{-2}{\cal O}_i$ and $q^{-4}{\cal O}_i$.

\section{Conclusions}
\label{sec:conclusions}
Assuming for WIMPs a Maxwellian velocity distribution in the Galaxy we
have explored in a systematic way the relative sensitivity of an
extensive set of 14 existing and 4 projected Dark Matter direct
detection experiments to each of the couplings that parameterize the
most general non-relativistic effective Hamiltonian allowed by
Galilean invariance for a contact interaction driving the elastic
scattering off nuclei of WIMPs of spin 1/2. We have performed our
analysis in terms of two free parameters: the WIMP mass and the ratio
between the WIMP-neutron and the WIMP-proton couplings $c^n/c^p$. For
each coupling we have provided contour plots in the
$m_{\chi}$--$c^n/c^p$ plane of the most stringent 90\% C.L. bound on
the WIMP--nucleon cross section and indicated with different shadings
the experiment providing the most constraining bound.
In~\ref{app:program} we will also introduce {\verb NRDD_constraints },
a simple interpolating code written in Python that allows to extract
the numerical value of the bound as a function of the WIMP mass
$m_{\chi}$ and of the coupling ratio $c^n/c^p$ for each NR coupling.

We found that 9 present experiments out of the total of 14 considered
in the present analysis provide the most stringent bound on some of
the effective couplings for a given choice of $(m_{\chi},c^n/c^p)$:
this is evidence of the complementarity of different target nuclei
and/or different combinations of count--rates and energy thresholds
when the search of a DM particle is extended to a wide range of
possible interactions.

In particular in our analysis we have adopted the approach of taking
all published result ``at face value'', and refrained from discussing
their relative robustness. It is however worth stressing out here
that, while the DarkSide 50 constraint appears from our analysis to be
quite competitive at low WIMP masses compared to other experiments, it
makes use of an ionization yield that has not been measured below 10
keV (see \ref{app:argon}).

In our analysis the lower part of the 2--sigma DAMA modulation
amplitude region in the $m_{\chi}$--$\sigma_p$ plane is included as if
it were an additional constraint, in order to locate in the parameter
space possible regions where the DAMA excess is compatible to other
constraints. DAMA does not appear as the most constraining bound in
any of the figures \ref{fig:c1_plane}--\ref{fig:c15_plane}, indicating
that an explanation of its annual modulation excess in terms of a WIMP
signals is in tension with the constraints of other experiments no
matter which of the effective operators among those in Eq.(\ref{eq:H})
is assumed to dominate in the WIMP--nucleus interaction. This result
is in agreement with the findings of Ref.~\cite{dama_2018_sogang}. While
in our analysis we assumed dominance of one NR operator at a time, we
have shown how our results can be used to estimate approximate limits
also in the case of interactions which depend on more than one NR
operator.

\section*{Acknowledgements}
This research was supported by the Basic Science
Research Program through the National Research Foundation of
Korea~(NRF) funded by the Ministry of Education, grant number
2016R1D1A1A09917964.

\appendix
\section{WIMP response functions}
\label{app:wimp_eft}

We collect here the WIMP particle--physics response functions introduced in Eq.(\ref{eq:squared_amplitude}) and adapted from \cite{haxton1,haxton2}:
\begin{eqnarray}
 R_{M}^{\tau \tau^\prime}\left(v_T^{\perp 2}, {q^2 \over m_N^2}\right) &=& c_1^\tau c_1^{\tau^\prime } + {j_\chi (j_\chi+1) \over 3} \left[ {q^2 \over m_N^2} v_T^{\perp 2} c_5^\tau c_5^{\tau^\prime }+v_T^{\perp 2}c_8^\tau c_8^{\tau^\prime }
+ {q^2 \over m_N^2} c_{11}^\tau c_{11}^{\tau^\prime } \right], \nonumber \\
 R_{\Phi^{\prime \prime}}^{\tau \tau^\prime}\left(v_T^{\perp 2}, {q^2 \over m_N^2}\right) &=& \left [{q^2 \over 4 m_N^2} c_3^\tau c_3^{\tau^\prime } + {j_\chi (j_\chi+1) \over 12} \left( c_{12}^\tau-{q^2 \over m_N^2} c_{15}^\tau\right) \left( c_{12}^{\tau^\prime }-{q^2 \over m_N^2}c_{15}^{\tau^\prime} \right)\right ]\frac{q^2}{m_N^2},  \nonumber \\
 R_{\Phi^{\prime \prime} M}^{\tau \tau^\prime}\left(v_T^{\perp 2}, {q^2 \over m_N^2}\right) &=& \left [ c_3^\tau c_1^{\tau^\prime } + {j_\chi (j_\chi+1) \over 3} \left( c_{12}^\tau -{q^2 \over m_N^2} c_{15}^\tau \right) c_{11}^{\tau^\prime }\right ] \frac{q^2}{m_N^2}, \nonumber \\
  R_{\tilde{\Phi}^\prime}^{\tau \tau^\prime}\left(v_T^{\perp 2}, {q^2 \over m_N^2}\right) &=&\left [{j_\chi (j_\chi+1) \over 12} \left ( c_{12}^\tau c_{12}^{\tau^\prime }+{q^2 \over m_N^2}  c_{13}^\tau c_{13}^{\tau^\prime}  \right )\right ]\frac{q^2}{m_N^2}, \nonumber \\
   R_{\Sigma^{\prime \prime}}^{\tau \tau^\prime}\left(v_T^{\perp 2}, {q^2 \over m_N^2}\right)  &=&{q^2 \over 4 m_N^2} c_{10}^\tau  c_{10}^{\tau^\prime } +
  {j_\chi (j_\chi+1) \over 12} \left[ c_4^\tau c_4^{\tau^\prime} + \right.  \nonumber \\
 && \left. {q^2 \over m_N^2} ( c_4^\tau c_6^{\tau^\prime }+c_6^\tau c_4^{\tau^\prime })+
 {q^4 \over m_N^4} c_{6}^\tau c_{6}^{\tau^\prime } +v_T^{\perp 2} c_{12}^\tau c_{12}^{\tau^\prime }+{q^2 \over m_N^2} v_T^{\perp 2} c_{13}^\tau c_{13}^{\tau^\prime } \right], \nonumber \\
    R_{\Sigma^\prime}^{\tau \tau^\prime}\left(v_T^{\perp 2}, {q^2 \over m_N^2}\right)  &=&{1 \over 8} \left[ {q^2 \over  m_N^2}  v_T^{\perp 2} c_{3}^\tau  c_{3}^{\tau^\prime } + v_T^{\perp 2}  c_{7}^\tau  c_{7}^{\tau^\prime }  \right]
       + {j_\chi (j_\chi+1) \over 12} \left[ c_4^\tau c_4^{\tau^\prime} +  \right.\nonumber \\
       &&\left. {q^2 \over m_N^2} c_9^\tau c_9^{\tau^\prime }+{v_T^{\perp 2} \over 2} \left(c_{12}^\tau-{q^2 \over m_N^2}c_{15}^\tau \right) \left( c_{12}^{\tau^\prime }-{q^2 \over m_N^2}c_{15}^{\tau \prime} \right) +{q^2 \over 2 m_N^2} v_T^{\perp 2}  c_{14}^\tau c_{14}^{\tau^\prime } \right], \nonumber \\
     R_{\Delta}^{\tau \tau^\prime}\left(v_T^{\perp 2}, {q^2 \over m_N^2}\right)&=& {j_\chi (j_\chi+1) \over 3} \left( {q^2 \over m_N^2} c_{5}^\tau c_{5}^{\tau^\prime }+ c_{8}^\tau c_{8}^{\tau^\prime } \right)\frac{q^2}{m_N^2}, \nonumber \\
 R_{\Delta \Sigma^\prime}^{\tau \tau^\prime}\left(v_T^{\perp 2}, {q^2 \over m_N^2}\right)&=& {j_\chi (j_\chi+1) \over 3} \left (c_{5}^\tau c_{4}^{\tau^\prime }-c_8^\tau c_9^{\tau^\prime} \right) \frac{q^2}{m_N^2}.
\label{eq:wimp_response_functions}
\end{eqnarray}

\section{Experiments}
  \label{app:exp}
In the present analysis we include an extensive set of constraints
that are representative of the different techniques used to search for
DM: XENON1T~\cite{xenon_2018}, PANDAX-II~\cite{panda_2017},
KIMS~\cite{kims_2014}, CDMSlite~\cite{cdmslite_2017},
SuperCDMS~\cite{super_cdms_2017}, COUPP~\cite{coupp},
PICASSO~\cite{picasso}, PICO-60 (using a $CF_3I$ target
~\cite{pico60_2015} and a $C_3F_8$ one \cite{pico60}) CRESST-II
\cite{cresst_II,cresst_II_ancillary}, DAMA (modulation data
\cite{dama_1998, dama_2008,dama_2010,dama_2018} and average count rate
\cite{damaz}), CDEX \cite{cdex} and DarkSide--50
\cite{ds50}. We also consider projected sensitivities of some future
detectors: LZ~\cite{LZ}, PICO-500 \cite{pico500} and COSINUS
\cite{cosinus}. In the following, if not specified otherwise we adopt
for the energy resolution a Gaussian form, ${\cal
  G}(E^{\prime},E_{ee})=Gauss(E^{\prime}|E_{ee},\sigma)=1/(\sqrt{2\pi}\sigma)exp(-(E^{\prime}-E_{ee})/2\sigma^2)$.
The quenching factor of bolometers (SuperCDMS, CRESST-II, COSINUS) is
assumed to be equal to 1.

\subsection{Xenon: XENON1T, PANDAX-II and LZ}

For XENON1T we have assumed 7 WIMP candidate events in the range of
3PE $ \le S_1 \le $ 70PE, as shown in Fig.~3 of Ref.~\cite{xenon_2018}
for the primary scintillation signal S1 (directly in Photo Electrons,
PE), with an exposure of 278.8 days and a fiducial volume of 1.3 ton
of xenon. We have used the efficiency taken from Fig.~1
of~\cite{xenon_2018} and employed a light collection efficiency
$g_1$=0.055; for the light yield $L_y$ we have extracted the best
estimation curve for photon yields $\langle n_{ph} \rangle /E$ from
Fig.~7 in~\cite{xenon_2018_quenching} with an electric field of
$90~{\rm V/cm}$.

On the other hand for PANDAX-II we implemented the combined result of
Run 9 and Run 10 with $\simeq$ 0.2 events after background subtraction
in the range 3 PE$\le S_1 \le$ 45 PE in the lower half of the
signal band, as shown in Fig.4, for a total exposure of 79.6+77.1 days
and a fiducial mass of 361.5 kg~\cite{panda_2017}. We have taken the
efficiency from Fig.16 and $L_y$ from Fig.13b of the supplemental
material provided in \cite{panda_2017}. To reproduce the published
PANDAX-II combined Run 9 and Run 10 result we adopted a photon gain
$g_1$=0.0557.

LUX--ZEPLIN (LZ) is a next generation dual--phase xenon DM direct
detection experiment which will operate with an active mass of 7
tonnes. We assumed an exposure of $5.6\times10^6$ kg days \cite{LZ}.
Its sensitivity to low mass WIMPs will depend strongly on the low
energy nuclear recoil efficiency. To obtain projections in
Ref. \cite{LZ} an extrapolation down to 0.1 keV following Lindhard
theory is used. Lacking any direct measurement of this quantity at low
energy we use the light yield of Fig.2 of \cite{LZ} with a hard cutoff
at 1.1 keV. We assume no signal in the lower half nuclear recoil band
below the red curve of Fig.7 in \cite{LZ} and a neutrino background of
12 events for $S_1<$ 4 PE.

For XENON1T, PANDAX-II and LZ experiments we have modeled the energy resolution
combining a Poisson fluctuation of the observed primary signal $S_1$
compared to $<S_1>$ and a Gaussian response of the photomultiplier
with $\sigma_{PMT}=0.5$, so that:

\begin{equation}
{\cal G}_{Xe}(E_R,S)=\sum_{n=1}^{\infty}
Gauss(S|n,\sqrt{n}\sigma_{PMT})Poiss(n,<S(E_R)>),
\label{eq:g_xe}  
\end{equation}

\noindent with $Poiss(n,\lambda)=\lambda^n/n!exp(-\lambda)$.

\subsection{Argon: DarkSide-50}
\label{app:argon}
The analysis of DarkSide-50 \cite{ds50} is based on the ionization
signal extracted from liquid argon with an exposure of 6786.0 kg
days. The measured spectrum for $N_{e^-}<$ 50 (with $N_{e^-}$ the
number of extracted electrons) is shown in Fig. 7 of \cite{ds50}, and
shows an excess for 4 $<N_{e^-}<$7 $N_{e^-}$ compared to a simulation
of the background components from known radioactive contaminants.
Following Ref.\cite{ds50} we have subtracted the background minimizing
the likelihood function: 

\be -2 {\cal L} =\sum_i \frac{(\sigma S_i+\rho
  b_i-x_i)^2}{\sigma_i^2},
\label{eq:bck_chi2}
\ee

\noindent where $i$ represents the energy bin, $x_i$ the measured
spectrum with error $\sigma_i$, while $\sigma S_i$ and $\rho b_i$ are
the DM signal and the background, respectively, with $\sigma$ and
$\rho$ arbitrary normalization factors ($\sigma$ is identified with
the effective WIMP-proton cross section $\sigma_p$). In particular we
obtain the 90\% C.L. upper bound on $\sigma_p$ by taking its profile
likelihood with $-2 {\cal L}- [-2 {\cal L}]_{min}=n^2$ and
$n$=1.28. We take $x_i$, $\sigma_i$ and $b_i$ from Fig.7
of~\cite{ds50}. The ionization yield of argon has been measured only
down to $\lsim$ 10 keVnr, while DS50 uses a model fit to calibration
data. We use the latter as taken from Fig. 6 of \cite{ds50} with a
hard cut at 0.15 keVnr, the lowest energy for which it is provided. We
take the efficiency from Fig. 1 of \cite{ds50}.  The signal/background
ratio in DS50 is at the percent level, so the background subtraction
procedure is sensitive to the details of its implementation. Following
the background subtraction procedure described above we reproduce
fairly well the published DS50 exclusion plot with the exception of
the mass range 4 GeV$\lsim m_{\chi}\lsim$15 GeV, where our procedure
is less constraining for a SI interaction by a factor up to
$\simeq$3.5 compared to the published one. Given the large
uncertainties involved we consider such result acceptable.
\subsection{Germanium: SuperCDMS, CDMSlite and CDEX}
The latest SuperCDMS analysis \cite{super_cdms_2017} observed 1 event
between 4 and 100 keVnr with an exposure of 1690 kg days. We have
taken the efficiency from Fig.1 of \cite{super_cdms_2017} and the
energy resolution $\sigma=\sqrt{0.293^2+0.056^2 E_{ee}}$ from
\cite{cdms_resolution}. To analyze the observed spectrum we apply the
optimal interval method \cite{yellin}.

For CDMSlite we considered the energy bin of 0.056 keV$<E^{\prime}<$
1.1 keV with a measured count rate of 1.1$\pm$0.2 [keV kg day]$^{-1}$
(Full Run 2 rate, Table II of Ref. \cite{cdmslite_2017}). We have
taken the efficiency from Fig.4 of \cite{cdmslite_2017} and the energy
resolution $\sigma=\sqrt{\sigma_E^2+B E_R+(A E_R)^2}$, with
$\sigma_E$=9.26 eV, $A$=5.68$\times 10^{-3}$ and $B$=0.64 eV from
Section IV.A of~\cite{cdmslite_2017}.

CDEX~\cite{cdex} uses a germanium target with an exposure of 737.1 kg
days. We analyze the residual excess events for 160 eVee
$<E^{\prime}<$ 2.56 keVee detected in the Anti-Compton Veto spectrum
of Fig. 7 in~\cite{cdex} and the efficiency from Fig.4
of~\cite{cdex}. For the quenching factor of germanium we use the
Lindhard formula \cite{lindhard}:
\begin{equation}
	Y(E_R) = \frac{k\cdot g{\left(\epsilon\right)}}{1+k\cdot g{\left(\epsilon\right)}},
    \label{eq:lindhard}
\end{equation}
where
$g{\left(\epsilon\right)}=3\epsilon^{0.15}+0.7\epsilon^{0.6}+\epsilon$,
$\epsilon=11.5E_R{\left(keV_{nr}\right)} Z^{-7/3}$, and $Z$ the atomic
number. For germanium, $k=0.157$ and $Z=32$.

\subsection{Fluorine: PICASSO, PICO-60 and PICO-500}

Bubble chambers are threshold experiments for which we employ the
nucleation probability:

\begin{equation}
{\cal P}_T(E_R)=1-\exp\left [-\alpha_T\frac{E_R-E_{th}}{E_{th}} \right ].
\label{eq:nucleation_probability}
\end{equation}

\noindent The PICASSO experiment \cite{picasso} uses $C_4 F_{10}$ as a
target and operated its runs with six energy thresholds. For each
threshold we provide the corresponding number of observed events and
statistical fluctuations in Table \ref{table:picasso} (extracted from
Fig. 4 of Ref.~\cite{picasso}). For the nucleation probability we used
Eq.(\ref{eq:nucleation_probability}) with $\alpha_C$=$\alpha_F$=5.

\begin{table}[t]
\begin{center}
{\begin{tabular}{@{}|c|c|c|c|c|@{}}
\hline
$E_{th}$ (keV) & Event rate (events/kg/day) & Fluctuation \\
\hline
1.0 &   -1.5 & 3.8 \\
1.5 &    -0.2 & 1.0 \\
2.7 &     0.3 & 0.8 \\
6.6 &     -0.8 & 1.8 \\
15.7 &    -1.4 & 2.3 \\
36.8 &     0.3 &1.0 \\
 \hline
\end{tabular}}
\caption{Observed number of events and 1--sigma statistical
  fluctuations (extracted from Fig. 4 of
  Ref. \cite{picasso}) for each operating threshold used in PICASSO.
  \label{table:picasso}}
\end{center}
\end{table}

The target material of PICO-60 and PICO-500 is $C_3F_8$. For
PICO-60~\cite{pico60} only the threshold $E_{th}$=3.3 keV is analyzed
with a total exposure of 1167.0 kg days and no event detected. We have
assumed the nucleation probability in Fig. 4 of \cite{pico2l}.

PICO-500 is a projected future extension of PICO-60~\cite{pico500} with
250 liters of fiducial volume. We have assumed an exposure of 6 months
at the energy threshold $E_{th}$=3.2 keV and an exposure of one year
at the energy threshold $E_{th}$=10 keV. We assumed no candidate
events, and the same nucleation probabilities of PICO-60.

\subsection{Fluorine+Iodine: COUPP, PICO-60 and PICO-500}
COUPP is bubble chamber using a $CF_3I$ target.  For each operating
threshold used in COUPP the corresponding exposure and number of
measured events are summarized in Table \ref{table:coupp}. For
fluorine and carbon we use the nucleation probability of
Eq.(\ref{eq:nucleation_probability}) with $\alpha$=0.15. For iodine we
adopt instead a step function with nucleation probability equal to 1
above the energy threshold.

\begin{table}[t]
\begin{center}
{\begin{tabular}{@{}|c|c|c|c|@{}}
\hline
$E_{th}$ (keV) & exposure (kg day) & measured events  \\
\hline
7.8 & 55.8  & 2 \\
11 & 70   & 3 \\
15.5 & 311.7  & 8 \\
\hline
\end{tabular}}
\caption{The operating thresholds with corresponding exposures and measured events for COUPP. \label{table:coupp}}
\end{center}
\end{table}

PICO-60 can also employ a $CF_3I$ target. For the analysis of
Ref.\cite{pico60_2015} we adopt an energy threshold of 13.6 keV and an
exposure of 1335 kg days. The nucleation probabilities for each target
element are taken from Fig.4 in~\cite{pico60_2015}.

PICO-500 is also planned to use a $CF_3I$ target~\cite{pico500}. As in
the case of $C_3F_8$ we adopt an exposure of 6 months for $E_{th}$=3
keV and of one year for $E_{th}$=10 keV, with no candidate events. We
have taken the nucleation probabilities from Fig.4 of \cite{pico60_2015}.

\subsection{Sodium Iodide: DAMA, KIMS and COSINUS}
For DAMA we consider both the upper bound from the average count rate
(DAMA0) and the latest result for the annual modulation amplitudes. In
particular we include the lower part of the 2--sigma modulation
amplitude region in the $m_{\chi}$--$\sigma_p$ plane in the analysis
of the most stringent bound of Section \ref{sec:analysis} as if it
were an additional constraint, in order to locate possible regions of
compatibility between the DAMA excess and other constraints in the
parameter space. For DAMA0 we have taken the average count rates from
\cite{damaz} (rebinned from 0.25-keVee- to 0.5-keVee-width bins) from
2 keVee to 8 keVee. We use the DAMA modulation amplitudes normalized
to kg$^{-1}$day$^{-1}$keVee$^{-1}$ in the energy range 1 keVee
$<E^{\prime}<$ 8 keVee from Ref.\cite{dama_2018}. In both cases we
assume a constant quenching factors $q$=0.3 for sodium and $q$=0.09
for iodine, and the energy resolution $\sigma$ = 0.0091
(E$_{ee}$/keVee) + 0.448 $\sqrt{E_{ee}/{\rm keVee}}$ in keV.

The KIMS collaboration operated caesium iodine scintillators with an
exposure of 24524.3 kg days \cite{kims_2014}. We obtained the energy
resolution extrapolating the two calibration peaks in Fig.2
of~\cite{kims_2014} at lower energy using the energy dependence
$\sigma$=$\sqrt{a+b E_{ee}}$, while we have used the efficiency from
Fig. 1(a) and the measured spectrum and background estimate in the
region of interest 2 keVee $<E^{\prime}<$ 4 keVee from Fig.1(b) of the
same paper. We have adopted the quenching factors for both targets
from \cite{kims_quenching}. We have applied background subtraction
using the same procedure described for DarkSide-50 and the likelihood
of Eq.~(\ref{eq:bck_chi2}).

COSINUS\cite{cosinus} is a next--generation cryogenic scintillating
calorimeter using the same targets as DAMA with discrimination between
electron and nuclear recoils to suppress the background. We follow the
analysis of \cite{cosinus2} assuming 5 events with exposure 105 kg
days and an energy threshold of 1 keV. We have also assumed the energy
resolution $\sigma$ = 0.2 keV and taken the nuclear recoil detection
efficiency from Eqs. (2.8) and (2.9) of \cite{cosinus2}.  
\subsection{$CaWO_4$: CRESST-II}
CRESST-II measures heat and scintillation using $CaWO_4$ crystals. We
considered the Lise module analysis from \cite{cresst_II} with energy
resolution $\sigma$=0.062 keV and detector efficiency from Fig. 4
of~\cite{cresst_II_description}. For our analysis we have selected 15
events for 0.3 keVnr$<E_R<$ 0.49 keVnr with an exposure of 52.15 kg
days.

\section{Nuclear response functions for Caesium and Tungsten}
\label{app:nuclear_response_functions}

In the case of Caesium in KIMS and of Tungsten in CRESST-II a shell
model calculation for the nuclear response functions
$W^{\tau\tau^{\prime}}_{l}$ is not available from
Refs.\cite{haxton2,catena}. As far as $W^{\tau\tau^{\prime}}_{M}$ is
concerned we simply assume a nuclear cross section scaling with the
square of the target mass number and a Helm form factor $F_{Helm}(qr)$
\cite{helm} (we take the parametrization of the nuclear radius $r$
from \cite{helm_r1}), which, with the normalization conventions of
Ref.~\cite{haxton2} corresponds for target $T$ to
$W^{\tau\tau^{\prime}}_{M}(q)$=$(2j_T+1.)/(16\pi)A_T^2 F_{Helm}(qr)$.
On the other hand, in the case of
$W^{\tau\tau^{\prime}}_{\Sigma^{\prime\prime}}$ and
$W^{\tau\tau^{\prime}}_{\Sigma^{\prime}}$ we assume
$W^{\tau\tau^{\prime}}_{\Sigma^{\prime}}= 2
W^{\tau\tau^{\prime}}_{\Sigma^{\prime\prime}}$ and use a Gaussian
approximation for the $q^2$ dependence. In particular, combining the
usual spin--dependent scaling law written as \cite{engel_spin}:

\begin{equation}
S(0)=\frac{1}{\pi}\frac{(2 j_T+1)(j_T+1)}{j_T}\left (a_p <S_p>+a_n <S_n> \right )^2
\end{equation}

\noindent with the Gaussian form factor \cite{spin_belanger}:

\begin{equation}
\frac{S(q^2)}{S(0)}=e^{-q^2 R^2/4},\,\,\,\, R=\left(0.92 A_T^{1/3}+2.68-0.78\sqrt{(A_T^{1/3}-3.8)^2+0.2} \right)\,\,\mbox{fm},
\end{equation}

\noindent implies:

\begin{eqnarray}
  W^{\tau\tau^{\prime}}_{\Sigma^{\prime\prime}}(q^2)&=& \frac{4}{3\pi}\frac{(2 j_T+1)(j_T+1)}{j_T} <S^{\tau}><S^{\tau^{\prime}}>e^{-q^2 R^2/4} \nonumber\\
  W^{\tau\tau^{\prime}}_{\Sigma^{\prime}}(q^2)&=& \frac{8}{3\pi}\frac{(2 j_T+1)(j_T+1)}{j_T} <S^{\tau}><S^{\tau^{\prime}}>e^{-q^2 R^2/4},
\label{eq:w_approx_cl}
\end{eqnarray}
\noindent with $<S^0>=(<S_p>+<S_n>)/2$ and
$<S^1>=(<S_p>-<S_n>)/2$. For $^{183}$W and $^{333}$Cs we take
$<S_p>$=0, $<S_n>$=-0.17, and $<S_p>$=-0.37, $<S_n>$=0.003,
respectively, both from Appendix C of
Ref.~\cite{nuclear_spin_averages}.

\section{The program}
\label{app:program}
The {\verb NRDD_constraints } code provides a simple interpolating
function written in Python that for a given NR effective coupling
calculates the most constraining limit among the experiments listed
in~\ref{app:exp} on the effective WIMP--nucleon cross section
$\sigma_{\cal N}$ defined in Eq.(\ref{eq:conventional_sigma_nucleon})
as a function of the WIMP mass $m_{\chi}$ and of the ratio
$r=c^{n}/c^{p}$.  The code requires the {\verb SciPy }
package and contains only four files, the code {\verb NRDD_constraints.py },
two data files {\verb NRDD_data1.npy } and
{\verb NRDD_data2.npy }, and a driver template {\verb NRDD_constraints-example.py }.
The module can be downloaded from
\begin{center}
{\verb https://github.com/NRDD-constraints/NRDD } 
\end{center} or cloned by 
\begin{Verbatim}[frame=single,xleftmargin=1cm,xrightmargin=1cm,commandchars=\\\{\}]
  git clone https://github.com/NRDD-constraints/NRDD
\end{Verbatim}
By typing:

\begin{Verbatim}[frame=single,xleftmargin=1cm,xrightmargin=1cm,commandchars=\\\{\}]
  import NRDD_constraints as NR
\end{Verbatim}

\noindent two functions are defined. The function {\verb sigma_nucleon_bound(inter,mchi,r) }
returns the upper bound
$(\sigma_{\cal N})_{lim}$ on the effective cross section of
Eq.(\ref{eq:conventional_sigma_nucleon}) in cm$^2$ as a function of
the WIMP mass {\verb mchi } and of the ratio {\verb r }=$r$ in the
ranges $0.1 \mbox{ GeV}<m_{\chi}<1000$ GeV, $-10^4 <r<10^4$, and
contains the same information of
Figs. \ref{fig:c1_plane}--\ref{fig:c15_plane}.  The {\verb inter } parameter
is a string that selects the interaction term and can be chosen in the
list provided by the second function {\verb print_interactions() }:


\begin{Verbatim}[frame=single,xleftmargin=1cm,xrightmargin=1cm,commandchars=\\\{\}]
  NR.print_interactions()\\     
  ['O1_O1','O3_O3', 'O4_O4', 'O5_O5', 'O6_O6',\\ 
   'O7_O7', 'O8_O8', 'O9_O9', 'O10_O10', 'O11_O11',\\
   'O12_O12', 'O13_O13', 'O14_O14', 'O15_O15'\\
   'O5_O5_qm4', 'O6_O6_qm4', 'O11_O11_qm4'] 
\end{Verbatim}

\noindent The list above includes also a few examples of long--range
interactions where the NR coupling is divided by a propagator term.














\end{document}